\documentclass[12pt]{article}
\usepackage[T2A]{fontenc}
\usepackage{xcolor}
\usepackage{epsf}
\usepackage[utf8]{inputenc}
\usepackage[russian,english]{babel}
\usepackage{amsmath}
\usepackage{amsfonts}
\usepackage{amssymb}
\usepackage{graphicx}
\usepackage{float}
\usepackage{changes}
\usepackage{comment}
\usepackage{authblk}
\usepackage{enumitem}
\usepackage{setspace} 
\usepackage{multicol}
\usepackage[
  a4paper,
  top=1cm,
  bottom=1cm,
  left=1cm,
  right=1cm,
  nohead,
  nofoot,
]{geometry}
\def\apj{ApJ}
\def\apjl{ApJL}
\def\apjs{ApJ Suppl. Ser.}
\def\apss{Astroph.Sp.Sci.}
\def\aap{A\&A}
\def\aaps{A\&A Suppl. Ser.}
\def\aapr{A\&AR}
\def\araa{ARAA}
\def\mnras{MNRAS}
\def\nat{Nature}
\def\pasa{PASA}

\def\pasp{PASP}

\def\zap{Zs. f. Ap.}

\def\beq#1{\begin{equation}\label{#1}}
\def\eeq{\end{equation}}
\def\beqa#1{\begin{eqnarray}\label{#1}}
\def\eeqa{\end{eqnarray}}

\def\myfrac#1#2{\left(\frac{#1}{#2}\right)}
\def\comment#1{\relax}

\def\apgt{\ {\raise-.5ex\hbox{$\buildrel>\over\sim$}}\ }
\def\aplt{\ {\raise-.5ex\hbox{$\buildrel<\over\sim$}}\ }
\newcommand{\ms}{$M_\odot$}
\newcommand{\msun}{$M_\odot$}

\newcommand{\ace}{\mbox {$\alpha_{ce}$}}
\newcommand{\al}{\mbox {$\alpha_{ce}\times\lambda$}}
\newcommand{\mbh}{\mbox {$M_\mathrm {BH}$}}    

\newcommand{\bux}{\mbox {$\mathrm{BHULX} $}}
\newcommand{\nux}{\mbox {$\mathrm{NULX} $}}
\newcommand{\te}{\mbox {$T_\mathrm{eff}$}}

\title{
{\bf POPULATIONS OF ULTRALUMINOUS X-RAY SOURCES IN GALAXIES: ORIGIN AND EVOLUTION}}

\author[1]{A.G. Kuranov \thanks{alexandre.kuranov@gmail.com}}
\author[1,2]{K.A. Postnov}
\author[3]{L.R. Yungelson}
\affil[1]{Sternberg Astronomical Institute, Lomonosov Moscow University}
\affil[2]{Kazan Federal University}
\affil[3]{Institute of Astronomy, Russian Academy of Sciences}
\begin{document}

\maketitle
\begin{abstract}
Employing hybrid population synthesis, a model of the population of
ultraluminous X-ray sources (ULX) in the binary systems with a black hole (BH)
accretors is computed. It is compared to the model of the population of ULX with
magnetized neutron stars (NS) that can be observed as pulsating ULX (Kuranov et
al. 2020). A model of formation of BH is considered, in which their mass is
determined by the mass of stellar CO core immediately before the collapse
($M_\mathrm {CO}$), as well as ``delayed'' and ``rapid'' collapse models (Fryer
et al. 2012). Possible transiency of ULX due to accretion disks instability is
taken into account The parameters and evolution of ULX are computed for the
galaxies with constant star formation rate (SFR) and for the ones formed by an
instantaneous star formation burst. The maximum number of ULX with BH ($\sim
10$) is reached in the galaxies with stationary $ SFR=10$\msun/yr in $\sim 1$
Gyr after beginning of star formation. ULX which are observed after the end of star
formation, are binaries, in which BH and/or NS formed before the completion of
star formation, while long-living donors with the mass $\sim$\,\msun\ continue
RLOF or even fill their Roche lobes later. In several Gyr after completion of
star formation the number of ULX in the galaxies with mass $M_G=10^{10}$\,\msun\
becomes less than 1 per 10 galaxies, most of them are ULX with NS. In ULX with
NS, regardless of the adopted SFR model, dominate persistent sources with the
donor overflowing Roche lobe. The number of transient sources is by more than an
order of magnitude lower. Wind-accreting ULX are by an order of magnitude
more rare than the sources with accretion via RLOF.
\end{abstract}
   
DOI: 10.31857/S0320010821120020

\onecolumn
\section*{Introduction}
\label{s:intro}

Ultraluminous X-ray sources (ULX) with apparent luminosity exceeding
Eddington's one for compact objects --- neutron stars (NS) and black holes (BH)
of stellar masses (hereinafter --- c.o.), have been in the spotlight of
astrophysical research for several decades. The interest in them is caused by
the need to understand such powerful electromagnetic radiation due to
accretion, which, in particular, may indicate the presence of unusually massive
($100-1000)\,M_\odot$ BH in binaries, the so-called ``intermediate-mass BH''
(Colbert and Mushotsky, 1999). Such BH are interesting from the viewpoint of the
origin and evolution of supermassive BH in the galactic nuclei (see discussion
in the reviews by Cherepashchuk (2016) and Volonteri et al. (2021)). ULX are
discovered in galaxies of all types (see, for instance, Bernadich et al., 2021)
and their lists are permanently growing. The most complete at the time of writing
catalog (Walton et al., 2021) contains 1843 ULX candidates in 951 galaxies.

Fabrika and Mescheryakov (2001) and King et al. (2001) independently suggested
that apparent super-Eddington luminosity of ULX results from  
beaming of the radiation of supercritical accretion disks around c.o. of stellar
mass. The discovery of pulsed X-rays from ULX (Bachetti et al. 2014)
confirmed that not only BH but also magnetized NS in close binary systems can
be accreting components of ULX. The nature of ULX is actively discussed, see, e.g.,
the reviews by Kaaret et al. (2017) and Fabrika et al. (2021).

This paper continues the study by Kuranov et al. (2020, Paper I), in which the
population of ULX with accreting magnetized neutron stars (NULX) in a spiral
galaxy similar to the Milky Way was considered. In the present paper, we discuss
ULX with accretors --- stellar-mass BH (BHULX, \mbh\ below several dozens of
\ms) and magnetized NS in the model galaxies with continuous and instantaneous
star formation (SF). Former galaxies can serve as the proxies for spiral
galaxies and the latter --- for elliptical ones. Thus, we have investigated
\emph{almost complete} models of ULX populations (except for the still
hypothetical sources with intermediate-mass BH and ULX in stellar clusters). In
the context of this paper, we consider all high-luminosity sources of X-ray
radiation with accreting NS and call them \nux, regardless of whether they can
be observed as pulsating ULX.

By the instant of the core collapse, massive stars in close binary systems (CBS)
almost completely lose their hydrogen and helium envelopes (e.g., Tutukov et
al., 1973; Laplace et al., 2021). Decrease of the mass of the collapsing core
due to the neutrino losses leads to the loss of hydrostatic equilibrium by
stellar envelopes and ejection of a fraction of their matter; the source of
energy is the recombination of hydrogen (``Nadezhin-Lovegrove effect'',
Nadezhin, 1980; Lovegrove and Woosley, 2013). Having in mind this effect, we
considered a model in which it was assumed that the gravitational mass of the
resulting black hole is 90\% of the baryonic mass of the CO
core of the presupernova (hereinafter referred to as model C). Taking into
account the existing uncertainties regarding the mechanism of the stellar
collapse, we also considered and compared ULX populations in which BH formation
occurs via ``delayed'' (model D) and ``rapid'' (model R) mechanisms (Fryer et
al., 2012). As well, we analyzed the influence of assumptions about the most
important parameters of the evolution of CBS with c.o. --- so-called
``efficiency of common envelopes'' and the magnitude of the natal kick, which
accompanies the formation of a c.o. Below, our main assumptions, computational
method and our results are presented and discussed. They are then compared to
the results of other authors. The appendix contains examples of typical
evolutionary tracks of CBS that pass through the ULX stage, computed by the
evolutionary code MESA (Paxton et al., 2011).

\section*{The method of computations}
\label{s:method}

Like in Paper I, we have implemented a hybrid approach to the population
synthesis --- a combination of the rapid  simplified calculation using analytical
formulas up to the stage of formation of a binary harboring a c.o. and a visual donor,
followed by the detailed calculation of the mass-transfer stage using full
evolutionary program. This approach allows much more accurately than the
population synthesis, investigate the nature of the mass-transfer in the CBS and
to determine the duration of the accretion stage and, therefore, more accurately
estimate the number of the model sources, their luminosity and other
characteristics. A similar hybrid method was already applied for simulations of
ULX, (e.g., Shao and Li, 2015 and Shao et al., 2019) and merger rates of binary
BH (Gallegos-Garcia et al., 2021). All our calculations were performed for stars with
metallicity $Z$=0.02.
\begin{figure}[ht]
\includegraphics[width=1.0\textwidth]{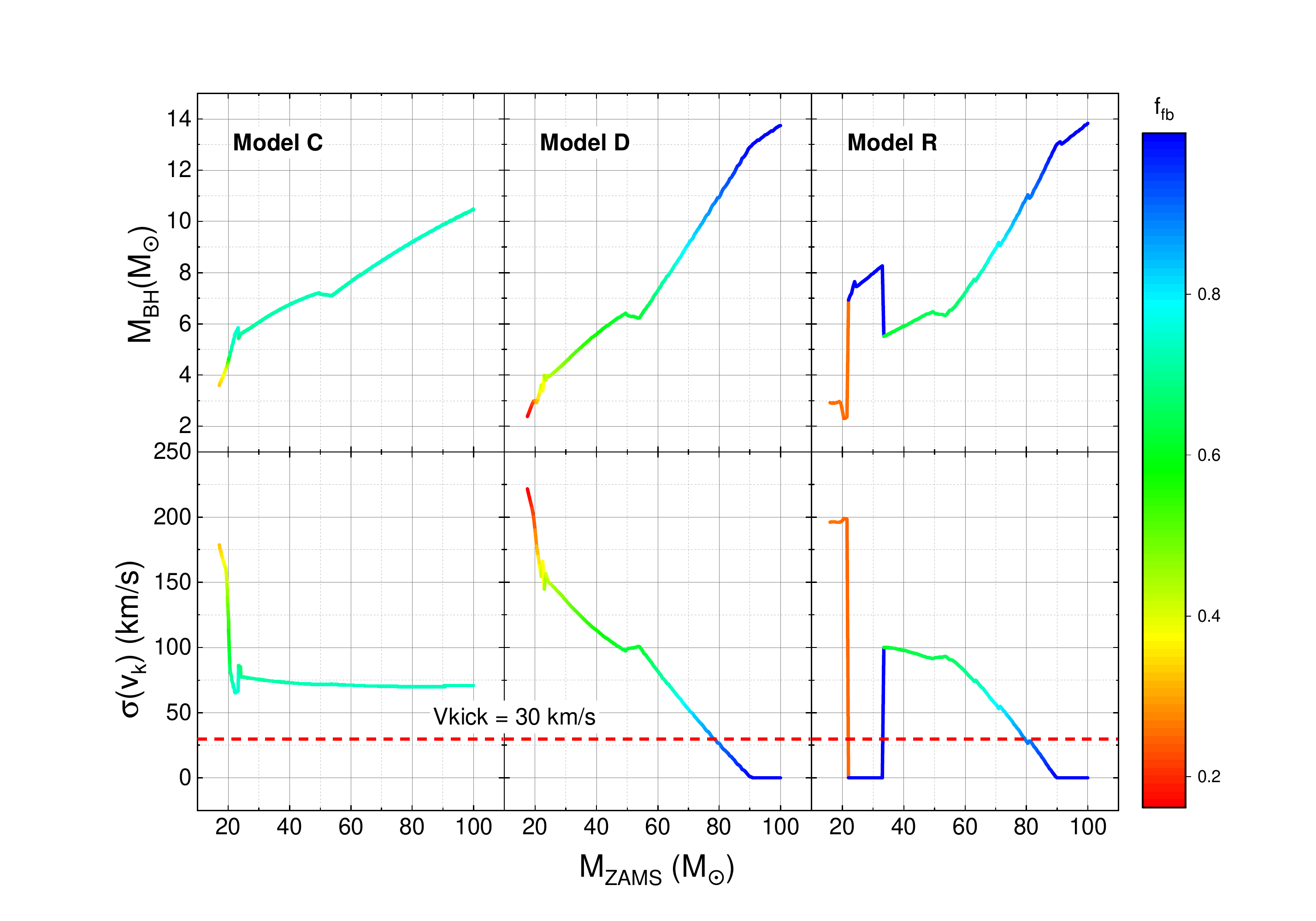}
\caption{\small Relation between ZAMS masses of BH progenitors, masses of BH, and natal kicks
The fraction of final stellar mass infalling onto proto-neutron star is shown
in color scale.}
\label{f:mzams_mbh3}
\end{figure}
Characteristics of the ULX population in a stellar system (galaxy) are
determined by the history of star formation, initial parameters of the close
binary systems (such as, for instance, the initial mass function of the primary
components) and evolutionary factors: mass-loss by stellar wind, the nature of
mass-loss upon RLOF experienced by the progenitor of the accretor, the
mechanism of the collapse of its core  and the size and
orientation of the natal kick received by nascent c.o.
Statistical distributions over the initial mass function of primary
components, mass ratios of components, their separation, orbital eccentricity
were taken into account. For BH, the above-mentioned collapse mechanisms and
various assumptions about the magnitude of the natal kick were considered.

In the first stage of computations, we have used an updated version of the BSE
code, (Hurley et al., 2002; see also Paper I)\footnote{Contributed by us changes
to the code make it, as the tests show, almost identical to the new author's
version of BSE (Banerjee et al., 2020).}. Further stellar evolution was
calculated using the code MESA (Paxton et al. 2011, revision 12778). For the
construction of the model population of ULX, the probability of formation of a
particular ULX was convolved with SFR and the period of the mass-transfer stage
in individual systems found by MESA calculations.

Angular momentum loss during the evolution of CBS was described by the equation
suggested by Soberman et al. (1997)\footnote{Eq. (12) in Paper I.}, that takes
into account mass transfer from donor to accretor, mass loss by the system from
the vicinity of accretor (re-ejection), and outflow of the matter through the
Lagrangian point $\mathrm{L_2}$. As in Paper I, it was assumed that 10\% of the
matter lost by the donor leaves the system through the vicinity of
$\mathrm{L_2}$ (parameter $\delta_ {mt}=0.1$); the radius of the co-planar
circumstellar disk was set to $\gamma_{mt}^2a$, where $\gamma_{mt}=\sqrt{3.0}$.

Mass-loss by O-B stars was calculated using option ``Vink'' in MESA, which is
based on algorithms suggested by Vink et al. (2000, 2001). Mass-loss by
Wolf-Rayet stars was described by the formulas of Nugis and Lamers (2000).
 
To determine the masses of c.o. in the cases of rapid and delayed collapse, we
used the parametrization of results of calculations by Fryer et al. (2012),
suggested by Giacobbo and Mapelli (2018). Relations between masses of BH and
ZAMS masses of their progenitors are shown in Fig.~\ref{f:mzams_mbh3}. These
relations are consistent with the claim of Smartt
(2015) based on observations of supernovae,
that the main fraction of BH progenitors have masses exceeding
(18 -- 20)\msun. The accuracy of relations suggested by Giacobbo and Mapelli is
sufficient for population synthesis, although we note that some
calculations of collapse models show that NS can have larger mass progenitors
(see, for example, Ertl et al., 2020). Giacobbo and Mapelli relations, in
principle, make it possible to estimate NS masses down to the theoretically
allowed maximum, but we assigned a mass of 1.4\msun\ to all NS and limited BH
masses from below by the value 2.15\,\msun, corresponding to the maximum
estimated mass of the observed pulsars M = $2.08\pm0.07$\,\msun\ (PSR~J0740 +
6620, Fonseca et al., 2021).

Note the increase of BH masses along the sequence of models D-R-C and the
absence of BH with masses below $\simeq 4$\,\msun\ in model C. Natal kicks in
model C are lower than in model D. These circumstances lead to the differences
in the number of ULX and differences of distributions of ULX populations over
components masses and orbital periods. The ``step'' in the distribution of BH
masses in the case of rapid collapse is associated with the existence of a range
of masses of stars that experience the direct collapse of CO cores (Fryer et al.
2012).

As a rule, CBS, which produce ULX, pass through the common envelopes. In the
``fast'' stage of calculations before the formation of the CBS with BH, it was
assumed, following BSE conventions, that common envelopes are formed as a result
of the loss of matter by donors -- red (super)giants with convective envelopes
in the dynamic time scale; the criteria for the formation of common envelopes
depend on the mass ratio of components and relative mass of the donor core. For
the computation of variation of separation of components in common envelopes,
$\alpha\lambda$ formalism of Webbink (1984) and de Kool (1990) was adopted,
which is based on the comparison of the orbital energy of the system and the
bounding energy of the donor envelope. We performed calculations for three
values of the so-called ``efficiency of common envelopes'' \ace --- 0.5, 1, and
4. The parameter $\lambda$, which characterizes the binding energy of the donor
envelope, was adopted after Loveridge et al. (2011). To a rough approximation,
initial and final separations of components in the common envelope, $a_0$ and
$a_f$, are related as $a_f\propto \al \times a_0$. If components of a CBS merged
in the common envelope, further calculations of the evolution of the system were
terminated. Note that the value of $\lambda$\ is one of the most uncertain
calculation parameters (Ivanova et al., 2013, 2020). In the stage of accretion
onto BH, formal criteria for the formation of common envelopes were not applied.
As have been shown by McLeod et al. (2018) and investigated in detail by Klencki
et al. (2021a), RLOF by a red super-giant with a convective envelope, a
companion the c.o., is followed by a stage of relatively stable mass-loss, which
transits into a stage of mass-loss in the dynamic time scale. If in the course
of calculation by MESA ``confinement'' of the donor in the Roche lobe became
impossible, which was numerically expressed as the divergence of the code (see
the lower right panel in Fig.~\ref{fig:trc_2.5} in the Appendix), we assumed
that the components merged. The time between the moment when accretor reached
$L_{\rm X} = 10^{39}$\,erg/s and the fusion of components or decline of the
luminosity below the limit specified above, if the components did not merge, was
taken as the lifetime of the system at the ULX stage.

Before the RLOF by the primary (initially more massive component, progenitor of
BH), its evolution was considered as quasi-conservative. Only the loss of mass and
angular momentum due to the stellar wind was taken into account (according to
Vink (2001) for $\te \geq 10000$\,K and de Jager et al. (1988) for the lower $\te $).
Tidal interaction of components was treated according to the algorithm adopted
in BSE (see Hurley et al., 2002, \S2.3).

It was assumed that BHs, like NSs, acquire isotropic natal kicks. The amplitude and
distribution of the additional velocity acquired during BH formation are the
parameters of the model. Currently, there is no consensus on this issue, see,
e.g., White and Van Paradijse (1996) arguments in favor of insignificant velocity
amplitudes and the opposite point of view in the later work of Atri et al.
(2019).

Two options were considered. First, the kick defiined as $v_k=(1-f_b)v$, where $v$ obeys
the Maxwellian velocity distribution with dispersion $\sigma=265$\,km/s
suggested for radio pulsars by Hobbs et al. (2005), while parameter $f_b$ is
determined by the fraction of the final mass of the star $M_{fin}$, which falls
back onto neutron proto-star with a mass of $ M_{prot}$:
$f_b=M_{fb}/(M_{fin}-M_{prot})$. Second, we considered Maxwellian velocity
distribution with $\sigma=30$ km/s (see
Fig.~\ref{f:mzams_mbh3}). 

Omitting fairly well-known details, we only note that in the ULX context it is
important that the mechanisms of collapse differ in the masses of the produced
BH. This circumstance and the magnitude of the natal kick lead, first of all, to
the differences in the fraction of binary systems, which remain bound in the
first supernova explosion in the system and, potentially, could, in the course
of further evolution, give rise to a ULX.

\section*{Calculation of X-ray luminosity}
\label{s:lx}

X-ray luminosity $L_\mathrm{X}$ of accreting c.o.  is computed based
of the capture rate of the matter $\dot{M}_\mathrm{X}$. If the donor overflows
Roche lobe, accretion occurs via disk, $\dot{M}_\mathrm{X} = 0.9 \dot M_\mathrm{O}$,
where $\dot M_\mathrm{O}$ is the rate of donor mass loss via $\mathrm
L_1$ obtained by evolutionary computations (see Fig.~\ref{fig:trc_2.5}).

If the visual star does not fill its Roche lobe and accretion occurs at the expense of
the stellar wind, the rate of the mass capture by c.o. is computed using
Bondi-Hoyle-Littleton formalism for spherically-symmetric wind.  
Thus for a circular orbit wits semi-axis $a$,
$\dot{M}_\mathrm{X}=\frac{1}{4}\dot{M}_\mathrm{O}\myfrac{R_B}{a}^2\sqrt{1+\myfrac{v_{x}}{v_w}^2}$,
where $R_B=\frac{2GM_\mathrm{X}}{v_w^2+v_x^2}$ -- is Bondi gravitational capture radius,
$M_\mathrm{X}$ -- the mass of the c.o.,
$v_w=v_p(M_\mathrm{O})(1-R_\mathrm{O}/a)^{1/2}$ -- wind velocity at the orbit of c.o.,
$v_x$ -- orbital velocity of the c.o.,
$v_p=\sqrt{2GM_\mathrm{O}/R_\mathrm{O}}$ -- escape velocity at the photosphere of the
visual star with mass $M_\mathrm{O}$ and radius $R_\mathrm{O}$.
For eccentric orbits, accretion rate depends on the phase of orbital motion.
In our computations we used averaged over orbital period $P$ accretion rate
$\langle \dot M_\mathrm{X}\rangle = 1/P\int \dot M_\mathrm{X}(t)
dt=1/P\int \dot M_\mathrm{X}(E) (dt/dE) dE $ ($E$ --- eccentric anomaly).
For the assumed dependence of variation of stellar wind with the distance from the star and  
$R_\mathrm{O}/a\lesssim 0.3$, the averaged accretion rate weakly increases with eccentricity.
Therefore, we neglect the dependence of averaged over orbital period accretion rate
on eccentricity. The average rate of Bondi accretion from stellar wind most
strongly depends on the ratio $R_\mathrm{O}/a$:
$\langle\dot{M}_\mathrm{X}\rangle\approx(1/64)\dot{M}_\mathrm{O}\myfrac{M_\mathrm{X}}{M_\mathrm{O}}^2\myfrac{R_\mathrm{O}}{a}^2/(1-\frac{R_\mathrm{O}}{a})^{2}$.

We did not consider possible geometrical focusing of the stellar wind, which can
change the efficiency of accretion, since it strongly depends on the assumed
mechanism of wind acceleration (El Mellah et al., 2019). We also neglected such
effects as wind Roche lobe overflow (Plavec et al., 1973), possible decrease in
the escape velocity at the visual star surface due to the tidal effects (Hirai
and Mandel, 2021).

In the case of disk accretion, X-ray luminosity
$L_\mathrm{X}=0.1\dot M_\mathrm {X}c^2$ ($c$ is the speed of light). For
magnetized NS we considered
the possibility of quasi-spherical subsonic accretion and modification
of ``standard'' Shakura-Sunyaev disks in the presence of a magnetic field (see
for details Paper I). As well, we took into account possible thermal-viscous
instability of
accretion disks. Dubus et al. (1999) criterion of disk stability was used:
a source was considered as transient if the rate of mass accumulation in the
disk $\dot M_\mathrm{O}$ was lower than a critical value $\dot M_{cr}$, which
depends on the masses of components and the radius of the outer edge of the
accretion disk. In quiescence, mass accumulates in the disk. Accumulated disk mass
is defined as $M_ {disk} = \dot M_\mathrm{O}(t) \times \Delta t$, where
$\Delta t = $ 30 yr is the average time between the outbursts, chosen by us
based on calculations by
Hameury and Lasota (2020); see also Coriat et al. (2012). Accretion rate onto
c.o. in the active state is assumed to be $\dot M_\mathrm {X} = \dot M_{\rm cr}$ 
(see Fig.~\ref{fig:trc_old_age} in Appendix). 
Duration of the outburst of a transient source ($\Delta t_{outb}$) is determined
by the time of loss of entire accumulated disk mass:
$\Delta t_{outb}=M_{disk}/\dot M_\mathrm{X}$. For the accretion rate below a
certain limit ($ 0.001\dot M_{cr}$) the disks were assumed to be stable.

The probability of detection of a transient source in an active state is determined
by the ratio of the time spent by the source in the outburst
$\Delta t_{outb}$ and duty cycle:
$p_{outb}=\Delta t_{outb}/(\Delta t_{outb}+\Delta t)$.

We assumed that the mechanism of disk accretion instability is the same for BH
and NS and for disks formed via RLOF and by the capture of stellar wind. For the
supercritical disk accretion onto BH, $\dot M_\mathrm {X}>\dot
M_\mathrm{Edd}(M_\mathrm {X})$, the beaming of radiation from the inner parts of
the disk was taken into account. Then the apparent luminosity of
spherically-symmetric radiation obtained from the observed X-ray flux and
distance to the source, $L_\mathrm{X} =\frac{1+\ln \dot {m_0}}{b}
L_\mathrm{Edd}(M_\mathrm{X})$, where beaming factor
$b=\mathrm{\max(10^{-3},73/\dot m_0^2)}$, $\dot m_0 = \dot {M}_\mathrm{X}/\dot
{M}_{\mathrm{Edd}}$ (King 2009). An example of an evolutionary track 
by the MESA code, with the phase of a transient ULX at the stage of RLOF by the
visual component experiencing case B of mass-exchange is shown in
Fig.~\ref{fig:trc_old_age} in the Appendix.

In the case of disk accretion onto NS, the relation between the radius of NS
magnetosphere and spherization radius, where the local energy release exceeds
Eddington limit and the outflow of matter begins, becomes important, see Paper I
and Grebenev (2017). When modeling these sources, it was taken into account that
the probability of discovery of a system with BH at the supercritical accretion
stage is inversely proportional to the beaming factor $b$. For supercritical
accretion onto magnetized NS, the beaming is defined by the geometrically thick
edge of the supercritical disk at the boundary of the magnetosphere only. Then
$H/R \sim 1$ and $ b\sim 1/2$, see Paper I, Mushtukov et al. (2021), and note
the estimate $b\sim 0.25$ from the observations of NGC~300~ULX-1 harboring a
neutron star (Binder et al., 2018).

\section*{Results of computations}
\label{s:results}
\begin{table*}[h!]
\begin{center}
\caption{\small The number of ULX at 10\,Gyr after the beginning of star
formation in a model galaxy with constant star formation rate 1\,\msun/yr for
different accretors and types of mass-transfer between components, various
models of c.o. formation, distribution of their natal kicks and common envelope
parameter \ace. In the systems B\_RLOF and NS\_RLOF, accretion occurs via flow
of matter through the vicinity of $\mathrm{L_1}$, in the systems BH\_wind and
NS\_wind stellar wind is accreted. The number of stationary (persistent) sources
is given in parenthesis.}
\label{tab:ulx2}
\vspace{5mm}
\scalebox{0.92}{
\begin{tabular}{l|c|l|l|c|c|c|c|c} \hline\hline
Model&Model of& $\sigma(v_{\rm k})$    & $\alpha_{ce}$  & N & N & N & N  & N \\
    & c.o. form. & (km/s) &    & ULX & $\rm BH\_RLOF$ & $\rm BH\_wind$  & $\rm NS\_RLOF$ & $\rm NS\_wind$   \\
\hline  
\hline
C265-05  & CO     &  265 &  0.5 &   0.88  &   0.40  &   0.03 &   0.44  &  0.007 \\
& & & &  (0.76) &   (0.32) &   ($<10^{-3}$) &  (0.44) &  ($2\times 10^{-4}$)\\
{\bf C265-1 } &      &  265 &  1.0 &   1.49  &   0.36  &   0.62 &   0.51 &  0.008 \\
 & & & &  (0.76) & (0.25) &   (0.005) &   (0.50) &   ($2.4\times 10^{-4}$) \\
 C265-4  &      &  265 &  4.0 &   3.38  &   0.02 &   0.10 &   3.15  &  0.101   \\
 &&&& (3.31) &(0.02) & (0.063) & (3.14) & (0.09) \\
\hline          
C30-05  &   &  30 &  0.5 &   1.58  &   1.08  &   0.05  &  & \\ 
&&&& (1.22) &(0.78) & (0.004) & & \\        
C30-1  &     &  30 &  1.0 &   1.84 &   0.72 &   0.61 & &\\ 
&&&& (0.96) & (0.44) & (0.017) & & \\
C30-4  &      &  30 &  4.0 &   3.51 &   0.04  &   0.22 & &\\
&&&&(3.43) &(0.03) & (0.172) &  & \\ 
\hline
D265-05  & Delayed     &  265 &  0.5 &   0.75&   0.11 &   0.19  &   0.44&  0.007\\
&&&&(0.54) &(0.10) & (0.001) &   ( 0.44) &($2\times 10^{-4}$ ) \\
{\bf D265-1 } &      &  265 &  1.0 &   0.78 &   0.07&   0.20&   0.51&  0.008 \\
&&&&(0.56) &(0.05) &(0.001) & (0.50) &  ($2.4\times 10^{-4}$) \\
D265-4  &     &  265 &  4.0 &   3.41  &   0.04 &   0.11&   3.15&  0.101  \\
&&&&  (3.27) &  (0.04) &(0.004) & (3.14) & (0.09) \\
\hline          
D30-05  &   &  30 &  0.5 &   0.89  &   0.15 &   0.29 & &\\ 
&&&& (0.54) & (0.10) &(0.00) & &\\
D30-1  &     &  30 &  1.0 &   0.88  &   0.07 &   0.29& &\\
&&&&(0.54) & (0.03) &(0.005) &
& \\
D30-4  &     &  30 &  4.0 &   3.49 &   0.04&   0.20 & &\\
&&&&(3.29) & (0.02) & (0.042) & &\\ 
\hline              
R265-1  & Rapid     &  265 &  1.0 &   0.81 &   0.07&   0.23&   0.51&  0.008 \\
&&&& (0.57) &(0.05) &(0.008) & (0.50) &  ($2.4\times 10^{-4}$) \\
R30-1  &      &  30 &  1.0 &   0.63   &   0.03&   0.09& &\\
&&&& (0.53) &(0.02) &(0.003) & &\\ 
\hline
\hline
\end{tabular}
}
\end{center}
\end{table*}
Table~\ref{tab:ulx2} lists the numbers of ULX for the model galaxy with a
constant rate of SFR 1 \msun/yr at the age of 10\,Gyr, depending on the model of
c.o. formation, natal kick, common envelope parameter and the mode of accretion.
For comparison, are included the numbers of NULX, computed in Paper I under
assumption of the log-normal distribution of magnetic momenta for typical
magnetic field strength log B=12.5 G. Since in our model all NS are formed with
the same mass of 1.4\,\msun\ and receive the same natal kicks with
$\sigma{(v_k)}$ = 265\,km/s, The cells of the Table for NS corresponding to
$\sigma{(v_k)}$ = 30\,km/s are left blank. However, such NULX are taken into
account in calculations of the total number of sources in the models.

Let us consider in more detail the evolution of distributions of masses and
orbital periods of CBS in which ULX are formed. As the main versions of
computations, we consider models with the collapse of stellar CO core and
delayed collapse with \ace=1 and Maxwellian distribution of natal kicks with
$\sigma(v_k)$=265 km/s (models C265-1 and D265-1 in the Table). In the series of
models ``D'' the kicks are scaled with account of fallback, as described above.
\begin{figure*} 
 \begin{minipage}{0.5\textwidth}
 \includegraphics[width=\textwidth]{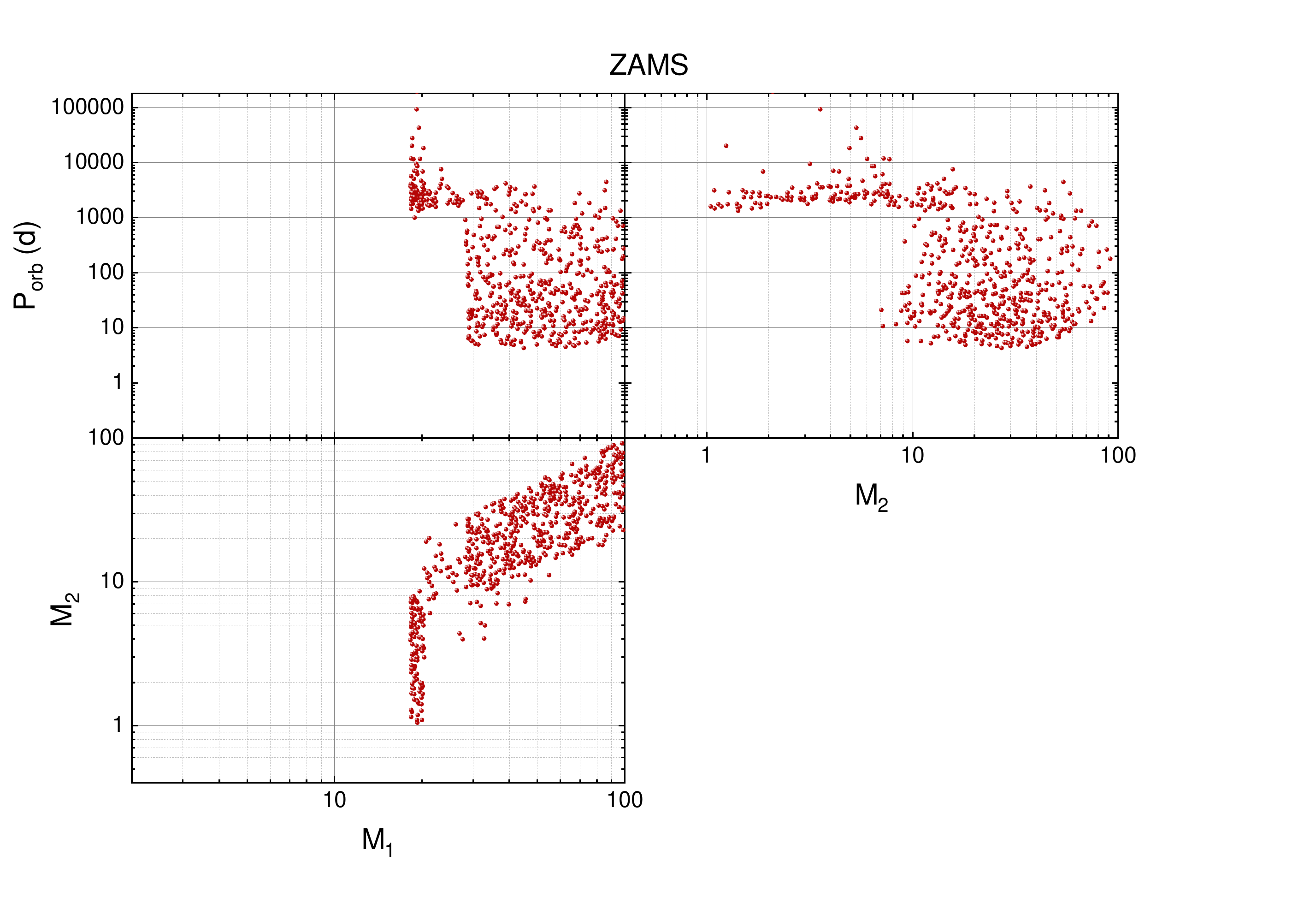}
 \vfill
\includegraphics[width=\textwidth]{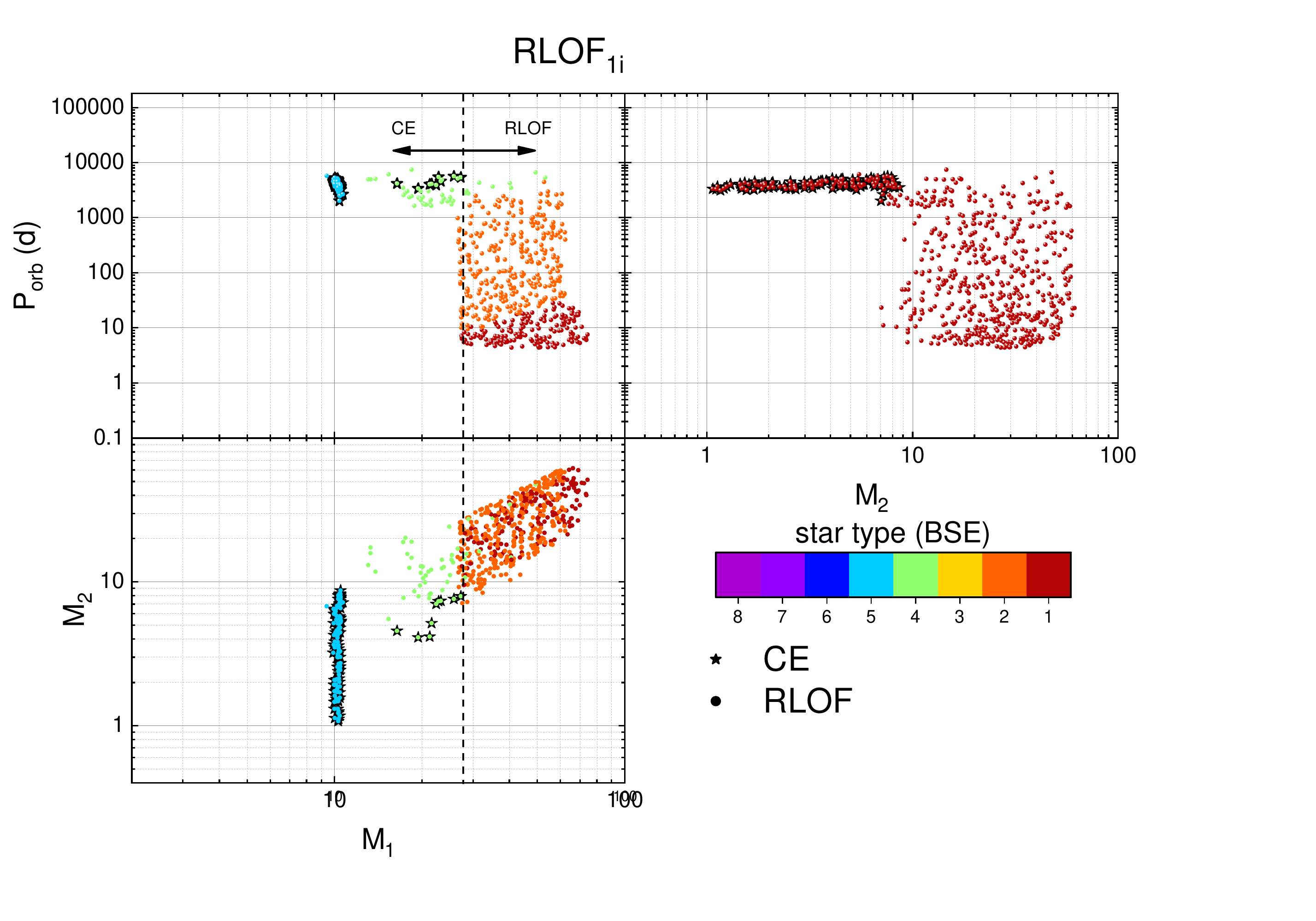}
\vfill
\includegraphics[width=\textwidth]{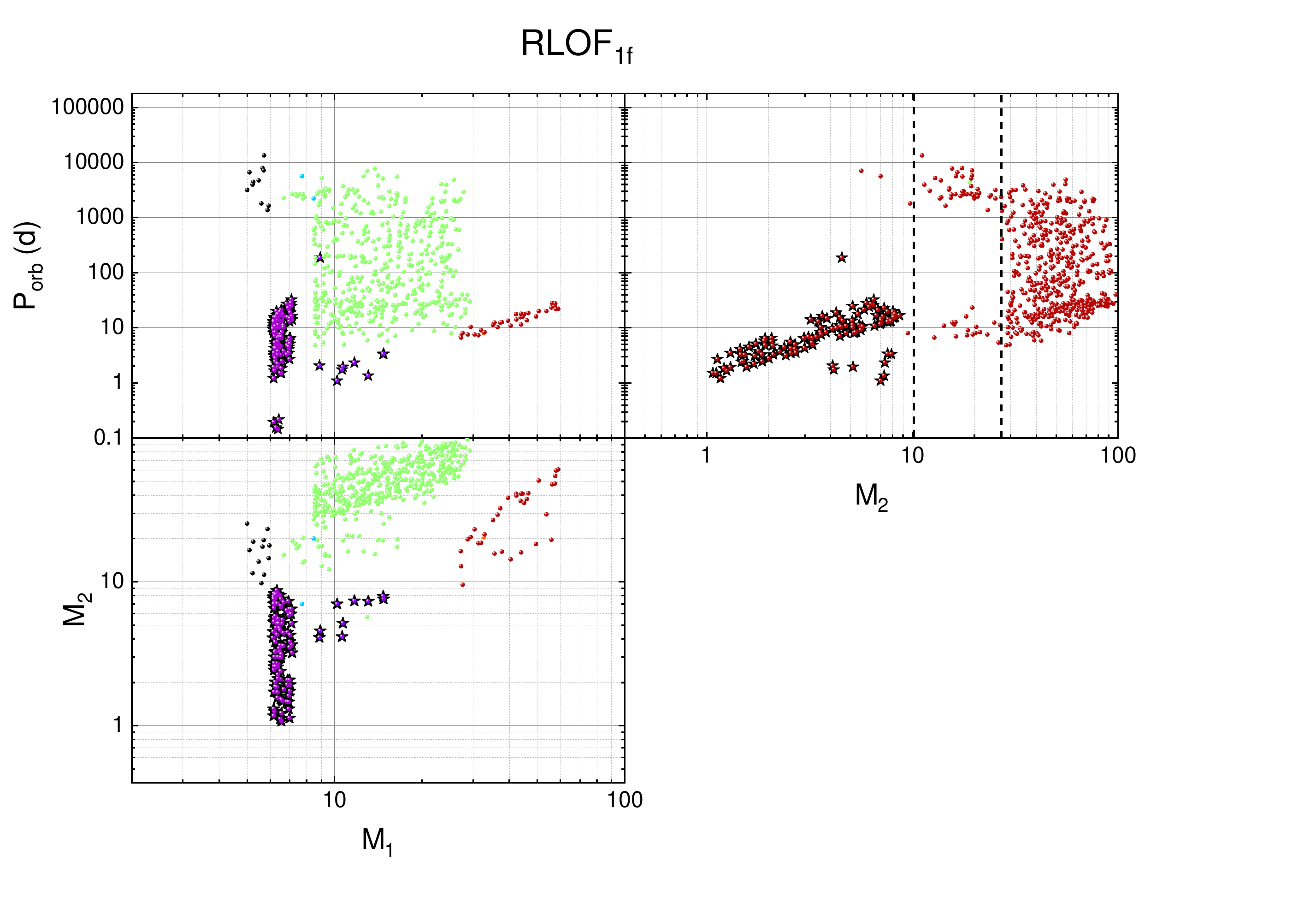}
\end{minipage}
\begin{minipage}{0.5\textwidth}
\includegraphics[width=\textwidth]{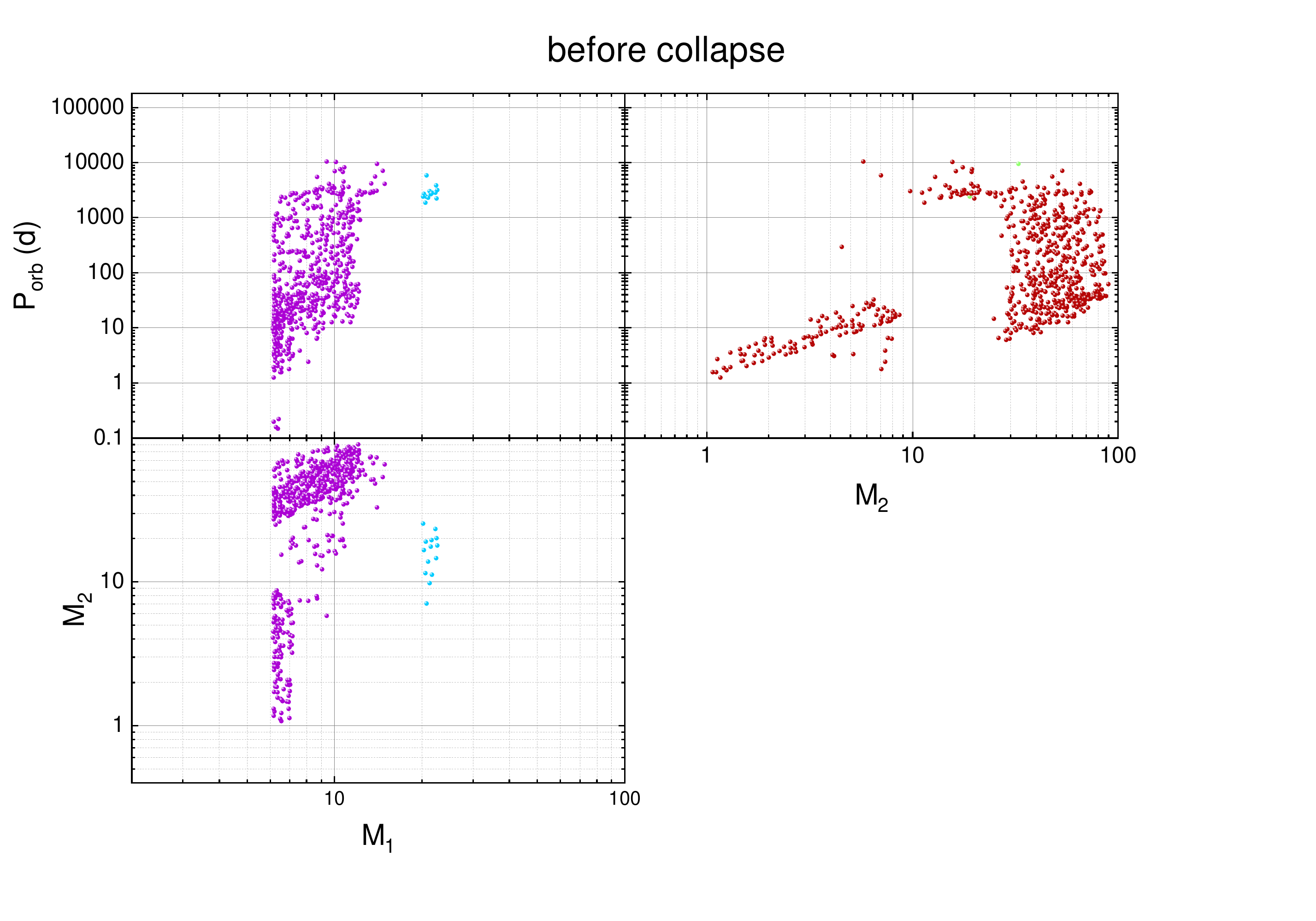}
\vfill
\includegraphics[width=\textwidth]{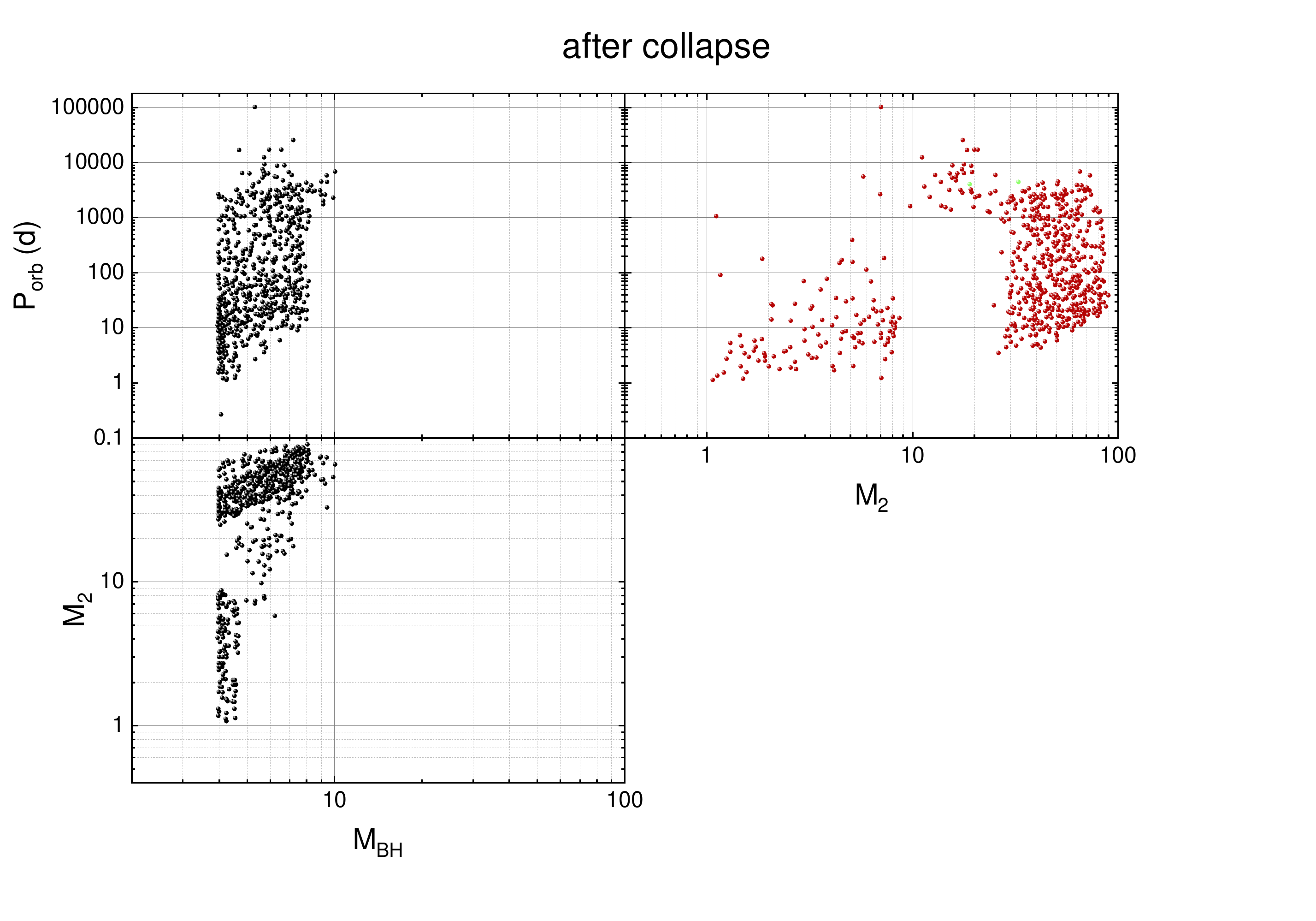}
\vfill
\includegraphics[width=\textwidth]{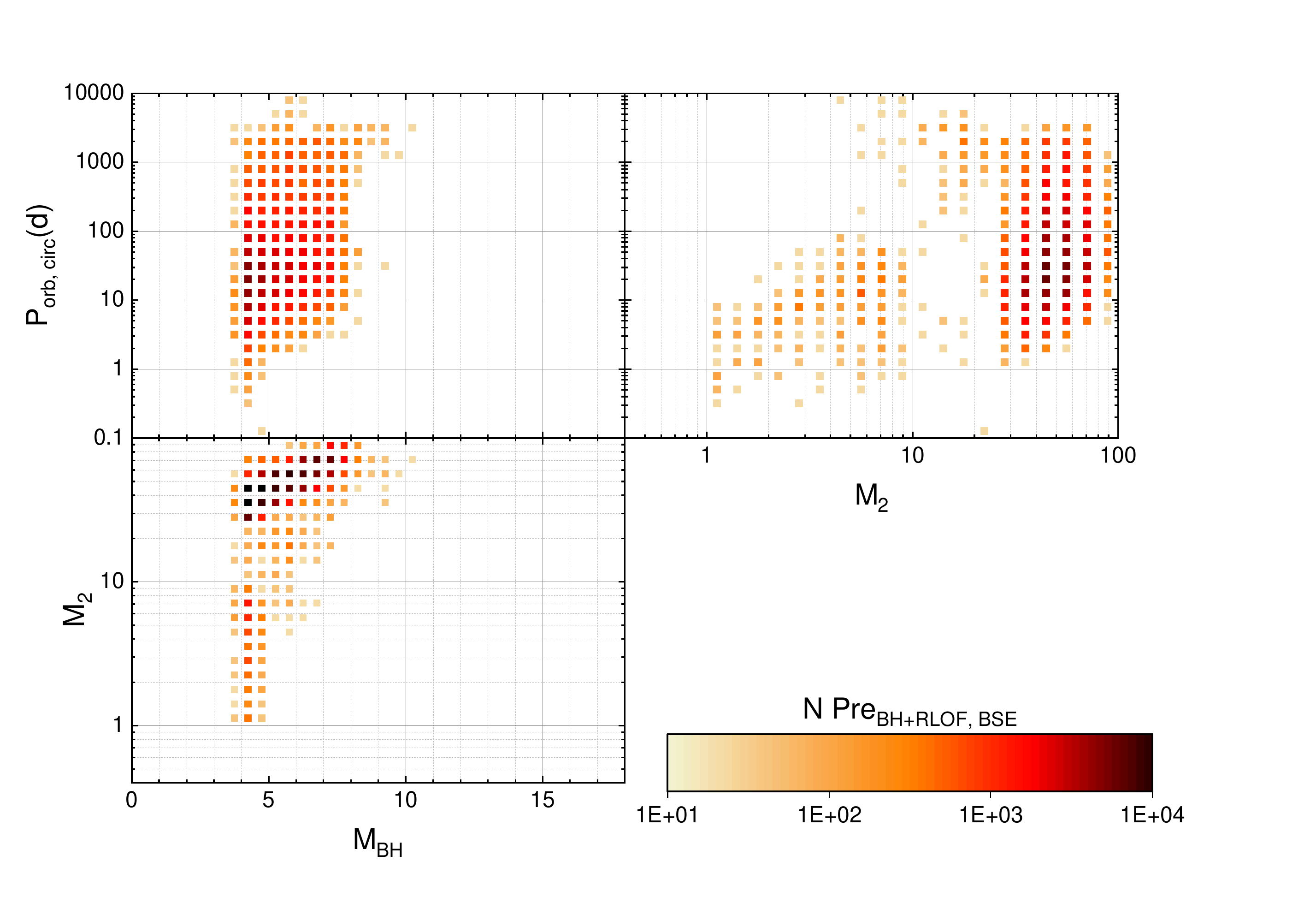}
\end{minipage}
\caption{\small Evolution of CBS resulting in formation of BH with Roche lobe
overflowing visual companions. The Figure corresponds to the model C265-1 with \ace=1
and scaled by fallback Maxwellian distribution of natal kicks with $\sigma = 265$ km/s. Figure
shows relations between the masses of components and orbital periods on ZAMS, at the beginning of
the first Roche lobe overflow
in the system (RLOF1i) and at the time of its completion (RLOF1f), before
collapse and after formation of c.o. (after collapse). Evolutionary stages of
the components are marked by colors
in accordance to BSE notation: 1 -- ZAMS, 2 --
Hertzsprung gap, 3 -- core He-burning stage, 4 -- first red giants branch, 5 --
early-AGB branch, 6 -- late-AGB branch, 7 -- He-remnant of the star after
envelope loss, 8 -- He-star in the Hertzsprung gap. Black dots mark c.o.
Circles and stars show CBS with stable loss and the ones with  common
envelopes, respectively.
Symbols in the Figure are not associated
with the birthrate of binaries with corresponding parameters and show only
``motion'' of the nodes of the computational grid and change of their evolutionary status.
Lower right panel shows the relationship between parameters of stars in the CBS at
the moment of RLOF by the BH companion; color scale corresponds to the number of
the systems in the $M_\mathrm{G}=10^{10}$\,\ms\ galaxy.}
\label{f:BSE_CO}
\end{figure*}
\begin{figure}[t!] 
 \begin{minipage}{0.5\textwidth}
\includegraphics[width=\textwidth]{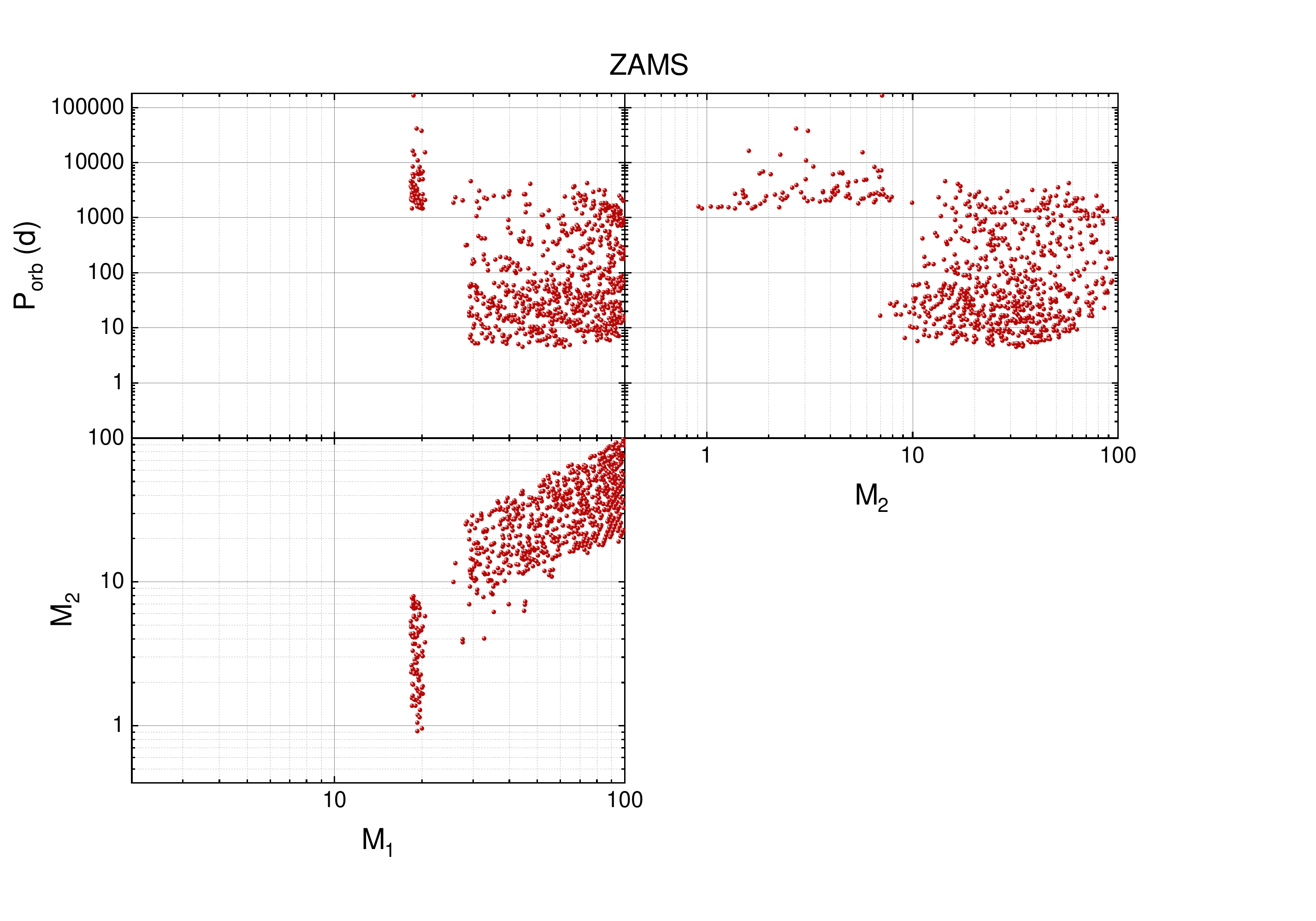}
\vfill
\includegraphics[width=\textwidth]{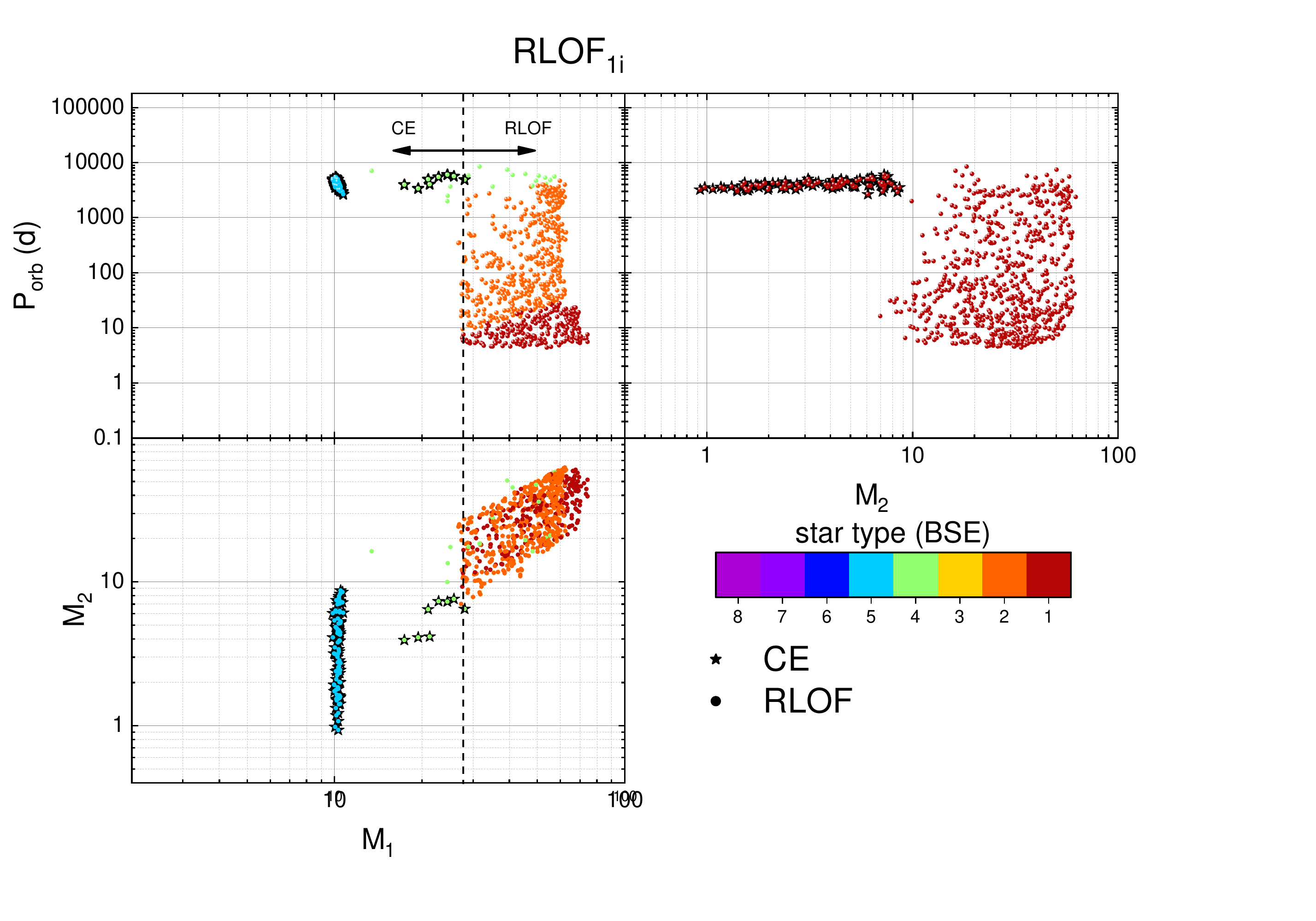}
\vfill
\includegraphics[width=\textwidth]{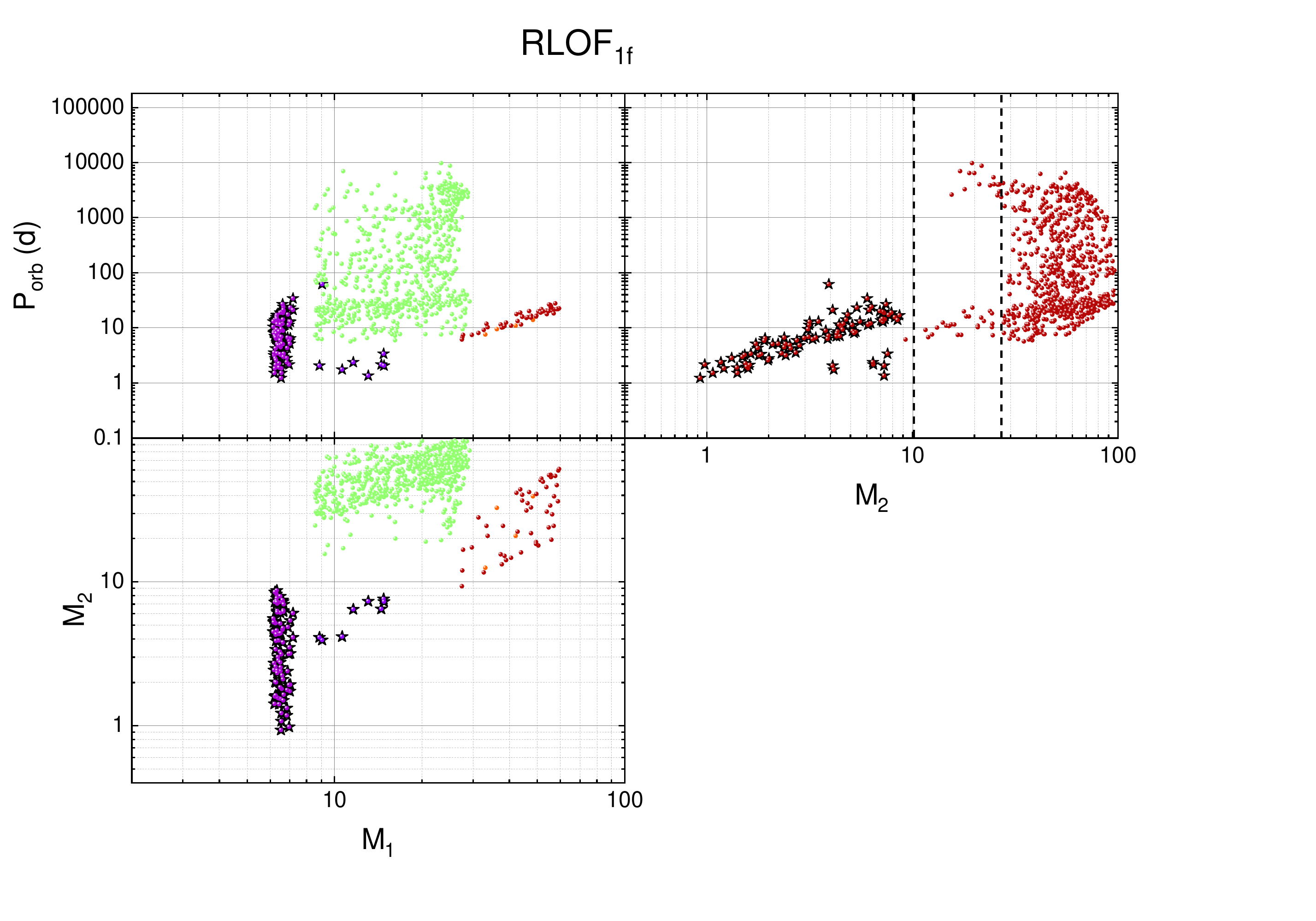}
\end{minipage}
\begin{minipage}{0.5\textwidth}
\includegraphics[width=\textwidth]{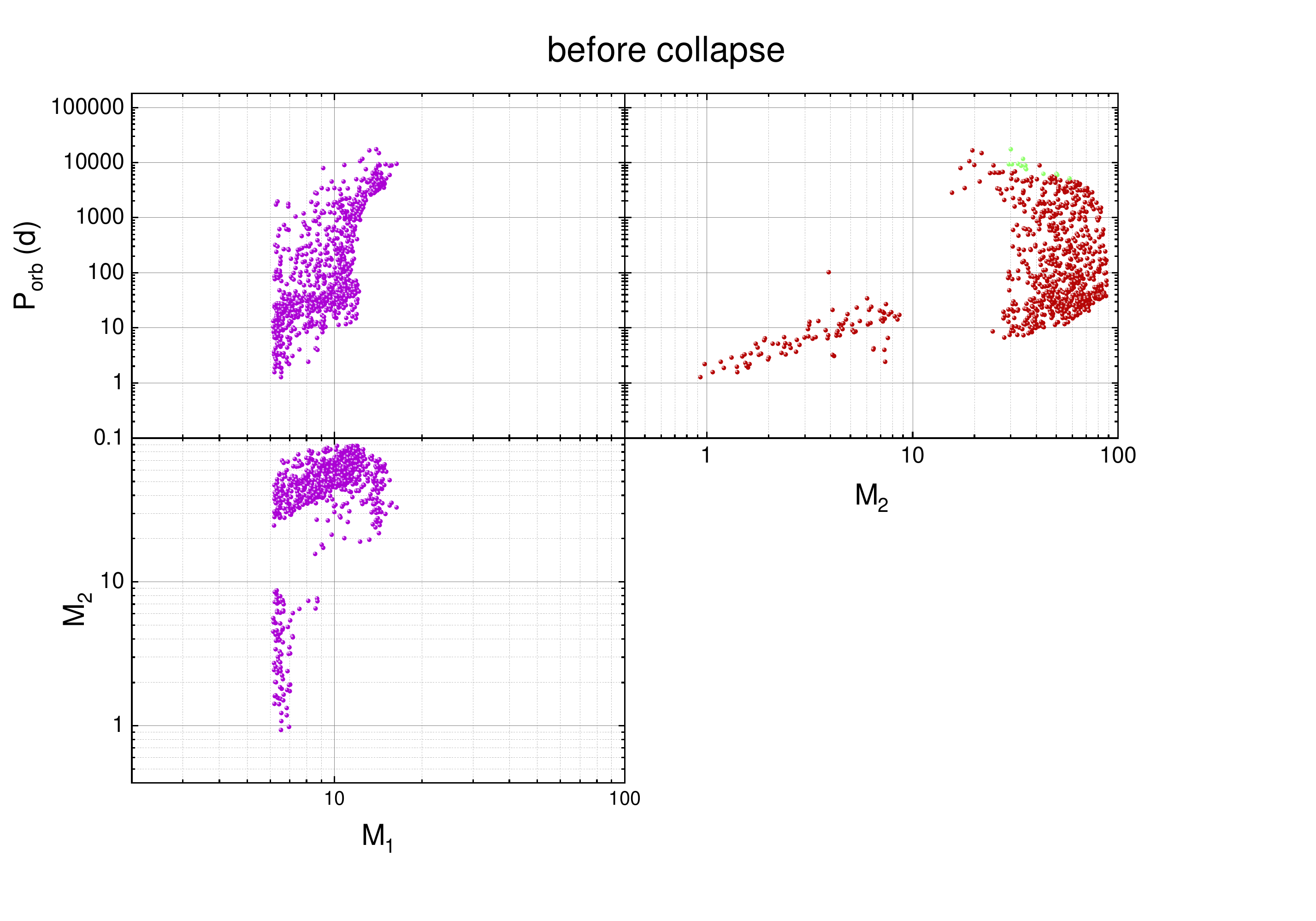}
\vfill
\includegraphics[width=\textwidth]{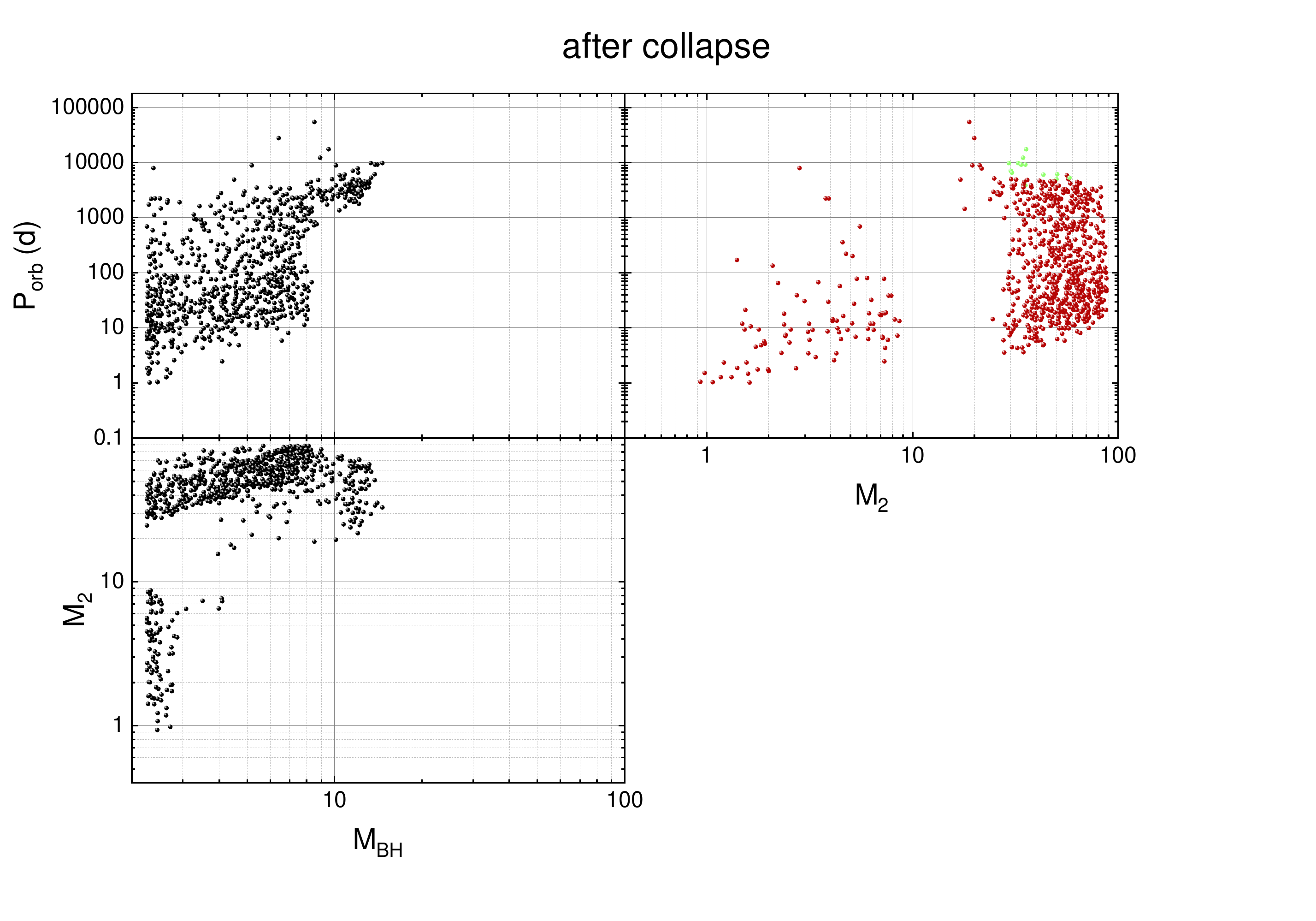}
\vfill
\includegraphics[width=\textwidth]{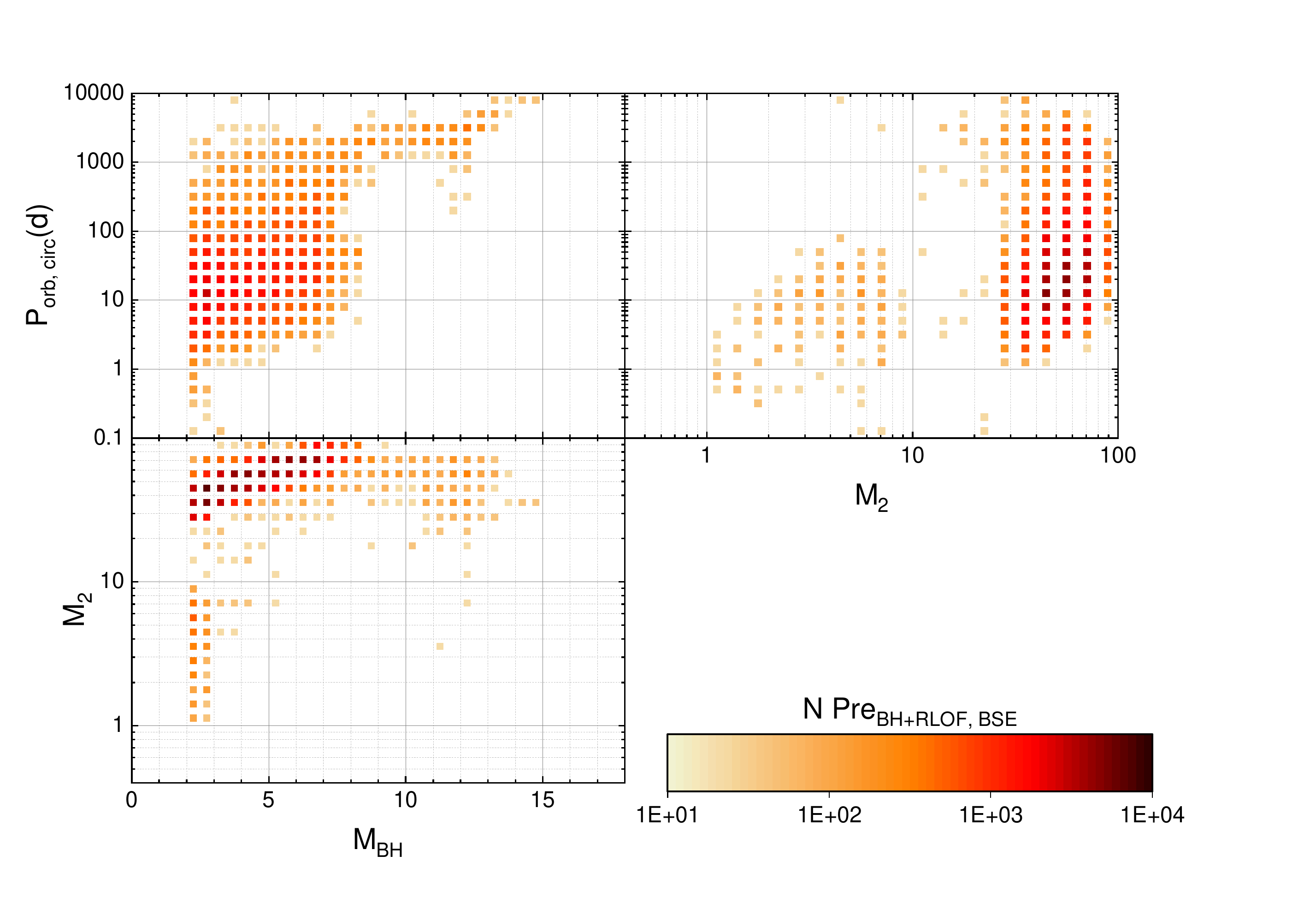}
\end{minipage}
\caption{\small Same as in Fig. \ref{f:BSE_CO} for the model D265-1.
}
\label{f:BSE_D}
\end{figure}
Figures~\ref{f:BSE_CO} and \ref{f:BSE_D} show, for the models C265-1 and D265-1,
under the assumption of instantaneous star formation, the sequential change of
the parameters of CBS --- progenitors of the systems in which in the course of
evolution companion to a BH overflows its Roche lobe and the system enters the
ULX stage. The systems were selected based on calculations by BSE. Their further
evolution is calculated using MESA, to assess possible lifetimes as ULX (X-ray
luminosity $L_\mathrm{X}>10^{39}$\,erg/s) to estimate the number of ULX in the
populations with different models of star formation (see as an example relevant
Eq. (11) for the Milky Way in Paper I). It was assumed that by the time the
companion to BH overflows its Roche lobe, the orbit had time be circularized
(using the formalization of this process in BSE). The dots in the left column of
the panels and in the upper two right panels correspond to the ``nodes'' of the
grid of initial parameters, which cover the entire evolutionary path from ZAMS
to BH (the points are not normalized and may represent different numbers of
systems). The main fraction of BH progenitors fills Roche lobes on the
main-sequence and in the Hertzsprung gap. As a result of RLOF, components of a
significant fraction of CBS merge. Surviving systems become tighter if they have
passed through a common envelope in which the masses of accretors practically do
not change. In the systems with mass transfer, separation of the components
increases, masses of accretors increase, primary components reach the stage of
giants or, after the loss of envelopes become helium stars (see RLOF1f panels).
In continuation of the evolution, giants, which are usually massive, also lose
the remains of the helium envelopes via stellar wind (see ``before collapse''
panel). Note a wide range of masses of satellites of future BH --- from $\sim$1
to $\sim$100\,\msun, as well as large orbital periods of a number of the systems.
Thanks to these circumstances, first, visual components of the ULX may become
red giants, as confirmed by observations (see, e.g., L{\'o}pez et al., 2020)
and, second, formation of ULX is possible after the completion of the star
formation process, enabling the existence of ULX in elliptical galaxies.

Comparison of Figs.~\ref{f:BSE_CO} and \ref{f:BSE_D} shows, in accordance with
Fig.~\ref{f:mzams_mbh3}, that in model D, BH are slightly more massive than in
model C. In model C, BH less massive than 3\,\msun\, are absent, due to the
assumption that in their progenitors collapses entire CO-core. In model D, BH
masses can extend down to maximum NS masses. Due to the larger natal kicks in
model D, there are fewer low-mass donors and maximum ULX periods are also lower.
We do not consider model R in detail, because in this model BH masses and natal
kicks are similar to model D (Fig.~\ref{f:mzams_mbh3}) resulting in a ULX
population similar to that in model D (see Table~\ref{tab:ulx2}).

\begin{figure}[t!] 
 \includegraphics[width=0.5\textwidth]{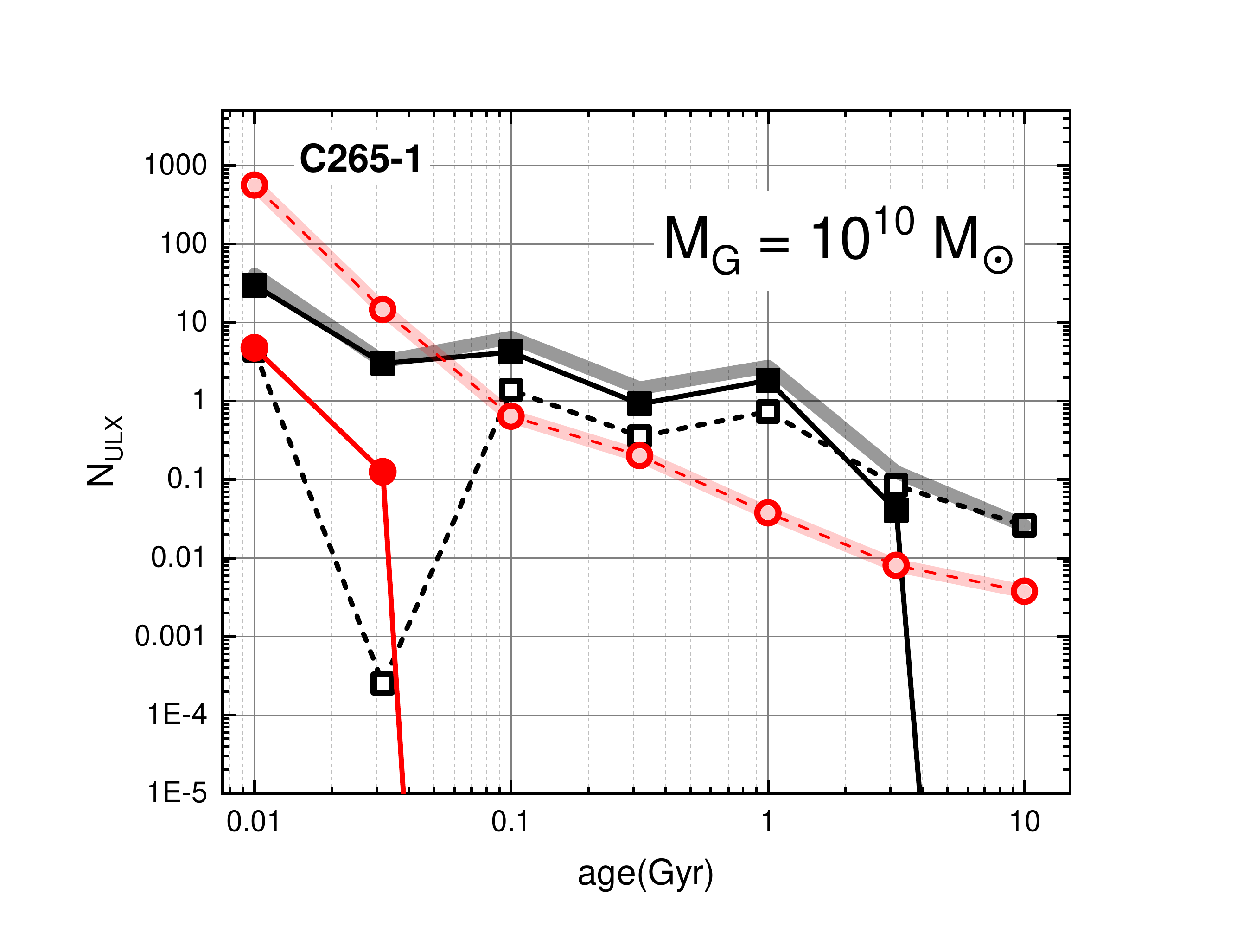}
 \includegraphics[width=0.5\textwidth]{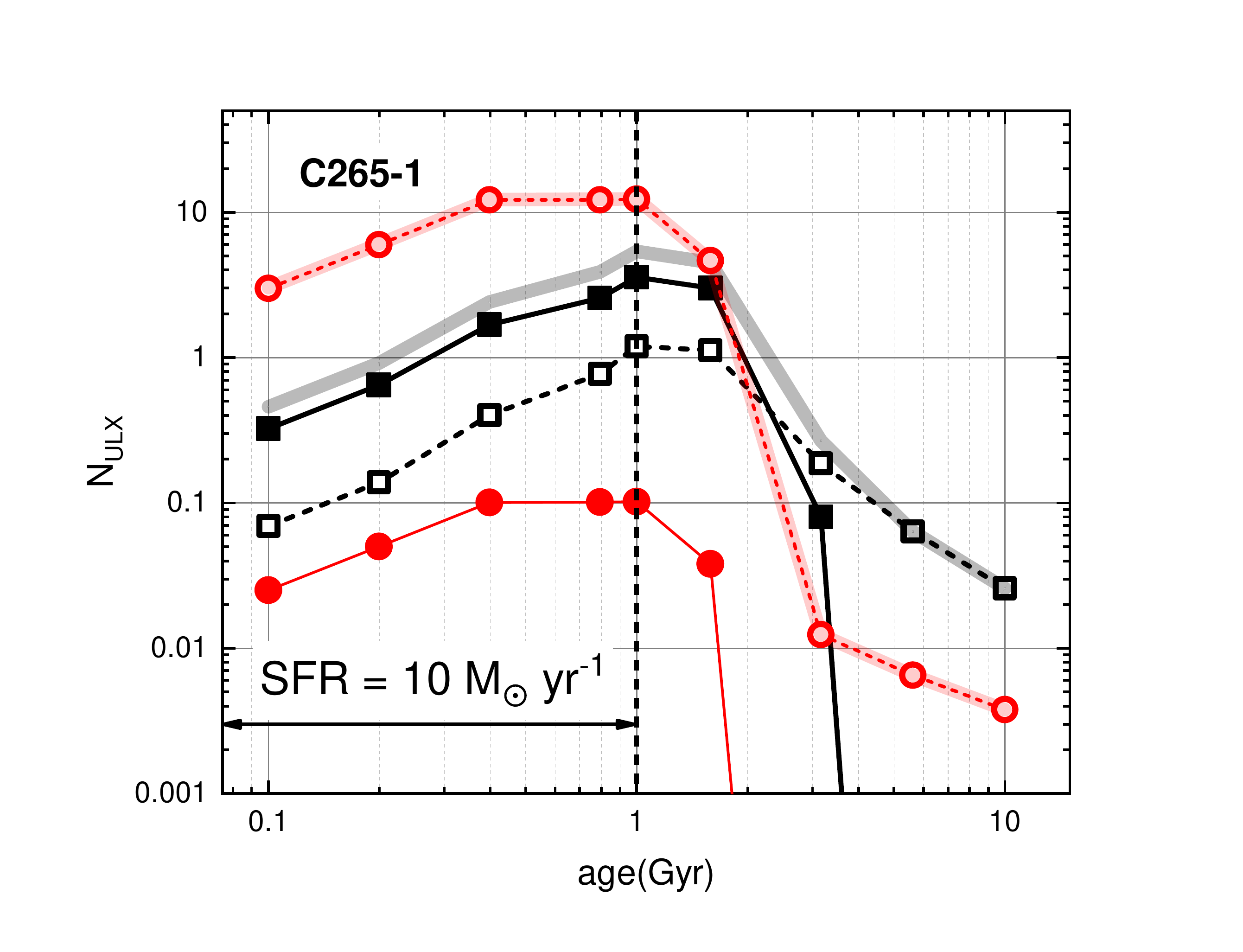}
\includegraphics[width=0.5\textwidth]{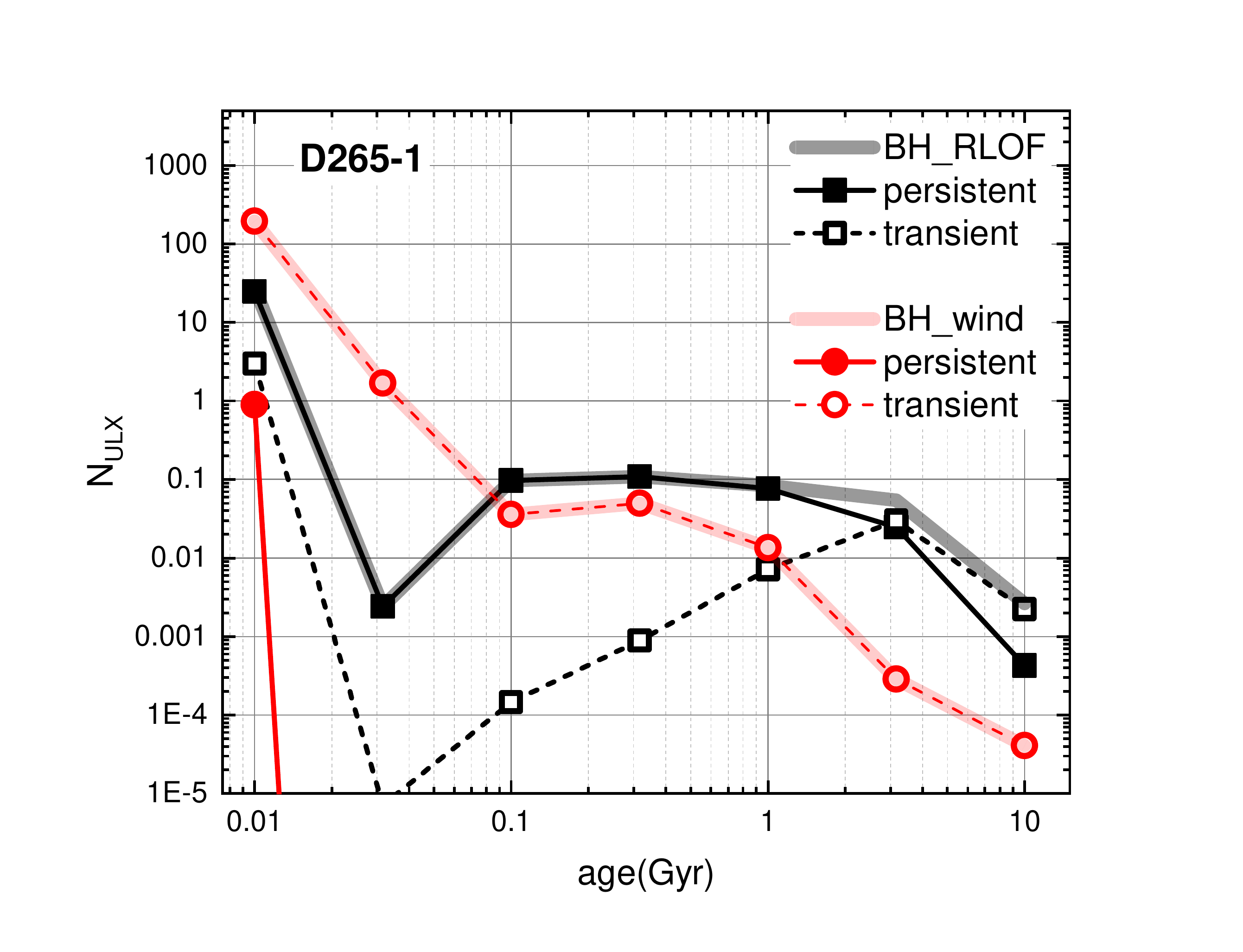}
\includegraphics[width=0.5\textwidth]{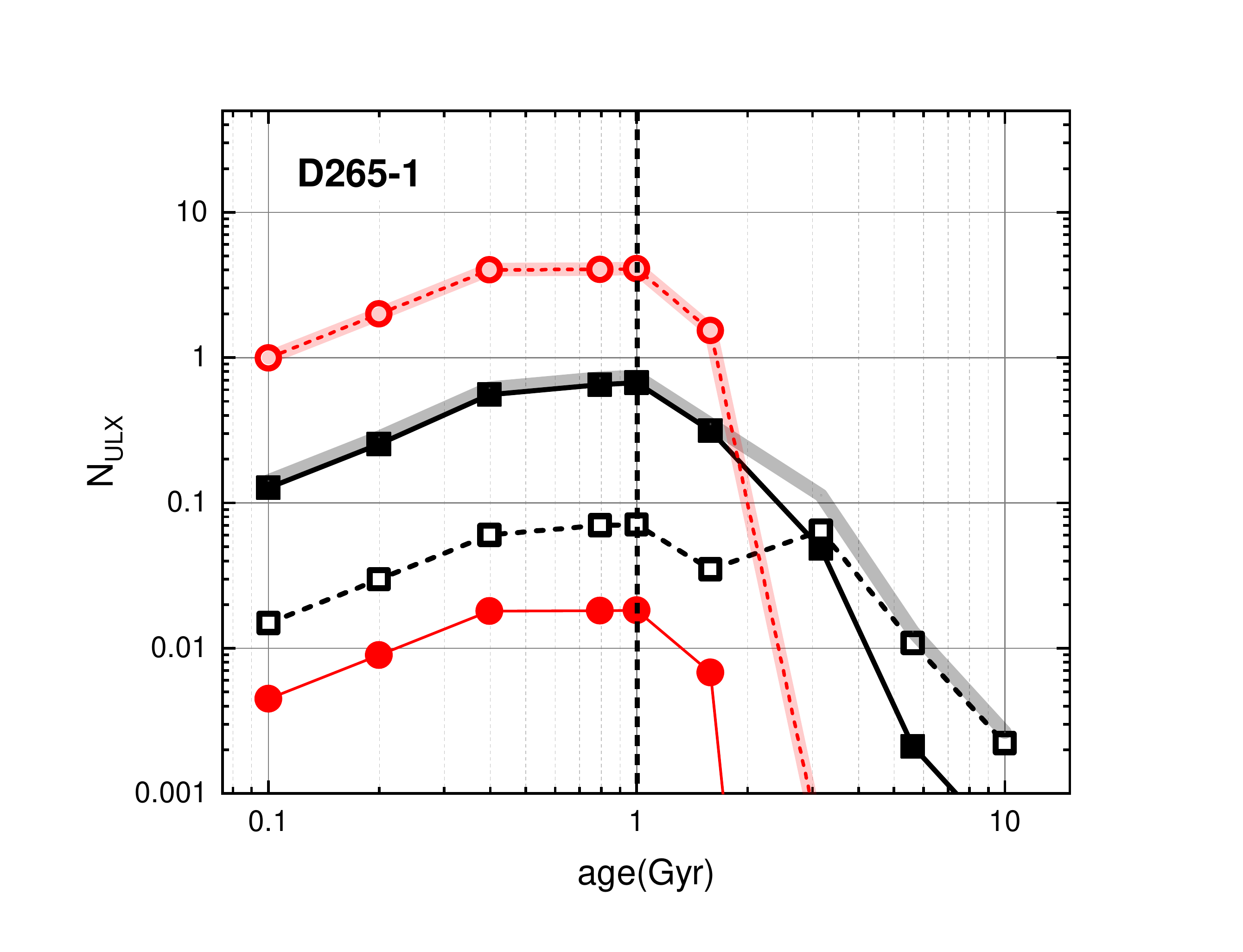}
\includegraphics[width=0.5\textwidth]{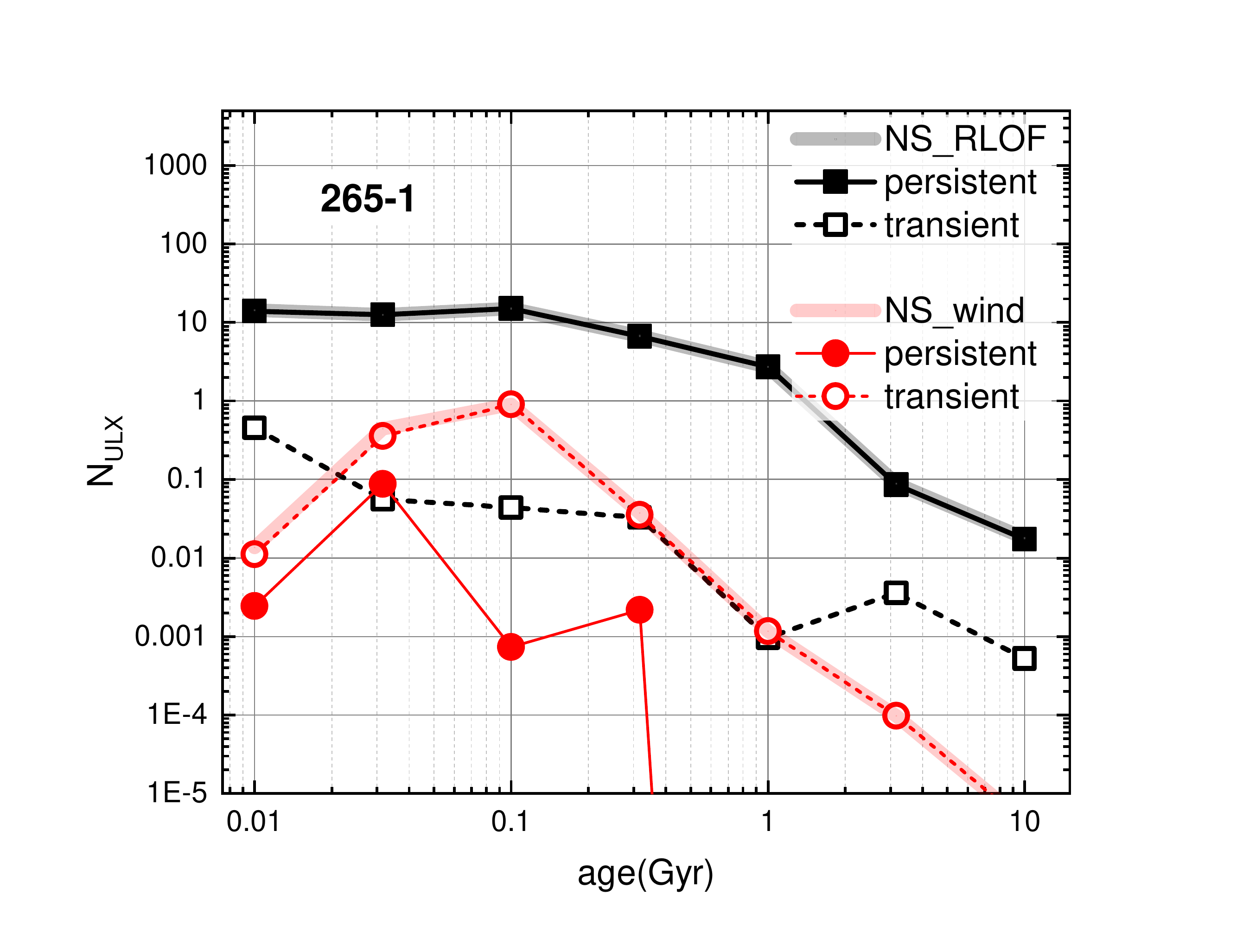}
 \includegraphics[width=0.5\textwidth]{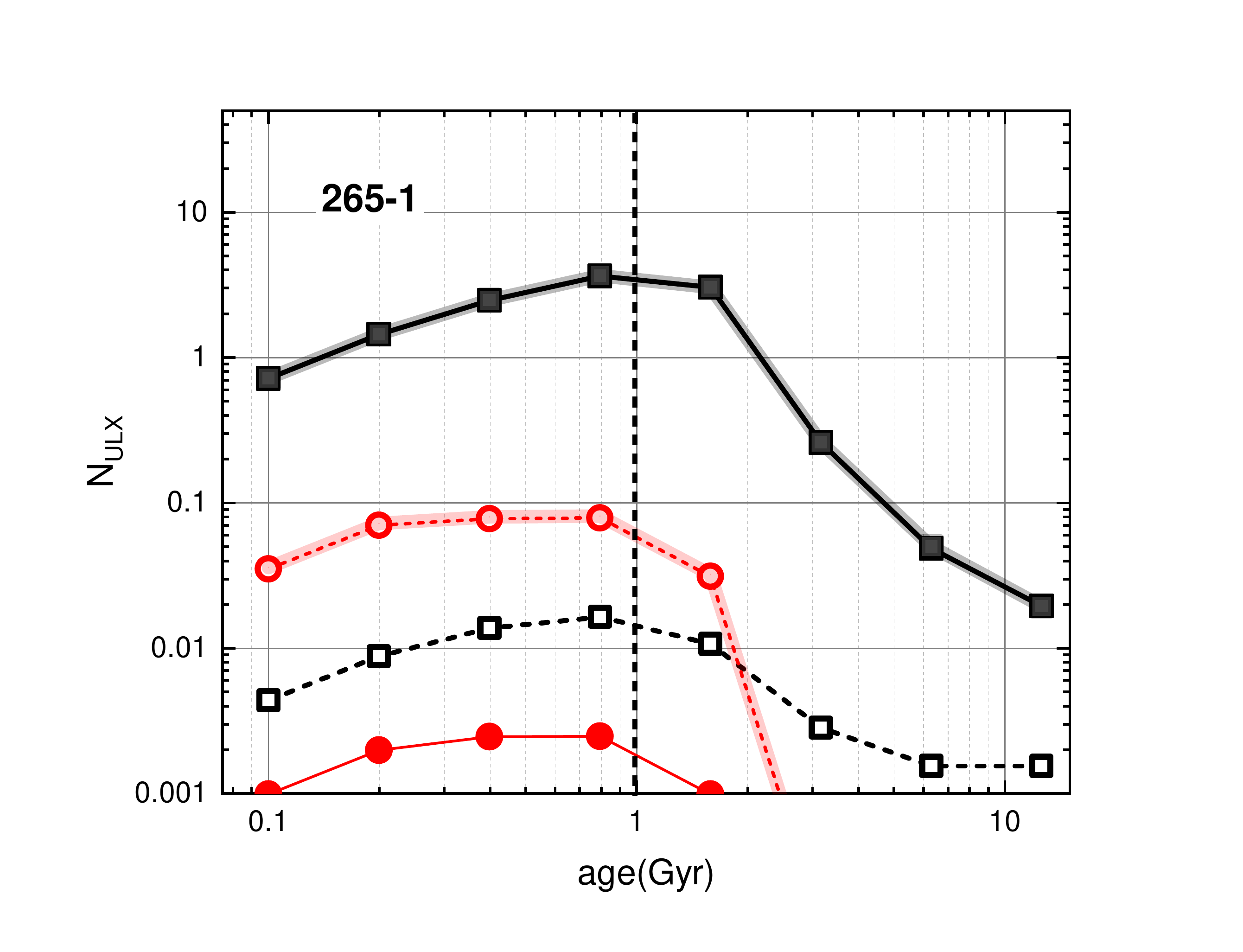}
\caption{\small Evolution of the number of ULX with persistent (solid lines) and
transient (dashed lines) disk accretion in the case of instantaneous mass formation
and star formation lasting for 10\,Gyr. Two upper rows of panels show evolution for models
C265-1 and D265-1. Panels in the lower row show the same for ULX with NS.  
Population mass $M_\mathrm{G}=10^{10}$\,\ms.}
\label{f:n_ulx_etransient}
\end{figure}

Table~\ref{tab:ulx2} shows the total number of ULX in our models. As mentioned
above, some of the sources can be transient due to the thermal-viscous
instability of accretion disks, which appears at low
accretion rates. In models C265-1 and D265-1,
approximately 70\% of all BH sources with donors overfilling their Roche lobes are
persistent. The number of sources with wind is about 2-3 times higher than those with RLOF, due to the constant replenishment of systems with
massive donors with a powerful wind. But for the same reason, practically all
sources with wind-accreting BH are transient. Among objects with NS with Roche
lobe overflowing donors, persistent sources dominate, while among
objects with wind transients dominate. As \ace\ increases, the fraction of
permanent sources, as a rule, increases, because the increase of \ace\
leads to an increase in the separation of components and the wind capture
does not yield a luminosity exceeding $10^{39}$~erg/s. The decrease of \ace\
also leads to the increase in the share of persistent sources with
overflowing Roche lobe donors; the number of persistent wind sources decreases because fewer systems reach evolutionary stages with a strong enough
wind.

Figure \ref{f:n_ulx_etransient} illustrates the evolution of the relationship
between persistent and transient sources in the cases of an instantaneous burst of
star formation and a burst lasting with a constant rate for 1\,Gyr.
$M_G$ is similar in both cases. In the first case, among the systems in which
the donor fills the Roche lobe, persistent sources dominate for about
3\,Gyr, i.e., while the donors are stars with masses somewhat larger than \msun.
For donors of lower mass, the rate of mass-loss (accretion onto BH) is too
small for stable disks to exist. When the binaries have donors that do not fill
their Roche lobes, persistent sources exist for the first $\simeq$30\,Myr only,
as long as there are donors with masses exceeding about 7\,\msun\ with a strong
enough wind in the giant stage. Entire rest of the evolutionary lifetime, except for
the first several tens of Myr, transient sources in wind-fed binaries
comprise the dominant population.

In model D265-1, the situation is qualitatively the same with some
differences, related to the difference in the masses of BH and donors. In the
case of  1\,Gyr long star formation, transient sources prevail over
persistent ones over the time of star formation and for about a similar time
after its cessation, thanks to the continuation of formation of ULX with massive
donors. Then the number of persistent and transient sources becomes comparable.
The existence of ULX after the end of star formation is possible due to the presence
of binary systems with low and moderate mass donors ($\lesssim 4$\,\msun)
overflowing Roche lobes much later after the end of star formation.

In the case of NS accretors, the systems with Roche lobe filling donors always dominate. In
these systems donor winds are usually not strong enough to provide
luminosity exceeding the threshold of $10^{39}$\,erg/s.

\begin{figure}[t!] 
\begin{minipage}{0.3\textwidth}
 \includegraphics[width=\textwidth]{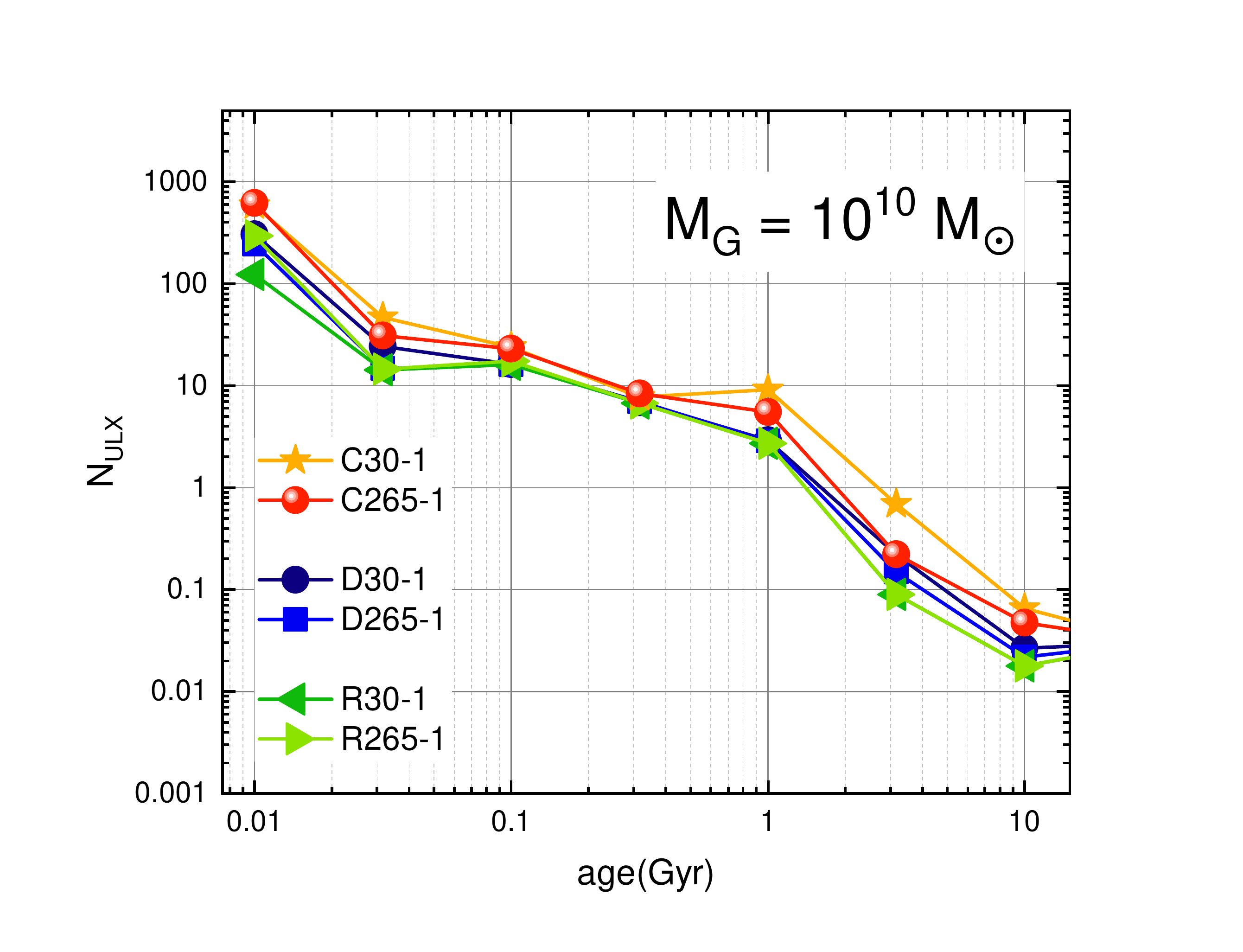}
 \vfill
\includegraphics[width=\textwidth]{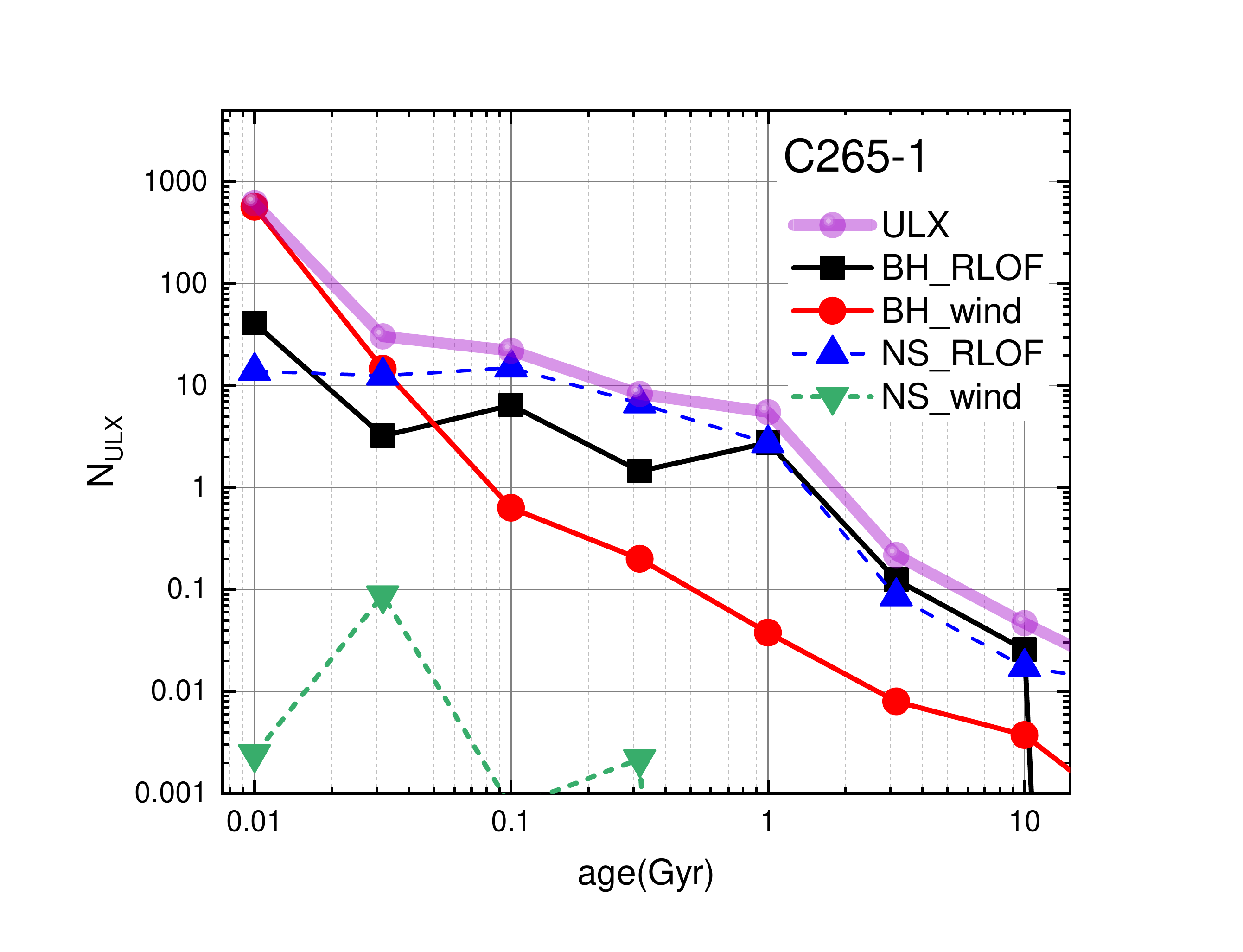}
\vfill
\includegraphics[width=\textwidth]{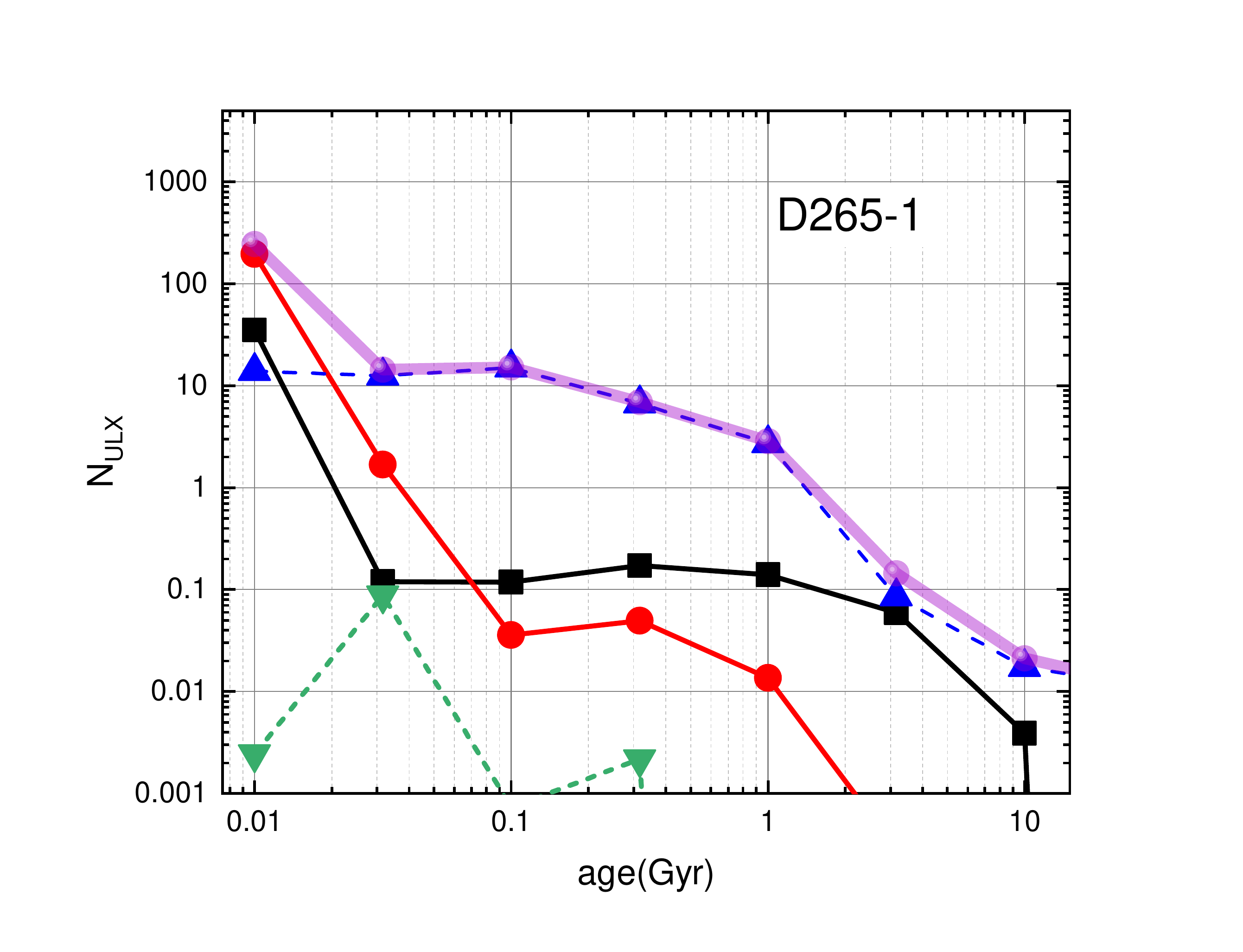}
\vfill
\includegraphics[width=\textwidth]{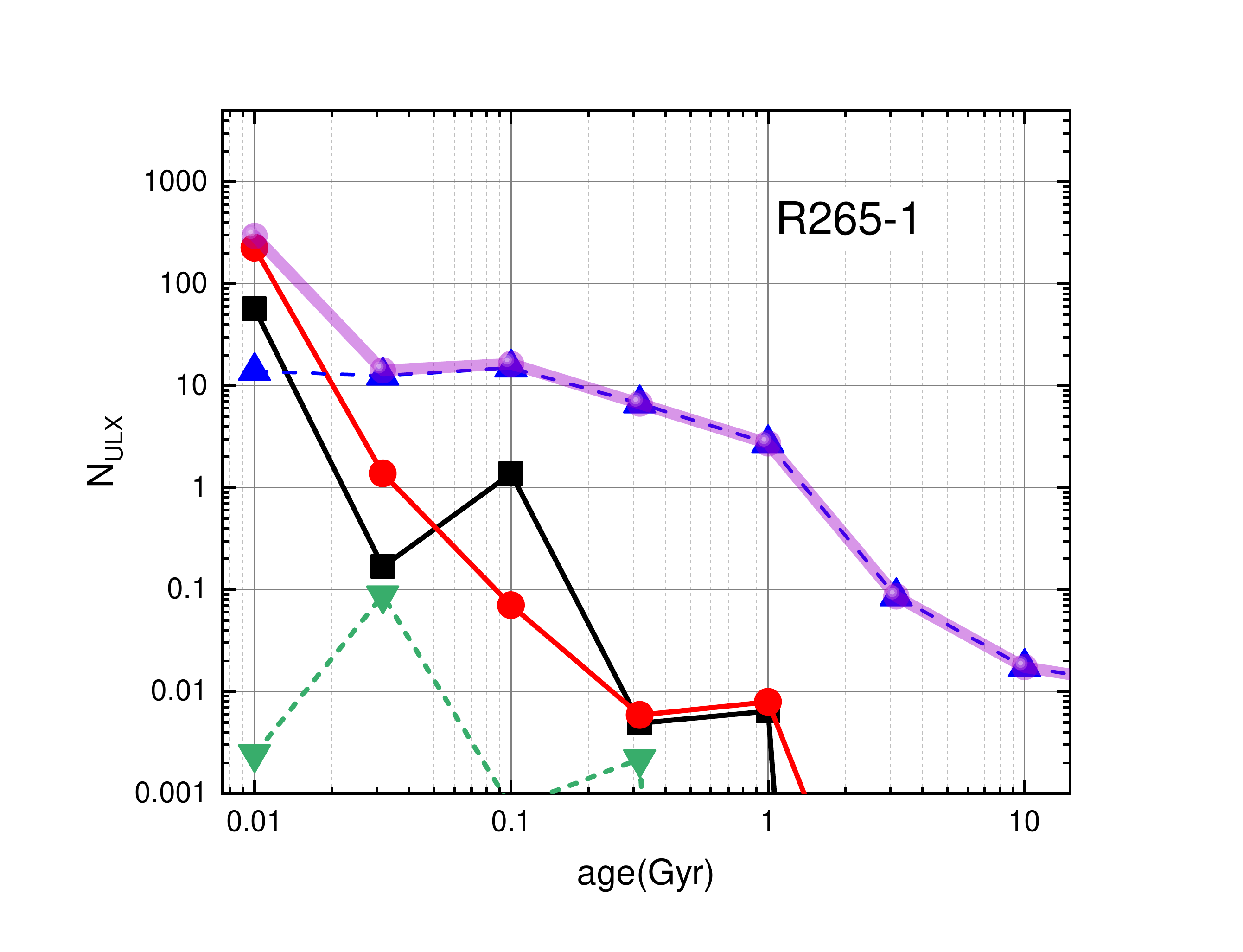}
\end{minipage}
\begin{minipage}{0.3\textwidth}
\includegraphics[width=\textwidth]{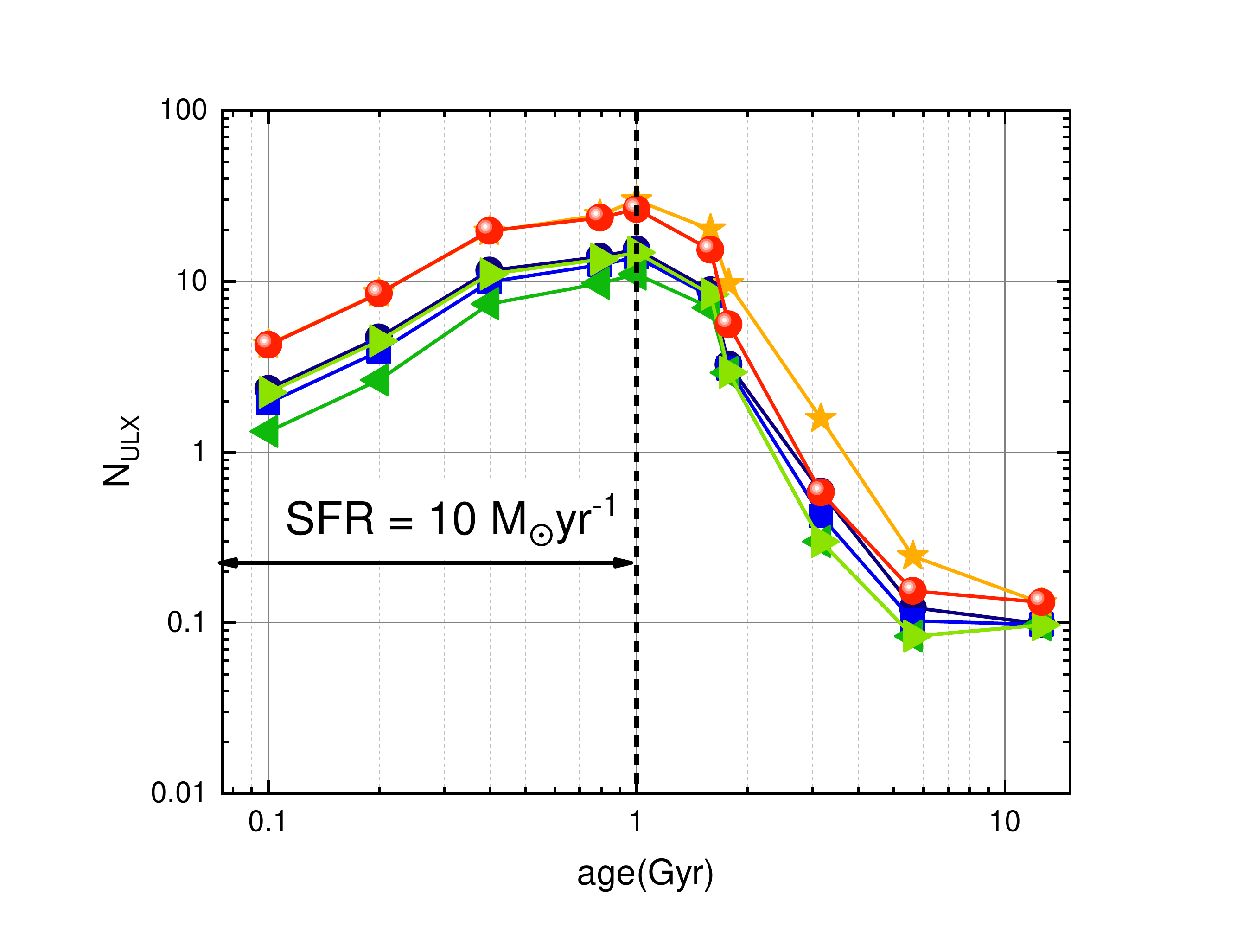}
 \vfill
\includegraphics[width=\textwidth]{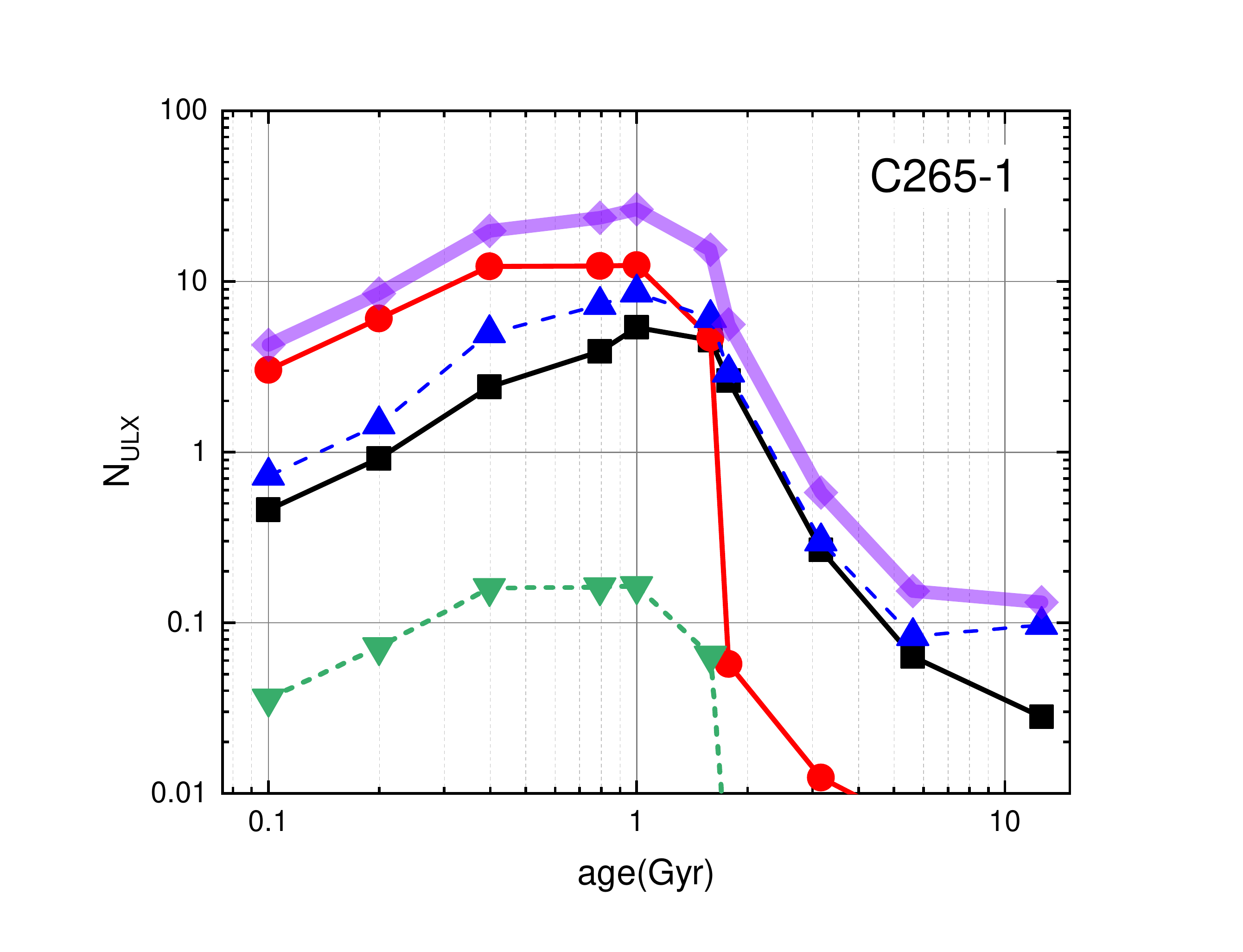}
\vfill
\includegraphics[width=\textwidth]{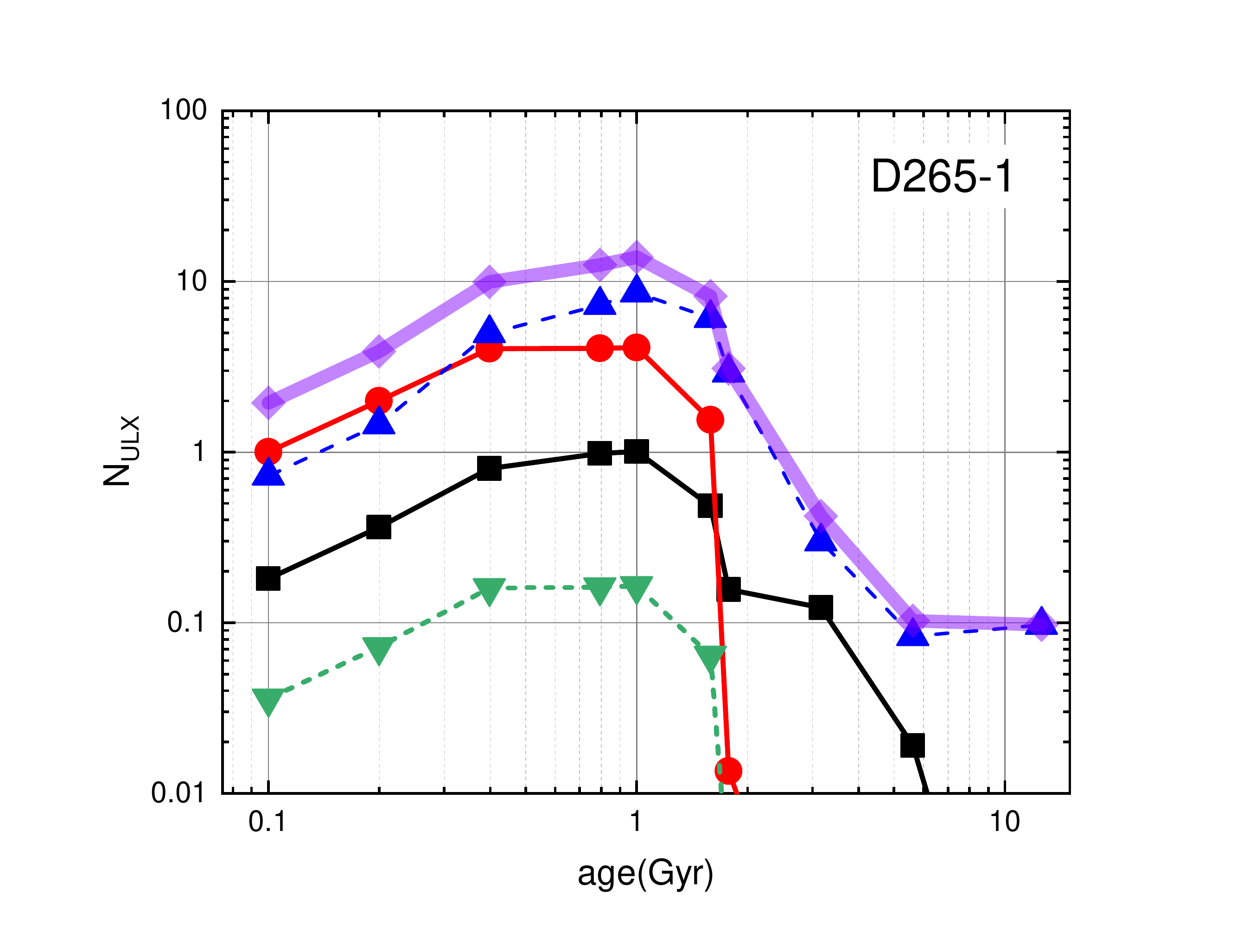}
\vfill
\includegraphics[width=\textwidth]{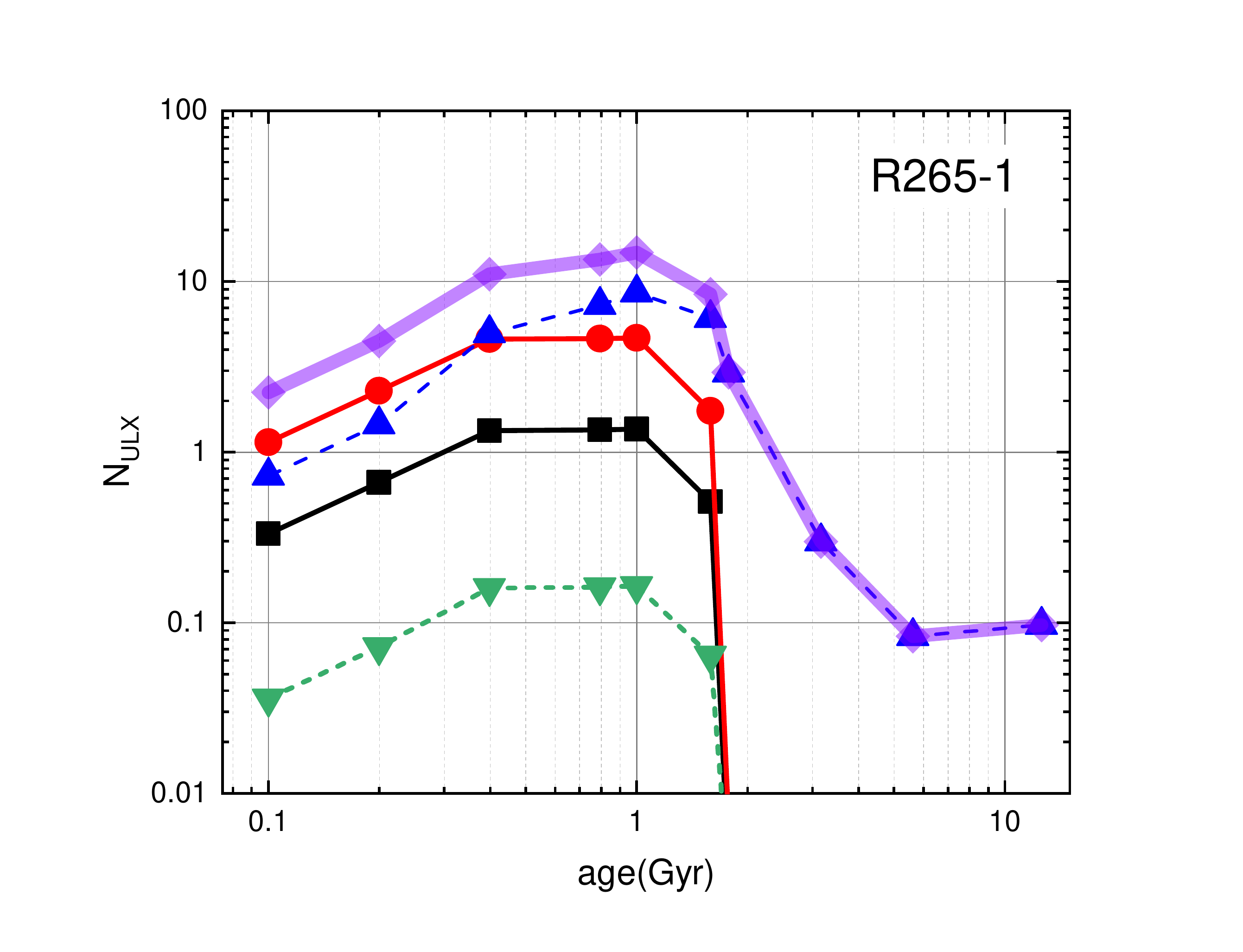}
\end{minipage}
\begin{minipage}{0.3\textwidth}
\includegraphics[width=\textwidth]{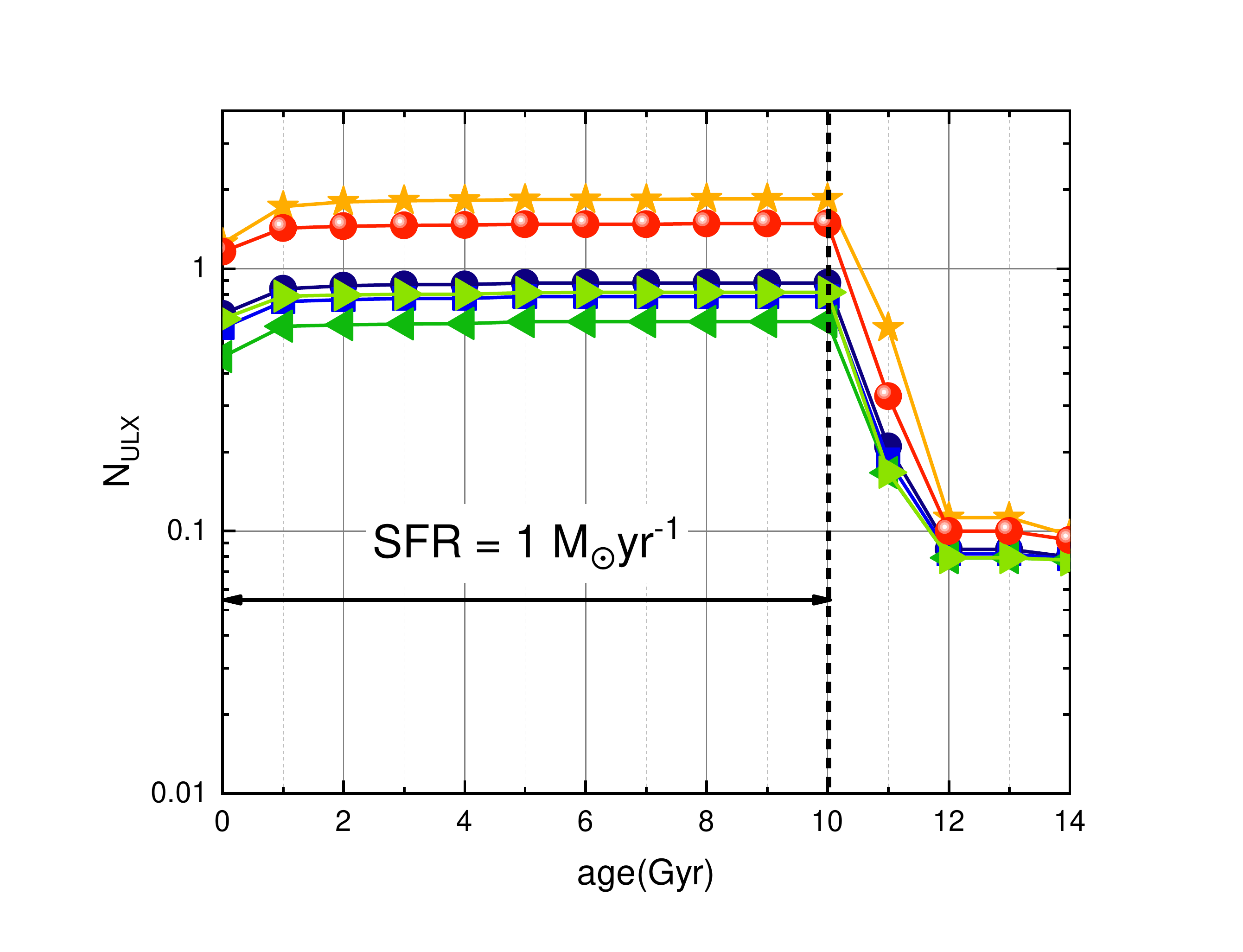}
 \vfill
\includegraphics[width=\textwidth]{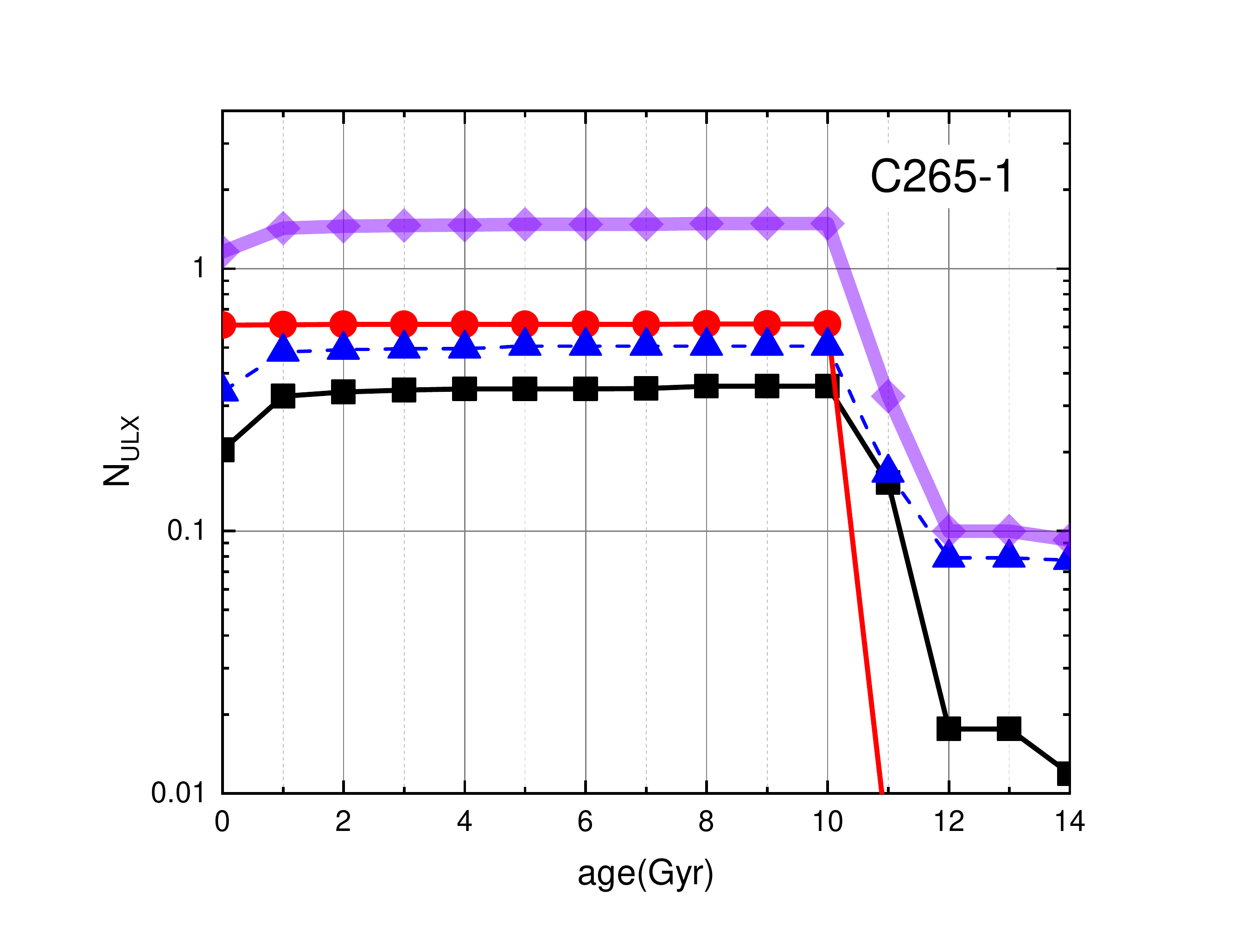}
\vfill
\includegraphics[width=\textwidth]{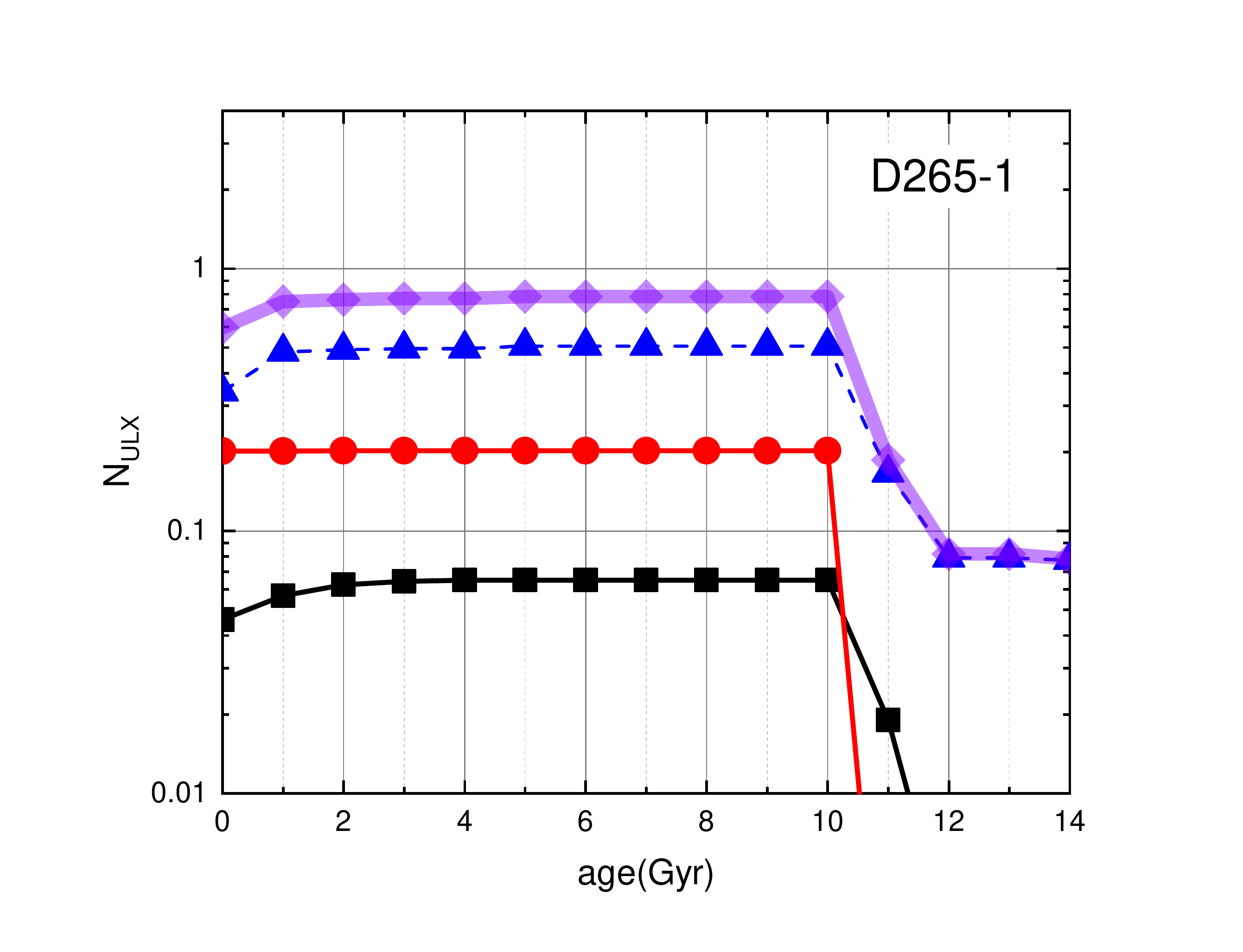}
\vfill
\includegraphics[width=\textwidth]{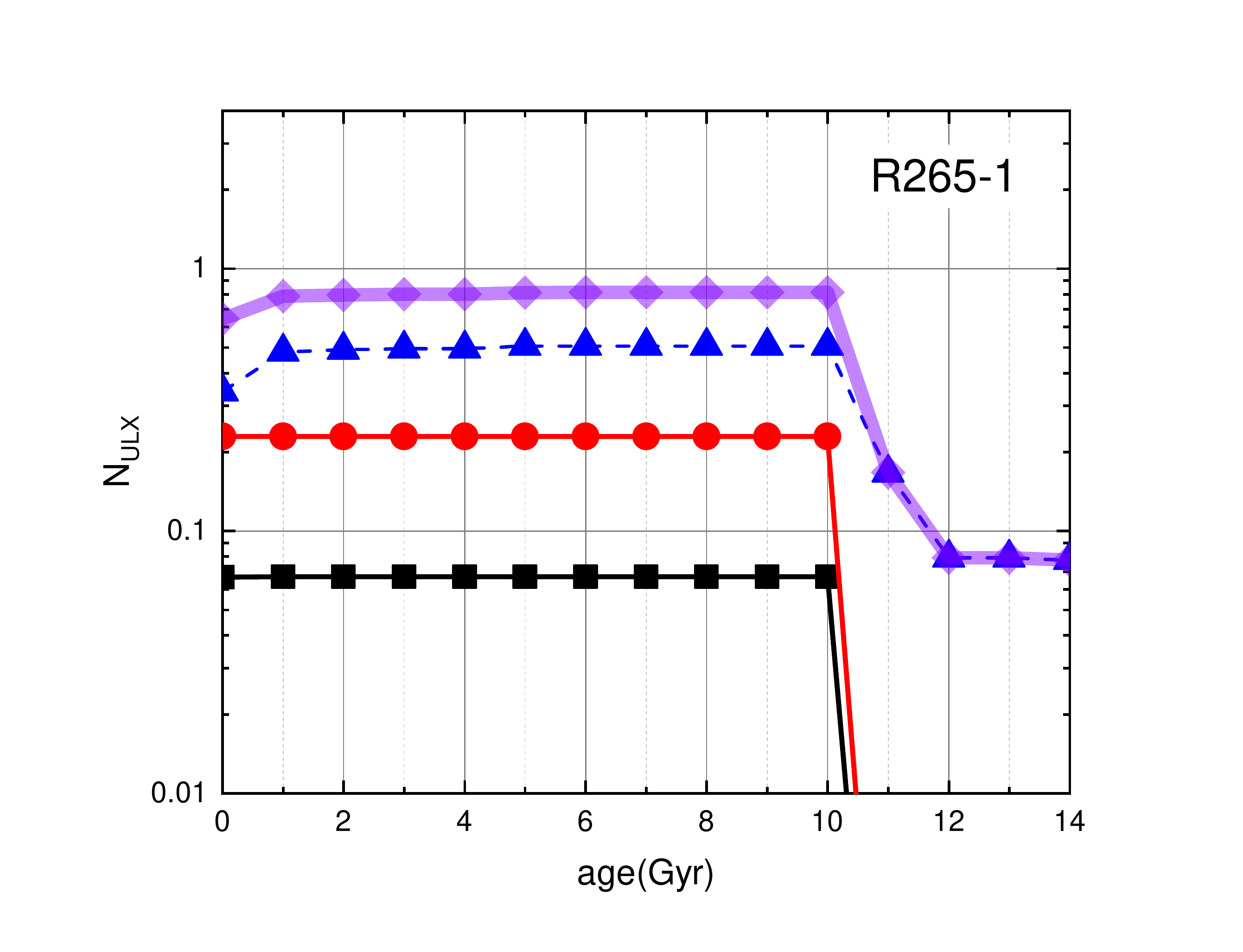}
\end{minipage}
\caption{\small
Evolution of the ULX number over time for various c.o. formation models
(C265-1, D265-1, R265-1, top
row) and accretion modes (three bottom
rows). Left column --- instantaneous star formation (normalization to the galaxy mass
$M_\mathrm{G} = 10^{10}$\msun); middle column is for constant SFR over 1\,Gyr
(normalization to SFR = $10M_\odot$/yr);
right column is for constant rate star formation over 10\,Gyr (normalization to
SFR = $1M_\odot$/yr).  }
\label{f:n_ulx_summ}
\end{figure}
\subsection*{Dependence of ULX parameters on the age of population}

As it is mentioned in the Introduction, ULX are observed both in early and late
galaxies.
However, their characteristics differ (Walton et al. 2021, Bernadich
et al. 2021). In this regard, it is interesting to find out how the ratio of
NULX and BHULX changes for different models of BH formation and modes of mass
flow (RLOF or wind-accretion), depending on the parameters of star formation and the
age of the population. The same is related to the problem of the kind of
sources prevailing in the galaxies of different ages -- BHULX or NULX? As above,
we consider in detail BH formation models C and D.  
 
Formation of ULX population was compared for $10^{10}$\,\msun\  galaxies
with instantaneous star formation, star formation with
constant rate of 10\,\msun/yr over 1\,Gyr; star formation with a constant rate
1\,\msun/yr over 10\,Gyr. In the latter case, we took the estimate of the
formation time of the thin Galactic disk from Miglio et al. (2021).

Figure~\ref{f:n_ulx_summ} shows results of the calculations of variation of the
number NULX and BHULX with time for the three considered star formation
histories and different mechanisms of BH formation. For comparison, we selected
population models in which the common envelope parameter \ace = 1 was used in
the calculations of evolution leading to the formation of ULX.

All models with different SFH have a common feature --- rapid emergence of ULX
and also a rapid decrease of their number with the termination of star
formation. For the model C265-1 with star formation over 10\,Gyr (in the right column
in Fig.~\ref{f:n_ulx_summ}, this is illustrated by the extension of evolution
until the age of 14\,Gyr). In the model with an instantaneous star formation,
the first BHULX begin to form after a few million years since the outburst,
with the collapse of the most massive stars and reaching by their companions the
stage of giants. The ULX stage is generally short ($\lesssim 10^7$ yr, see
examples in the Appendix).

 By 10\,Gyr, the initial number of sources
decreases by 4-5 orders of magnitude. Thus, if we would take for all model
galaxies instantaneous burst of star formation and similar mass, at the moment
$t=10$\,Gyr there should be one ULX with BH per several dozen galaxies.

\begin{figure}[t!] 
\includegraphics[width=0.9\textwidth]{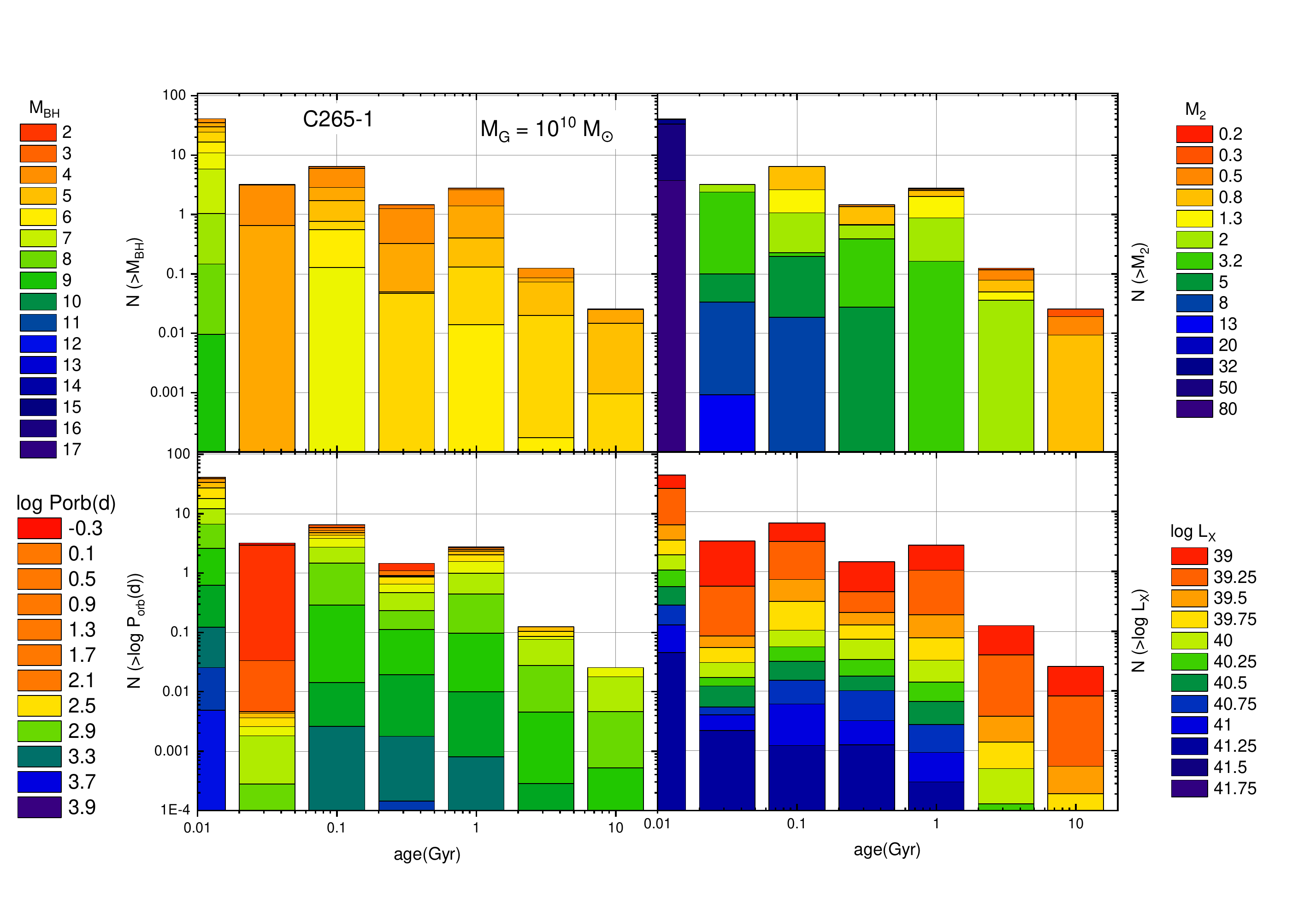}
\includegraphics[width=0.9\textwidth]{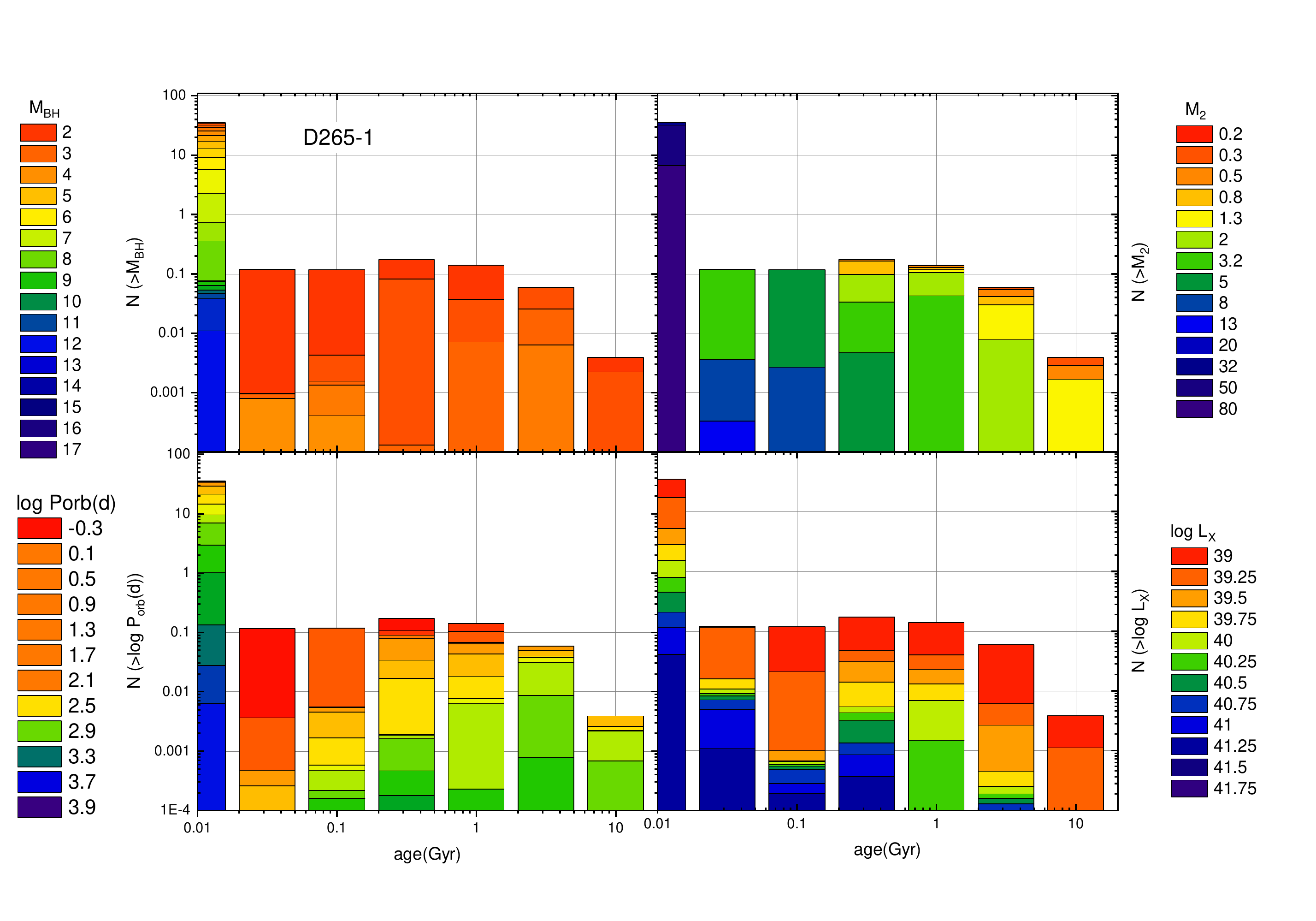}
\caption{\small Distributions of ULX with BH and a donor filling the Roche lobe,
by the mass of components (top panels), orbital periods (lower
left panel) and X-ray luminosity (lower right panel) depending
on the time after the star formation burst. The upper plot is
for the C265-1 model, the lower one is for the D265-1 model. The
plots are normalized to the galaxy mass $ M_\mathrm{G}=10^{10}$\,\msun.}
\label{f:M4_M11_BH_RLOF1}
\end{figure}

As it is shown by Fig.~\ref{f:n_ulx_etransient}, ``long-living'' sources with BH
are transient. The difference in the mechanisms of BH formation leads to the
difference in the BHULX number which at any time does not exceed a factor $\sim
10$. Noteworthy is a rather similar evolution of the total population of ULX for
c.o. formation models C and D.

\begin{figure}[t!] 
\includegraphics[width=1.0\textwidth]{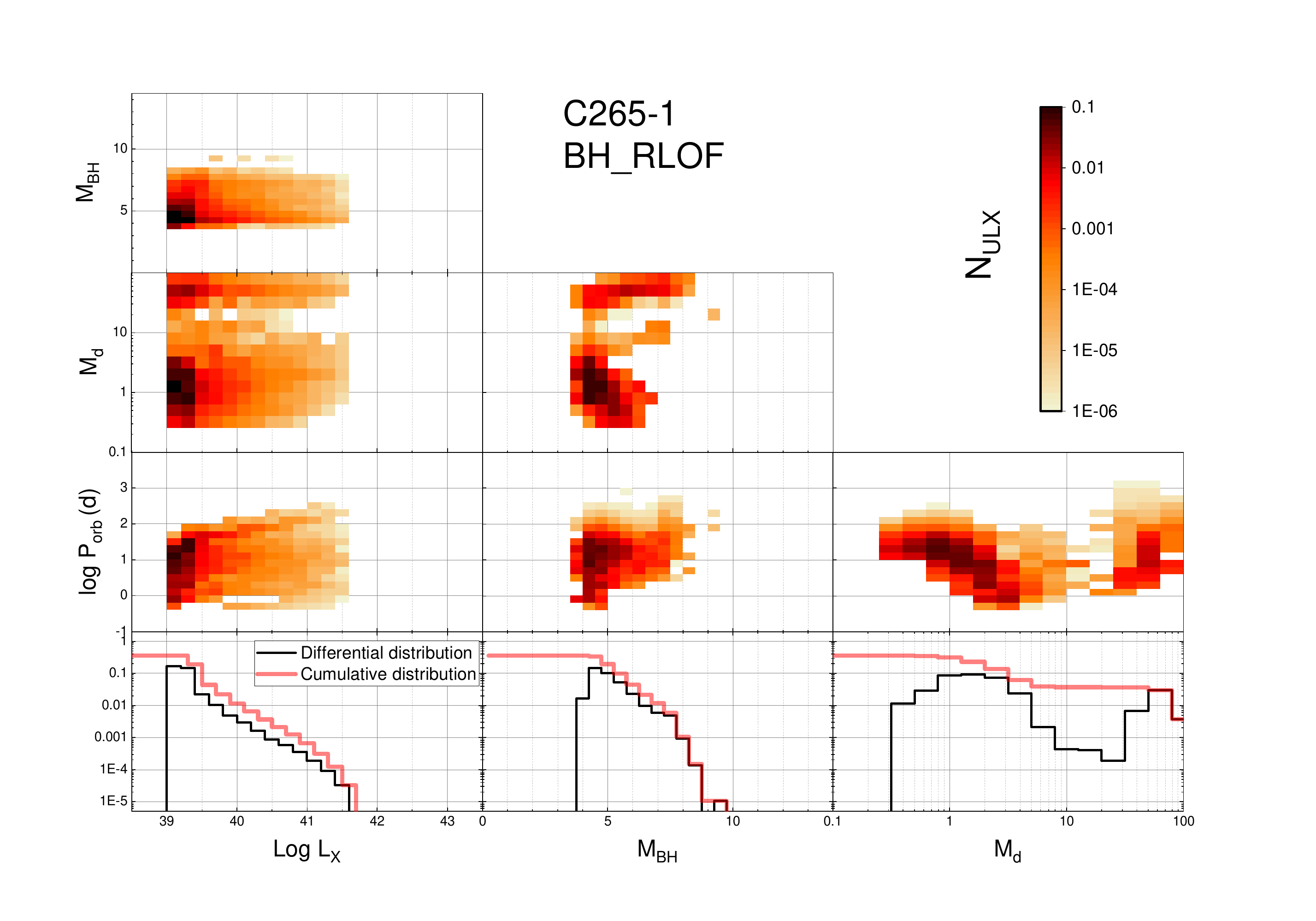}
\caption{\small Distributions of BHULX in model C265-1 with the Roche lobe
overflowing donors over X-ray luminosity,
orbital periods, masses of components in the population with a constant
SFR=1\msun/yr at $t=$\,10~Gyr  (see Table 1).
}
\label{fig:ULX_3D_model_CO_RLOF}
\end{figure}

\begin{figure}[t!] 
\includegraphics[width=1.0\textwidth]{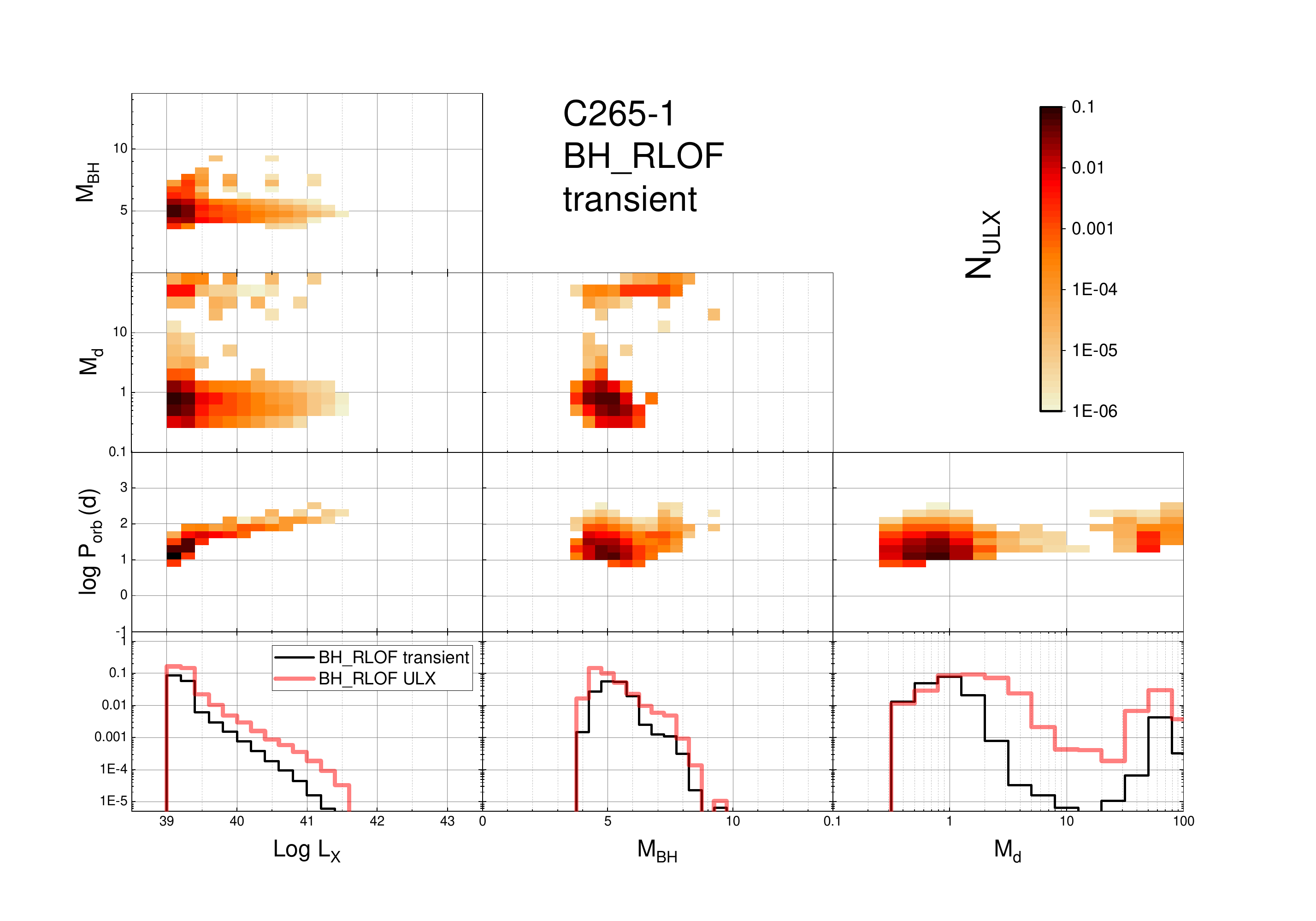}
\caption{\small Same as in Fig.~\ref{fig:ULX_3D_model_CO_RLOF} for transient   BHULX with the Roche lobe
filling donors  at the outburst stage
(in active state).  
In the lower panels the distributions of transients are compared to distributions
of total population of BH\_RLOF in model C265-1. }
\label{fig:ULX_3D_model_CO_RLOF_tr}
\end{figure}

\begin{figure}[ht] 
\includegraphics[width=1.0\textwidth]{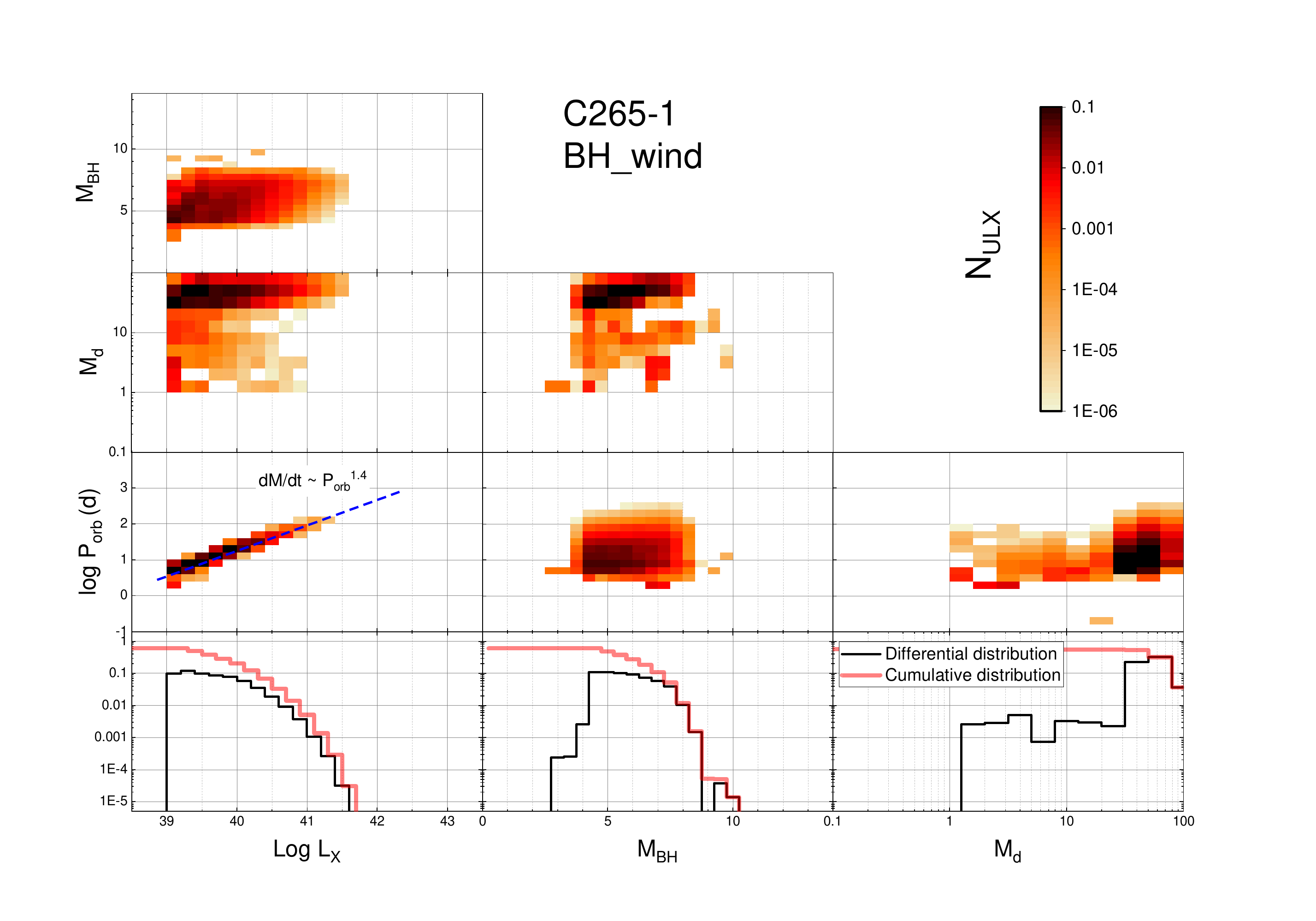}
\caption{\small Same as in Fig.~\ref{fig:ULX_3D_model_CO_RLOF} for the systems with wind-accretion
and disk formation around BH. Model  C265-1. Most of the sources are transient.
Dashed line --- relation between luminosity ($L_{\rm X,Edd}$) of disks
during outbursts and periods of sources (Dubus et al.,
1999). }
\label{fig:ULX_3D_model_CO_wind}
\end{figure}

\begin{figure}[t!]  
\includegraphics[width=1.0\textwidth]{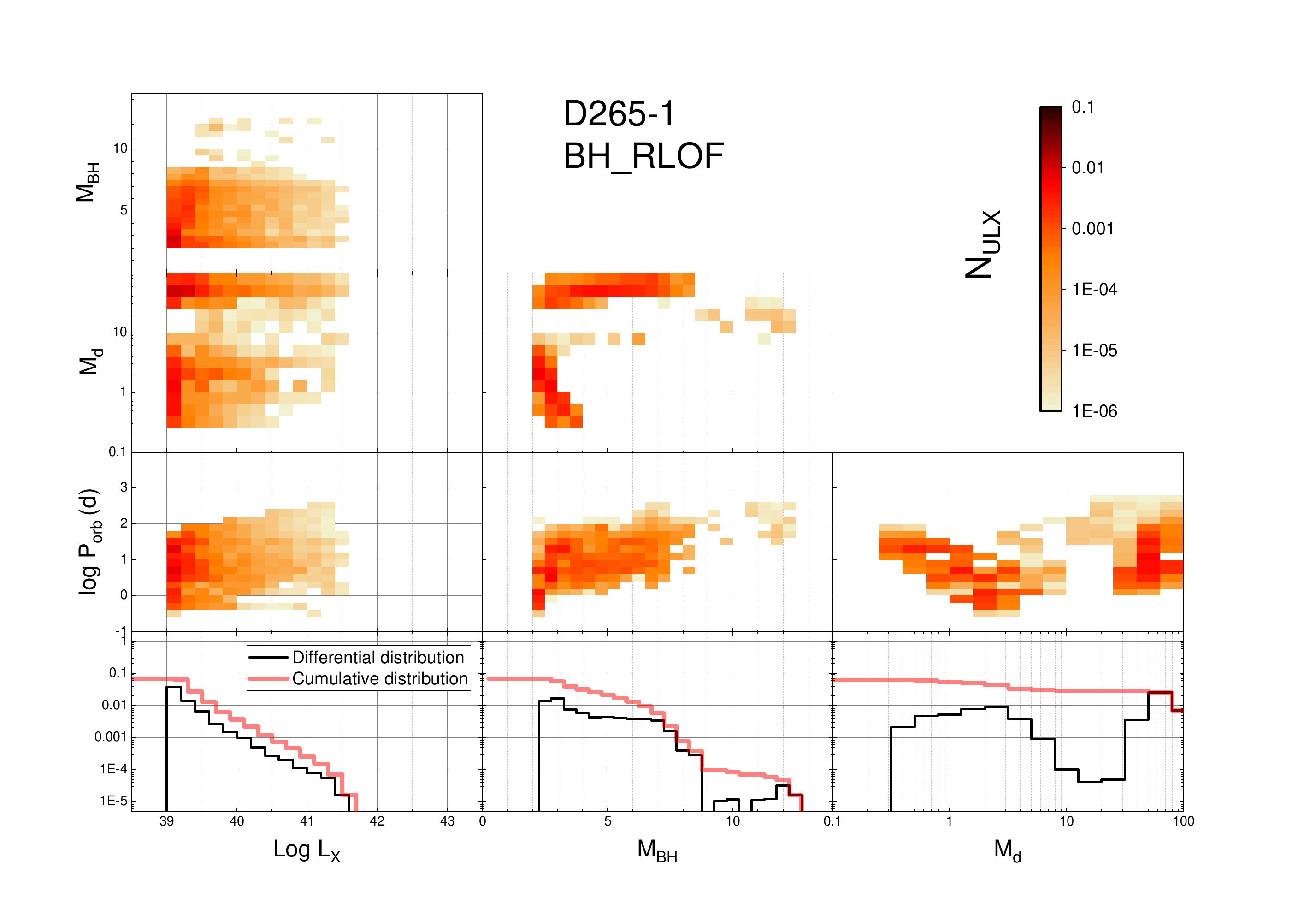}
\caption{Same as in Fig.~\ref{fig:ULX_3D_model_CO_RLOF} for the model D265-1 with Roche lobe
overflowing donor.}
\label{fig:ULX_3D_model_D}
\end{figure}

\begin{figure}[t!] 
\includegraphics[width=1.0\textwidth]{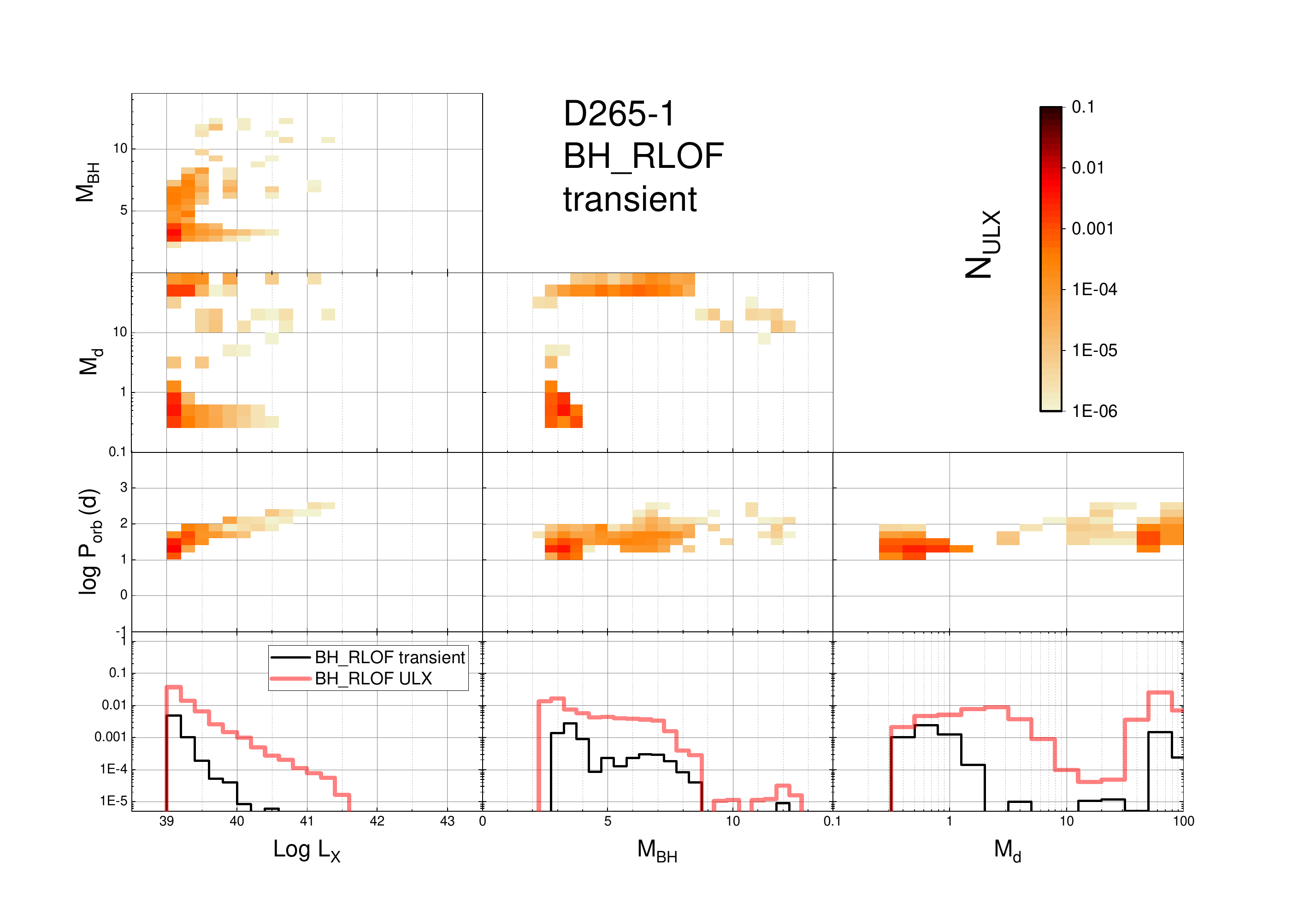}
\caption{Same as in Fig.~\ref{fig:ULX_3D_model_CO_RLOF} for transient BHULX with Roche lobe overflowing
donors at the outburst (in active state). Model D265-1.}
\label{fig:ULX_3D_model_D_RLOF_tr}
\end{figure}

\begin{figure}[t!] 
\includegraphics[width=1.0\textwidth]{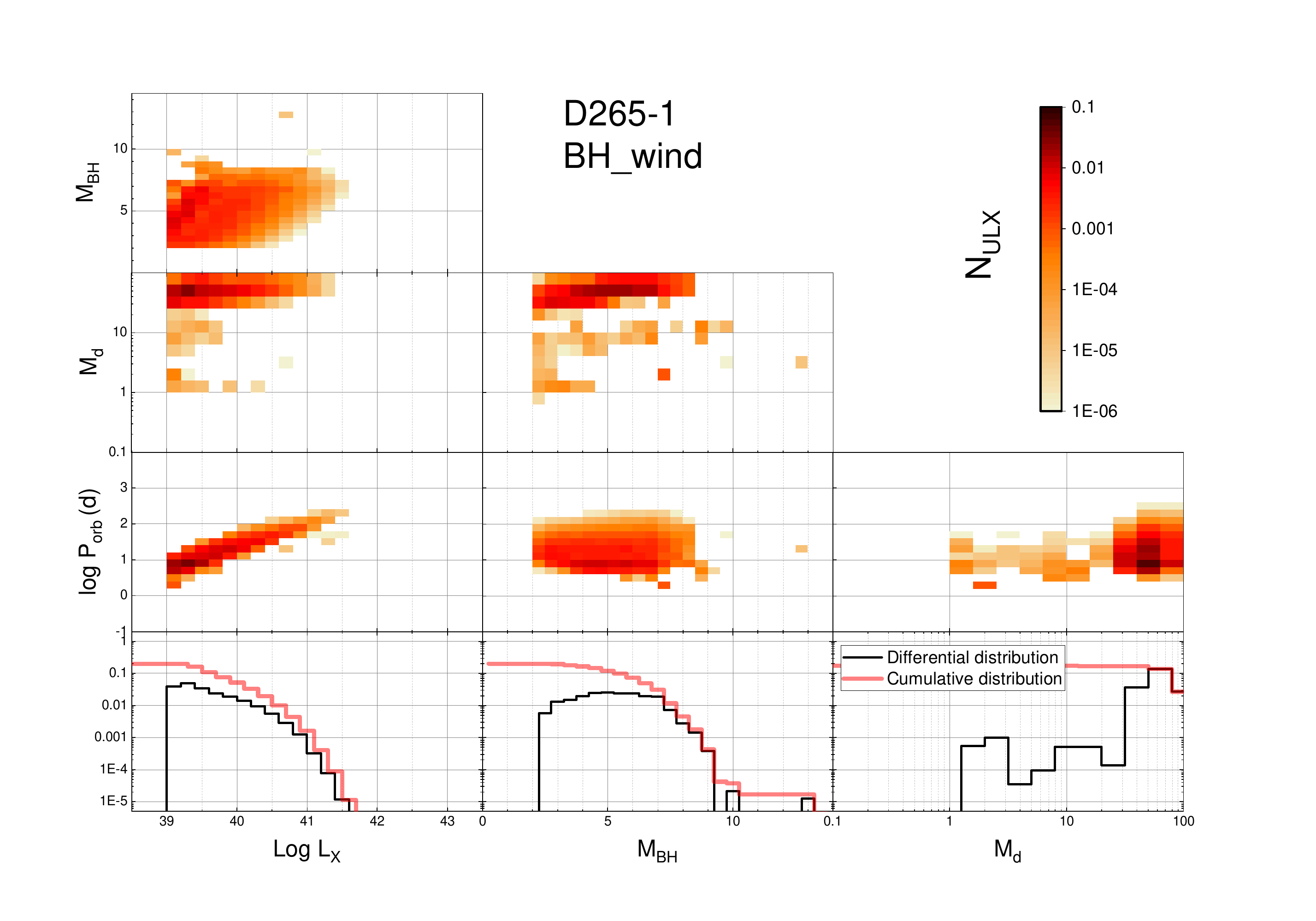}
\caption{Same as in Fig.~\ref{fig:ULX_3D_model_CO_RLOF} for the model
D265-1 with wind-accretion. Most of the systems are transient.}
\label{fig:ULX-3D_model_CO_wind}
\end{figure}

In the more realistic cases of star formation that lasts for 1 and 10\,Gyr, the
number of BHULX at the star formation stage somewhat increases: the sources that
have completed the ULX stage are replaced by newly formed sources with similar
characteristics and, in addition, the sources with donors of lower masses
formation of which takes a longer time to replenish the population. After completion
of the star formation stage, the number of ULX remains almost constant for some
time, due to the existence of semidetached objects, in which BH companions
are low-mass stars ($\sim 1$\,\msun) experiencing mass exchange in case B.
Then the ULX population drops (see.
Figs.~\ref{fig:ULX_3D_model_CO_RLOF} -- \ref{fig:ULX-3D_model_CO_wind}).
 
Figure~\ref{f:M4_M11_BH_RLOF1} illustrates the evolution of BH masses $M_{\rm
BH}$, donor masses $M_2$, orbital periods $P_{\rm orb}$ and X-ray luminosity
$L_{\rm X}$, in the models C265-1 and D265-1 after an instantaneous burst of star
formation. Models with a prolonged star formation can be considered as the sum of
similar outbursts, but on a smaller scale.

In the model C265-1, BH with masses exceeding $\simeq$9\,\msun\ are absent
and BH with $M_{\rm BH}\approx$(3-5)\,\msun\ dominate (see
Fig.~\ref{f:mzams_mbh3}). Model D265-1 is dominated by BH with masses $\sim
3$\,\msun. Black holes of a higher mass (up to $\approx$14\,\msun,
Fig.~\ref{f:mzams_mbh3}) are also formed, but the lifetime of systems with large
BH masses is short.

Model C265-1 after $t\approx$ 100~Myr, is dominated by BHULX with
$M_2<1.5$\,\msun, which, as a rule, must be persistent
(Fig.~\ref{f:n_ulx_etransient}). Throughout the entire time of evolution, \bux\
with orbital periods $\lesssim$\,300 days prevail. The same should be observed
in the model with a continuous star formation. Luminosity of \bux\ are mainly
confined to the interval $10^{39} -- 2\times 10^{39}$\,erg/s.

In model D265-1, BH masses are slightly lower than in model C265-1. In
terms of donor masses, model D265-1 practically does not differ from C265-1 ---
the donors with masses $M_2<1.5$\,\msun\ also dominate. But typical orbital
periods in this model are $\aplt$100 day. X-ray luminosity of ULX is slightly lower than in model C265-1, they only slightly exceed $10^{39}$\,erg/s, due
to lower accretion rates in closer systems.


\begin{figure}[t!] 
\includegraphics[width=1.0\textwidth]{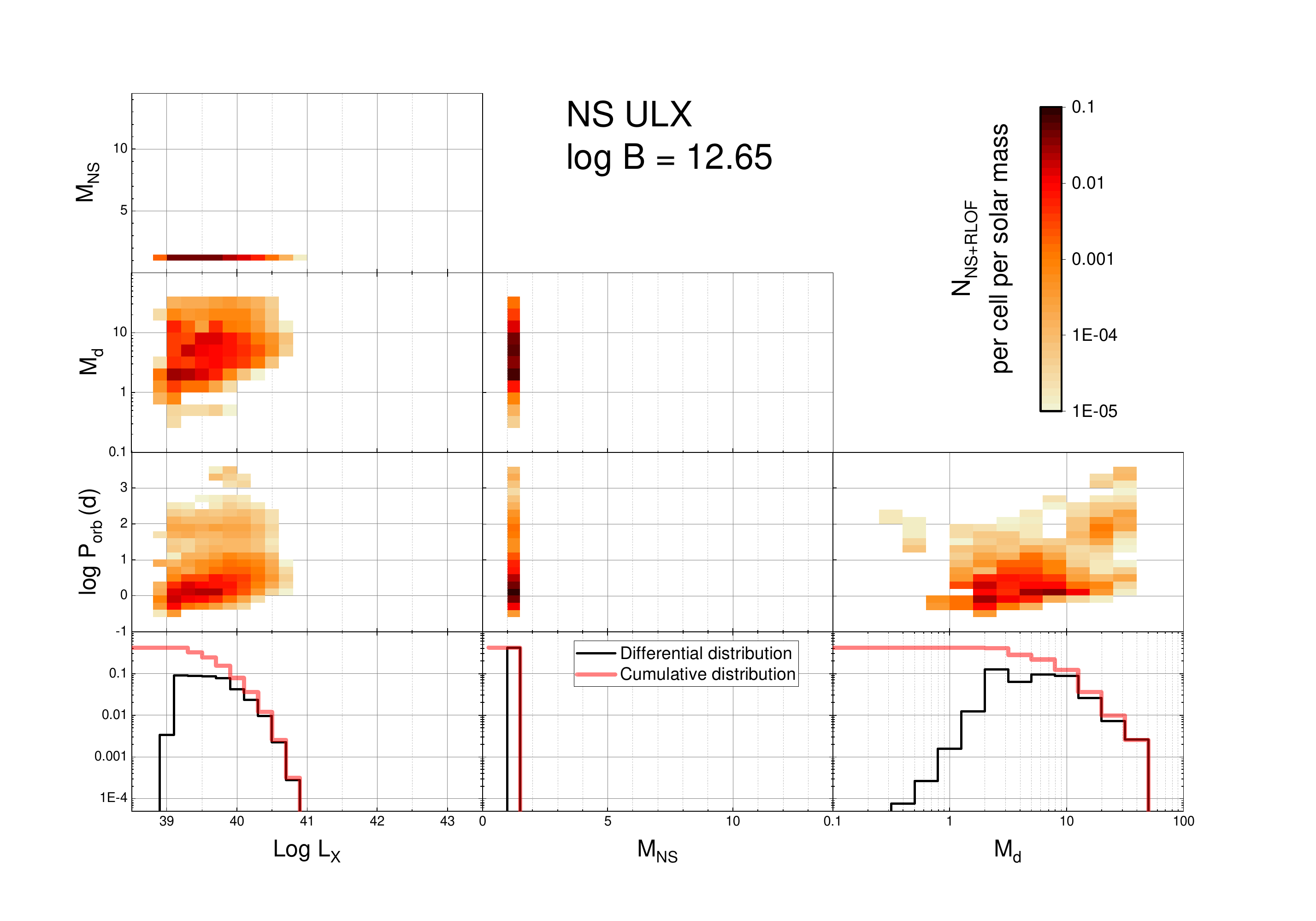}
\caption{Model distributions of ULX with NS and Roche lobe overflowing donor over X-ray luminosity,
orbital periods, and component  masses in a galaxy with constant SFR at $t=$10\,Gyr. The distributions
are normalized to SFR=1\,\msun/yr.}
\label{fig:ULX_3D_model_NS_B12}
\end{figure}

In the Figs.~\ref{fig:ULX_3D_model_CO_RLOF} --- \ref{fig:ULX-3D_model_CO_wind}
models C265-1 and D265-1 are compared for the case of a constant rate of star
formation SFR = 1\,\msun/yr over 10\,Gyr. For each model, we compare
distributions of parameters for systems with RLOF (BH\_RLOF) and accretion from
the wind (BH\_wind). Essentially, these are characteristics of the ULX
population with BH in a spiral galaxy. Figure~\ref{fig:ULX_3D_model_NS_B12}
shows for comparison the relations of distributions for the same parameters for
NULX, obtained in Paper I for a model with the standard characteristic magnetic
field $\log B$ = 12.65.

Characteristic features of the models are as follows. For the model C265-1 for
BH\_RLOF concentration of the masses of BH and donors to the minimum values and
low values of orbital periods is inherent. Such a combination ensures a stable
exchange of matter. The most massive BH allowed by model C (up to $\approx
11$\,\msun) are absent in the `` observed '' BH\_RLOF ULX due to their short
lifespan.  

For ULX with accretion from the wind, a smaller spread of BH masses with a
smaller concentration around the minimum values is typical. In this case, the
orbital periods are, predominantly, from $\simeq$\,10~day to $\simeq$\,50~day,
but can exceed 100~day. Donor masses can reach $\simeq$\,70\,\msun. The latter
means the possibility of the existence of donors-(super)giants, which is consistent
with observations (see detailed discussion in Fabrika et al. (2021) review).
Note the existence of donors with masses 1-2\,\msun\ and large X-ray
luminosity achieved during outbursts. Most sources should be transient, which
is consistent with conclusions based on Fig.~\ref{f:n_ulx_etransient}.

The luminosity of most of BH\_RLOF sources is in the range $(1-3)\times
10^{39}$\,erg/s, i.e. at the  threshold luminosity over which X-ray sources
are classified as ULX. The bulk of BH\_wind sources
reaches luminosity $ 10^{40}$\,erg/s, but it should be borne in mind that this
is the luminosity of unstable accumulated disks in outbursts.

Models D265-1 with RLOF and with wind accretion are characterized by a more
uniform distribution of parameters. This is related to the larger than in model
C spread of the BH masses and natal kicks. Although ULX with accretors formed in
model D may have lower than in the models C BH masses, the upper limit of their
masses is also close to 8~\msun.
Among the donors of BH\_RLOF models, there is no clearly expressed concentration
to masses $\lesssim$3\,\msun. As well as in models C, there are donors
with masses up to 100\,\msun, thus, the population should contain ULX with
super-giant components. In this ``family'', a significant fraction of the sources
have comparable masses, enabling a stable flow of matter. There are donors with
masses $\aplt$\,\msun. The sources harbouring them must be transient. Just like
among the models C265-1 BH\_RLOF, the overwhelming fraction of ULX has
a luminosity not exceeding approximately $3\times 10^{39}$\,erg/s, but
there is a ``tail'' stretching up to $3\times 10^{41}$\,erg/s.

Models D265-1 BH\_wind, in essence, differ from models C265-1 BH\_wind, like
models D265-1 BH\_RLOF, by a more even distribution of parameters. BH masses do not show a pronounced concentration to (4-7)\,\msun\ and are
evenly distributed in the range (3 --  8)\,\msun. The fraction of systems with $M_d\lesssim$\,25\,\msun\ is insignificant, donors with masses below 20\,\msun\
are practically absent. The range of orbital periods in D265-1 BH\_wind models
is the same as in models C265-1 BH\_wind, but the latters demonstrate a lower
concentration to relatively short periods. As a result, D265-1 BH\_wind systems
are less crowded in the $L_{\rm X}10^{39} -- 10^{40}$\,erg/s range. The
highest luminosity is the same --- $(3 - 4)\times 10^{41}$\,erg/s.

\section*{Discussion of results}
\label{s:disc}

Results presented above depend on model assumptions usually used
in the population synthesis studies.

\begin{itemize}
    \item Parameters of compact object formation.
 As noted above, the formation of
          ULX with BH is determined by the process of formation of BH resulting
          from the stellar core collapse (which defines the BH mass, possible
          natal kick), and the rate of accretion onto c.o. The masses of BH
          being formed in all the considered models do not exceed 15\,\msun. In
          our calculations, standard assumptions on the evolution of stars of
          solar chemical composition were used. In alternative scenarios (e.g.,
          chemically homogeneous evolution of massive stars in close binary
          systems, Marchant et al., 2017), in stars with low metallicity, BH
          masses can reach 60\,\msun. At this, the number of bright ULX with
          $L_\mathrm{X}>4\times 10^{39}$ erg/s in galaxies with stationary
          star formation with the rate 1\,\msun/yr can be about 0.13. This
          number is comparable to our result for solar $Z$ (see the bottom left
          panel in Fig.~\ref{fig:ULX_3D_model_CO_RLOF}).    
    \item Parameters of stellar evolution prior to the formation of the compact object.
One of the most significant uncertainties in the treatment of the evolution of
massive stars is associated with stellar wind mass loss. As
Figs.~\ref{f:BSE_CO} and \ref{f:BSE_D} show, by the time the progenitors of BH
fill their Roche lobes, a significant fraction of them are giants and
super-giants. There are observational and theoretical arguments that
$\dot{M}_{wind}$, determined from observations and calculated theoretically that
are usually used in the population synthesis and evolutionary programs, are
overestimated (see, e.g., Beasor et al., 2021; Fink, 2021). This also applies to
the codes BSE  and MESA. In such a case, the distances between components after the
end of the mass-loss via $\mathrm{L_1}$ must be larger than computed ones, while
for common envelopes the situation is opposite. In both cases, this can
decrease the total number of \bux. Also, this may increase the number of
wind-accreting sources which are mainly transient.
As it was noted by Kippenhahn and Weigert (1967) in their pioneering work on the evolution
 of close binaries, CO cores of helium remnants of stars after the loss of matter
during RLOF  are less massive
    than the cores of single stars of the same initial mass.
    The reason is that in single stars the masses of helium cores increase as a result of hydrogen
burnout in the shell source, while in the remnants of components
of binary systems, convective cores decrease as a result of wind mass loss due \footnote {Laplace et al. (2021) studied this issue in detail, but
for the mass interval corresponding to the NS progenitors only. They also
 drew attention to the fact that the isotopic composition of pre-supernovae in these cases is different.}.  
 The difference in the core masses can lead to the difference in the values of mass
that delimit NS and BH progenitors in single stars and components of binary systems.
 It should be noted that
in population synthesis codes, as a rule, the relations $M_{in}-M_{fin}$ for single stars are used.
  \end{itemize}.

\subsection*{The number of ULX}
\begin{table}[b!]
\vspace{6mm}
\begin{center}
\caption{\small The ratio of numbers of  \bux\ and \nux\
in a model galaxy with constant SFR=1\,\msun/yr at $t=10$\,Gyr for different models
of BH formation.}
\label{t:ratio}
\begin{tabular}{l|c|c|c|c|c|c}
\hline\hline
$\ace$        & C265 & C30  & D265 & D30  & R265 & R30 \\
\hline
 0.5 & 0.96 & 2.51 & 0.67 & 0.07 &  -  &  -   \\
 1   & 1.89 & 2.61 & 0.52 & 0.69 & 0.58 & 0.24     \\
 4   & 0.04 & 0.04 & 0.04 & 0.07 &  -  &  -    \\
\hline\hline
\end{tabular}
\end{center}
\end{table}
Figure \ref{f:n_ulx_summ} shows that by the time $t$=10\,Gyr in a galaxy with a
constant SFR=1\,\msun/yr for 10\,Gyr the total number of ULX should be
approximately 300 times larger than the number of ULX in a galaxy of the same
mass with instantaneous star formation. For a galaxy with a starburst lasting
1\,Gyr, the ratio is close to 15. However, these ratios may turn out to be not
entirely correct if we take into account that metallicity in old galaxies, as a
rule, is significantly lower than solar. Unfortunately, the evolution of the CBS
with $Z\ll Z_\odot$ has not been systematically investigated. In a single study
of the evolution of CBS with donor masses up to $\approx 53$\,\msun,
$Z_{\mathrm{Fe}}/Z_{\mathrm{Fe,\odot}}\approx 0.2$  and 0.36
 and a fixed initial mass ratio
of the donor and c.o. equal to 0.6, Klencki et al. (2020, 2021b) found that the
transfer of matter, which begins at the core He-burning stage, can occur both
in a long nuclear time scale ($\simeq 10^5$\,yr) and in a fast thermal ($
<10^5$\,yr) time scale. As the duration of mass transfer increases, the
probability of the existence of semi-detached ULX with massive giant donors
increases. In both cases, the hydrogen envelope is not lost completely. It is important to note that the ``new'' modes of mass loss do not significantly increase the
mass threshold of BH progenitors (up to $\approx$\,25\,\msun) and, as a
result, the total number of \bux\ should change only slightly.

Since one of the main objectives of our study was to consider the impact of
parameters of the evolutionary scenario on the relative number of BHULX and NULX,
in Table ~\ref{t:ratio}, we present the ratios of the numbers of  \bux\ and \nux\
for various c.o. formation models. and \ace\ for a galaxy with a constant
SFR = 1\,\msun/yr for 10\,Gyr, since the results
obtained for Z = 0.02, can be considered as the most justified.

Table~\ref{t:ratio} suggests that for \ace=0.5 and 1 the ratio of the numbers of
BHULX and NULX
in model C is $\approx$(0.5 -- 2.5). The models
with \ace=4 fall out of the general trend, for them the ratio is a few hundredths.
However, such \ace\ means that the energy required to eject the common envelope
should be much larger than the orbital energy of the binary. Possible additional
energy sources, for example, the release of recombination energy and other
processes accompanying the formation and ejection of common envelopes are actively
studied, but so far do not allow definite conclusions;
numerical modeling of this process is still beyond the current possibilities (see Ivanova et al. (2020)
and references therein). Therefore, taking into account all the uncertainties and
simplifications of the population
synthesis, it can be argued that the abundance of ULX with accretors --- BH or NS in
the galaxies with a constant star formation rate are comparable. The same can
be stated for models D265-1 and R265-1 (Fig.~\ref{f:n_ulx_summ}).

Concerning transient sources, it should be borne in mind that the duty cycle
is, primarily, the function of the orbital period of the system
and the accretion rate and may be larger or smaller by a factor  2--3  
from the value of 30
yrs accepted by us. Therefore, the number of transient
sources is rather uncertain. As noted by Hameury and Lasota (2020),
for long duty cycles, which may last, according to their calculations,
$\simeq 60$\,yrs, some of the sources that are observed as persistent, in
reality can be transients caught in quiescence.

We note also that classification of an X-ray  source as  NULX is based either on observations of
coherent pulsations of X-ray radiation (Bachetti et al., 2014) or
the presence of the cyclotron line scattering feature in the X-ray
spectrum (Walton et al., 2018). Both features are sufficient, but not
necessary, since accreting NSs can have magnetic fields in the
range from $\sim 10^8$ to $\sim 10^{14} $~G. Pulsations of an accreting
magnetized NS can be suppressed in the propeller mode (Tsygankov et al., 2016) or
get blurred during the interaction of radiation with the matter at the stage of supercritical
accretion onto the NS magnetosphere (S.A.~Grebenev, {\it in preparation}). The cyclotron
feature in the X-ray spectrum of ULX can be observed in a certain range of
magnetic fields of NS only, and its origin in radiation-dominated
accretion columns requires additional research. These selection effects
can underestimate the observed ratio of NULX and BHULX, leaving the most reliable
criterion of large mass c.o. as a sign of BHULX in non-pulsating sources.

A significant uncertainty factor for the ratio of \bux\ and \nux\ is the rate of
NS formation as a result of electron captures in stellar cores (ECSN, see
Poelarends et al., 2017 and references therein). This phenomenon should be
accompanied by low natal kicks (Dessart et al., 2006). This significantly
reduces the frequency of disintegration of binaries compared to the
formation of NS accompanied by a ``standard'' Maxwellian natal kick with
$\sigma{(v_k)}$=265\,km/s.    

\subsection*{Comparison to other studies}

Theoretical studies of the populations of ULX are not numerous.
Let compare our results with other studies employing population synthesis.

In the calculations by the StarTrack code (Wiktorowicz et al., 2017, 2019), it was
found that for $\mathrm{Z=Z_\odot}$ the  number of NULX begins to exceed the number of
BHULX at 100\,Myr since the star-formation burst.
Wiktorowicz et al. did not use the hybrid method to calculate the rate of mass loss
by the optical component overflowing Roche lobe. 
Moreover, in these papers, it was assumed that the beaming factor is the same for supercritical
accretion onto
BH and NS. This is not the case for magnetized NS (see our Paper I and the analysis of Mushtukov
et al., 2021). Up to the accepted normalization, the total number of ULX in our
calculations (see Fig.~\ref{f:n_ulx_summ}) generally agrees with the results of Wiktorowicz
et al. for $Z=Z_\odot$ (see Fig.~2 in
Wiktorowicz et al. (2017) and Fig.~1 in Wiktorowicz et al. (2019)).

In a recent paper by Wiktorowicz et al. (2021),
accretion from the stellar wind was considered. In this paper, the interpretation
of Bondi-Hoyle-Littleton accretion for
elliptical orbits is, however, significantly different from that in our study.
Averaged over an orbital period, accreted mass should be  practically independent of
the eccentricity of the orbit, while in the approximate formula (2) in the paper by
Wiktorowicz et al. (2021) the fraction of donor wind accreted by c.o. depends on $e$
as $\sim 1/\sqrt {1-e^2}$. In some
models considered by Wiktorowicz et al. (2021), the number of wind-accreting ULX exceeds
the number of
ULX accreting via RLOF by the optical companion. However, no such case
was found in our models (see Fig.~\ref{f:n_ulx_summ}, top row). Apparently, this difference
is related to the different treatment of the rate of accretion from stellar winds.

After the discovery of the first pulsating source M82 X-2 (Bachetti et al., 2014), Shao
and Li (2015) considered the model of ULX model with NS accretors, but, differently with our study, they did not take into account the specific effects due to the interaction
of NS magnetosphere with an accretion disk. They assumed the same beaming
factor $b=0.1$ for all NS. Model systems were selected by the criterion
$L_{\rm X} > 10^{39}$\,erg/s. Thus, by accepting the certain history of star formation,
Shao and Li (2015) model can be considered as a
sample of the population of X-ray sources with NS, which reached a high luminosity due to
a fixed geometric factor. Taking the rate of Galactic
star-formation to be 3\,\msun/yr over 13\,Gyr, Shao and Li estimated
that in the Galaxy at present there should exist about 30 ULX with NS (with
donor masses exceeding 2\,\msun), which slightly exceeds our estimate
of the order of 1 system per 1\msun, even taking into account that, according to Paper I, we
took the beaming factor equal to $\approx 0.3$.
 
The same authors (Shao and Li, 2020) found that in the Galaxy (with the star formation history described above) $\sim 10$\ ULX with BH can exist. They used
Raithel et al. (2018) model in which the helium core of presupernova collapsed
(taking into account the Nadezhin-Lovegrove effect). Natal kicks were modeled
using Hobbs et al. (2005) distribution, with a scaling factor 3\msun/$M_{\rm
BH}$. Taking into account the differences in the assumptions about the star
formation, pre-supernovae masses and natal kicks, this model
is roughly consistent with our model C265-1 for \bux.

In contrast to the papers of Wiktorowicz et al., Shao and Li have used, like us,
the hybrid population synthesis and applied for elliptical orbits the averaged over
orbital period accretion rate from the stellar wind onto c.o.., but do not
consider the formation of transient ULX via accretion from unstable disks around
c.o.

Thus, the main differences between our calculations and above-mentioned studies
is an account for the possibility of transient accretion onto c.o., which may lead
to the formation of ULX both during RLOF by the visual star and during accretion
from stellar wind and the interpretation
of accretion onto magnetized NS (see details in Paper I). Results of independent
calculations of the ULX population by different groups under similar assumptions
about the formation of c.o. and parameters of the evolution of CBS are,
generally, consistent. Therefore, it is extremely important to measure the
parameters of ULX in various galaxies to get their observed distributions with the
purpose of clarifying the ways of formation of ULX.

\section*{Conclusion}
\label{s:concl}

Currently, supercritical accretion onto compact objects (neutron
stars and black holes), first considered by Shakura and Sunyaev (1973),
is observed as a phenomenon of ultraluminous X-ray sources. In Paper I, applying
hybrid population synthesis, we analyzed in detail the evolution of
ULX with magnetized NS at the stage of supercritical accretion and
have shown that they reproduce the range of parameters of pulsating sources
(PULX). In the hybrid method, the stages with accretion onto compact stars
are calculated in detail taking into account the evolution of the visual
component.  In this study, we continued Paper I
studying the formation of ULX with BH in massive binaries in
galaxies with different star formation histories (proxies for late-type galaxies
with ongoing star formation and for old elliptical galaxies).

We have considered several models of BH formation during the collapse of
massive stars cores: model C, in which the mass of BH is determined by the
mass of the CO-core of the star before the collapse and models D and R --- delayed and rapid
BH formation upon the collapse with a fallback onto
proto-NS (Fryer et al., 2012), which are often used in the literature. We
assumed that the nascent BH get natal kicks  with a
Maxwellian distribution and a characteristic velocity of 265~km/s, scaled
by the fraction of the collapsing core mass falling onto  proto-NS  (models
C, D, R265) and models with Maxwellian distribution and a characteristic velocity of  30~km/s (without scaling
by the fraction of in-falling) matter (models C, D, R30).
In the calculations of the evolution of the orbits of
binaries, three values of the parameter of efficiency of common envelopes were tested:
\ace = 0.5, 1, 4. In the calculations
of X-ray luminosity generated by accretion of matter onto c.o.
the possible transient character of disk accretion due to thermal-viscous
instability was taken into account after Dubus et al. (1999). Observed X-ray
luminosity $L_\mathrm {X}$ from supercritical accretion disks around BH
was scaled by the beaming factor according to King (2009) prescription.

Results of calculation of the number of ULX in a model galaxy with a constant rate of
SFR = 1$M_\odot$/yr at the time of 10\,Gyr are summarized in Table 1. In the
parentheses
the number of persistent sources is shown. Table 2 shows the ratio of
BHULX and NULX numbers for a galaxy with a stationary star formation for various models
of c.o. formation. The number of ULX with BH is comparable to or prevails over the number of
ULX with NS in model C (except for the value of the common envelope parameter \ace = 4) and,
conversely, it is lower than the number of NULX in D and R models.

Two models (C265-1 and D265-1) with parameter \ace=1 were studied in more detail
 (model R only slightly differs from model D):
\begin{itemize}
\item
Figures \ref{f:BSE_CO} and \ref{f:BSE_D} exemplify the evolution of CBS
leading to the formation of systems with BH and optical components overflowing
Roche lobes. Calculations were made by modified BSE code (Hurley et al. 2002).
\item
Figure \ref{f:n_ulx_etransient} shows evolution of the number of
persistent and transient ULX with BH and for comparison --- ULX with NS --- in
the sources with RLOF and wind-accretion after an instantaneous  star-formation burst (left column of plots) and in a model with stationary
SFR = $10M_\odot$/yr for 1\,Gyr (right column of plots). In models C265-1
with accretors-BH, by the time $t=10$\, Gyr transient sources completely dominate
This concerns both models with RLOF and models with wind accretion. It is
remarkable that the number of objects observed by $t=$10\,Gyr weakly depends on
the star-formation model.    
   
In models D-261, by the same time after the start of star-formation also
dominate transient sources with RLOF and in both models of star-formation their
abundance is comparable, like for the model of c.o. formation C265-1. In both
cases, this is explained by the fact that these transient sources are close
binaries, in which the donor with mass $\sim$\msun\ experiences case B of
mass-exchange (after leaving the main-sequence, see Fig.~\ref{fig:trc_old_age}).
For stars with M$\sim$\msun\ main-sequence lifetime is inversely
proportional to mass in the power 3 -- 4 and the difference in the star-formation
time of a Gyr is insignificant.    
In the models with NS, regardless of the adopted star-formation model,
persistent sources with RLOF dominate. The number of transient sources is by an order of
magnitude lower. The number of sources with wind accretion is by several orders
of magnitude below than sources with RLOF.    
\item In Fig.~\ref{f:n_ulx_summ} the evolution of the ULX number is compared for
different models of the BH formation (C, D, R) and star formation history
with the subdivision of sources with BH and NS accreting via RLOF and via
stellar wind from the visual component. The maximum number of ULX per galaxy
(about 10) is achieved in models C. The number of ULX with NS can be comparable
(and after the end of star-formation --- can exceed) the number of ULX with BH.
Note that the sources which are observed after the completion of star formation
are CBS in which BH were formed before the end of star formation, while
long-living donors with mass $\sim$\,\msun\ filled their Roche lobes after
the termination of star formation.
\item Figures \ref{f:M4_M11_BH_RLOF1} -- \ref{fig:ULX-3D_model_CO_wind} show
distributions of ULX sources with BH over the masses of the latter $M_{BH}$, masses of visual components $M_d$, orbital periods $P_{orb}$ and apparent
X-ray luminosity $L_\mathrm{X}$ in a model galaxy with a constant
SFR of 1\,$M_\odot$/yr at the time of 10 Gyr after the beginning of star
formation. The figures also present differential and cumulative
distributions of X-ray luminosity and masses of components. The Figures
 show separately the systems accreting via RLOF and via stellar wind. For
comparison, in Fig.~\ref{fig:ULX_3D_model_NS_B12} we show similar
 distributions for ULX with NS accreting via RLOF.
\end{itemize}

Examples of calculated rates of mass transfer through  $\mathrm{L_1}$ point
and the resulting accretion rates onto BH are shown in the Appendix.

Calculations of the number of ULX with BH made in this study for galaxies
with different star-formation histories can be used to clarify the formation channels of ULX which are a subject of discussion in the current literature.
The masses of the BH  candidate in specific ULX also remain uncertain. In
the BH formation models considered by us, $M_{\rm BH}$ do not exceed 15\,\msun\
(Fig.~\ref{f:mzams_mbh3}). This does not contradict the estimate of $M_{BH}$ in
P1~NGC~7793 by Motch et al. (2014). Some models of ULX allow BH masses like
30-50\,\msun\ (Ambrosia et al., 2021), although there are no reliable dynamical
determinations of BH masses in ULX as yet. Our work was aimed at a detailed study of
the contribution of various possible ULX progenitors to their total number. It
was limited to the evolution of $Z=Z_\odot$ stars. The consideration of the evolution
of stars of sub-solar chemical composition, which can lead to the formation of
BH of larger mass, is the subject of a separate study that we plan in the
future.

\section*{Acknowledgments}

This study was supported by RFBR grant 19-02-00790. A.G. Kuranov and K.A. Postnov
were supported by Interdisciplinary science and education school of
M.V. Lomonosov Moscow University "Fundamental and applied space studies".
L.R. Yungelson was partially supported by RFBR grant 19-07-01198.

\section*{REFERENCES}
\begin{enumerate}
\item E. Ambrosi, L. Zampieri, F. Pintore, A. Wolter, \mnras, 509, 4694 (2022).
\item P. Atri, J.C.A. Miller-Jones, A. Bahramian et al., \mnras, \textbf{489}, 3116 (2019).
\item M. Bachetti et al., \nat, {\bf 514}, 202 (2014).
\item S. Banerjee, K. Belczynski, C.-L. Fryer, et al., \aap, \textbf{639}, A41 (2020).
\item M.C. Bernadich,  A.~D. Schwope, K. Kovlakas et al., arXiv:2110.14562 (2021).
\item E.R. Beasor, B. Davies, N.  Smith,  \apj, 922, 55 (2021).
\item B. Binder, E.M. Levesque, T.  Dorn-Wallenstein,  \apj, \textbf{863}, 141 (2018).
\item T.A. Callister,  W.M. Farr, M. Renzo, \apj, {\bf 920}, 157  (2021).
\item A.M. Cherepashchuk, Physics Uspekhi, 59, 702 (2016).
\item E.J.M. Colbert, R.F. Mushotzky, ApJ, \textbf{519}, 89 (1999).
\item M. Coriat, R.P. Fender, G. Dubus, MNRAS, \textbf{424}, 1991 (2012).
\item C. de Jager,  H. Nieuwenhuijzen, K.A. van der Hucht, \aaps, \textbf{72}, 259 (1988).
\item M. de Kool, \apj, {\bf 358} , 189 (1990).
\item L. Dessart, A.  Burrows, C.-D. Ott et al., \apj, {\bf 644}, 1063 (2006).
\item G. Dubus, J.-P. Lasota, J.-M. Hameury et al.,  \mnras, 303, 139 (1999).
\item I. El Mellah, J.O. Sundqvist, R. Keppens, \aap, \textbf{622}, L3 (2019).
\item T. Ertl, S.E. Woosley, T. Sukhbold et al., \apj, \textbf{890}, 51  (2020).
\item S. Fabrika \& A.Mescheryakov, A., IAU Symp. 205, p. 268 (2001).
\item S.N. Fabrika, K.E. Atapin, A.S. Vinokurov et al., Astrophys. Bull., \textbf{76}, 6 (2021).
\item E. Fonseca, H.T. Cromartie, T.T. Pennucci et al.,  \apjl, \textbf{915}, L12 (2021).
\item C.L. Fryer, K. Belczynski,  G. Wiktorowicz et al.,   \apj, \textbf{749}, 91 (2012).
\item M. Gallegos-Garcia, C.P.L. Berry, P. Marchant, et al., \apj, 922, 110 (2021).
\item N. Giacobbo, M. Mapelli, \mnras, {\bf 480}, 2011 (2018).
\item J.-M. Hameury, J.-P. Lasota, 2020, \aap, {\bf 643}, A171 (2020).
\item R. Hirai \& I. Mandel,  \pasa, 38, e056 (2021).
\item G. Hobbs et al., \mnras, {\bf 360}, 974 (2005).
\item J. Hurley et al. \mnras, {\bf 329}, 897 (2002).
\item G.-L. Israel et al., \mnras, {\bf 466}, L48 (2017).
\item N. Ivanova, S. Justham,  X. Chen et al.\, \aapr, \textbf{21}, 59 (2013).
\item N. Ivanova, S. Justham, P. Ricker,  AAS-IOP Astronomy Book Series, IOP Publishing, Online ISBN: 978-0-7503-1563-0, Print ISBN: 978-0-7503-1561-6 (2020).
\item P. Kaaret, H. Feng, T. Roberts  et al., \araa, {\bf 55}, 303 (2017).
\item A. King et al., \apjl, {\bf 552}, L109 (2001).
\item А.R. King, \mnras,  {\bf 393}, L41 (2009).
\item R. Kippenhahn \& A. Weigert, \zap, {\bf 65}, 251 (1967a).
\item R. Kippenhahn,  K. Kohl \& A. Weigert, \zap, {\bf 66}, 58 (1967b).
\item J. Klencki, G. Nelemans, A.G. Istrate et al.,  \aap, {\bf 638}, A55 (2020).  
\item J. Klencki, G. Nelemans, A.G. Istrate et al. \aap, {\bf  645}, A54 (2021а).
\item J. Klencki,  A.G. Istrate, G. Nelemans et al.,  arXiv:2111.10271 (2021b).                                
\item A.G. Kuranov, K.A. Postnov  \& L.R. Yungelson, Astron. Lett., 46, 658 (2020, Paper I).
\item E. Laplace, S. Justham, M.  Renzo et al., arXiv:2102.05036 (2021).
\item K.M. L{\'o}pez, M. Heida,  P.~G. Jonker et al., \mnras, \textbf{497}, 917 (2020).
\item A.-J. Loveridge et al., \apj, {\bf 743}, 49, (2011).
\item E. Lovegrove, S.E. Woosley,  \apj, \textbf{769}, 109 (2013).
\item M. MacLeod, E.C. Ostriker, R. Stone, \apj,  {\bf 863}, 5 (2018).
\item P. Marchant et al., \aap, {\bf 604}, A55, (2017).
\item A. Miglio, C. Chiappini,  J.T. Mackereth, et al., \aap, \textbf{645}, A85 (2021).
\item C. Motch et al., \nat, {\bf 514} 198 (2014).
\item A.A. Mushtukov et. al., \mnras,  {\bf 501}, 2424 (2021).
\item D.K. Nadezhin, \apss, {\bf 69}, 115 (1980).
\item T. Nugis \& H.J.G.L.M. Lamers, \aap, {\bf 360}, 227 (2000).
\item B. Paxton et al., \apjs, {\bf 192}, 3 (2011).
\item M. Plavec,  R.K., Ulrich  \& R.S. Polidan,  \pasp, {\bf 85}, 769 (1973).
\item A.J.T. Poelarends, S. Wurtz, J. Tarka et al., \apj, \textbf{850}, 197 (2017).
\item C.A. Raithel, T. Sukhbold \& F. {\"O}zel), \apj,  {\bf 856},  35  (2018).
\item N.I. Shakura, R.A.~Sunyaev,  \aap, {\bf 24 }, 337 (1973).
\item Y. Shao, X.-D. Li, \apj, {\bf 802}, 131 (2015).
\item Y. Shao  et al., \apj, {\bf 886}, 118 (2019).
\item S.J. Smartt,  \pasa, \textbf{ 32},   e016 (2015).
\item G.E. Soberman et al., \aap, {\bf 327}, 620 (1997).
\item S.S. Tsygankov,  A.A. Mushtukov, V.F. Suleimanov et al.,  \mnras, \textbf{457}, 1101 (2016).
\item A. Tutukov, L. Yungelson \& A. Klayman, 1973, Nauchnye Informatsii, 27, 3 (1973).
\item J.S. Vink, arXiv:2109.08164 (2021).
\item J.S. Vink, A. de Koter \&  H.J.G.L.M. Lamers, \aap, {\bf 362}, 295 (2000).
\item J. Vink et al., \aap, {\bf 369}, 574 (2001).
\item M. Volonteri, M. Habouzit M., M. Colpi, Nature Reviews Physics, \textbf{3}, 732 (2021).
\item D.J.~Walton et al., \apjl, {\bf 857}, L3 (2018).
\item D.J. Walton, A.D.A. Mackenzie, H. Gully et al.,  \mnras, 509, 1587 (2022).
\item R.F. Webbink, \apj, {\bf 277}, 355 (1984).
\item N.E. White, J. van Paradijs, \apjl, \textbf{473}, L25 (1996).
\item G. Wiktorowicz et al., \apj, {\bf 846}, 17 (2017).
\item G. Wiktorowicz et al.,  \apj, {\bf 875}, 53 (2019).
\item G. Wiktorowicz, J.-P. Lasota, K. Belczynski et al., \apj,  {\bf 918},  60 (2021).
\end{enumerate}

\newpage
\section*{Appendix: examples of evolutionary tracks}
\begin{figure}[hb!]
\includegraphics[width=0.5\textwidth]{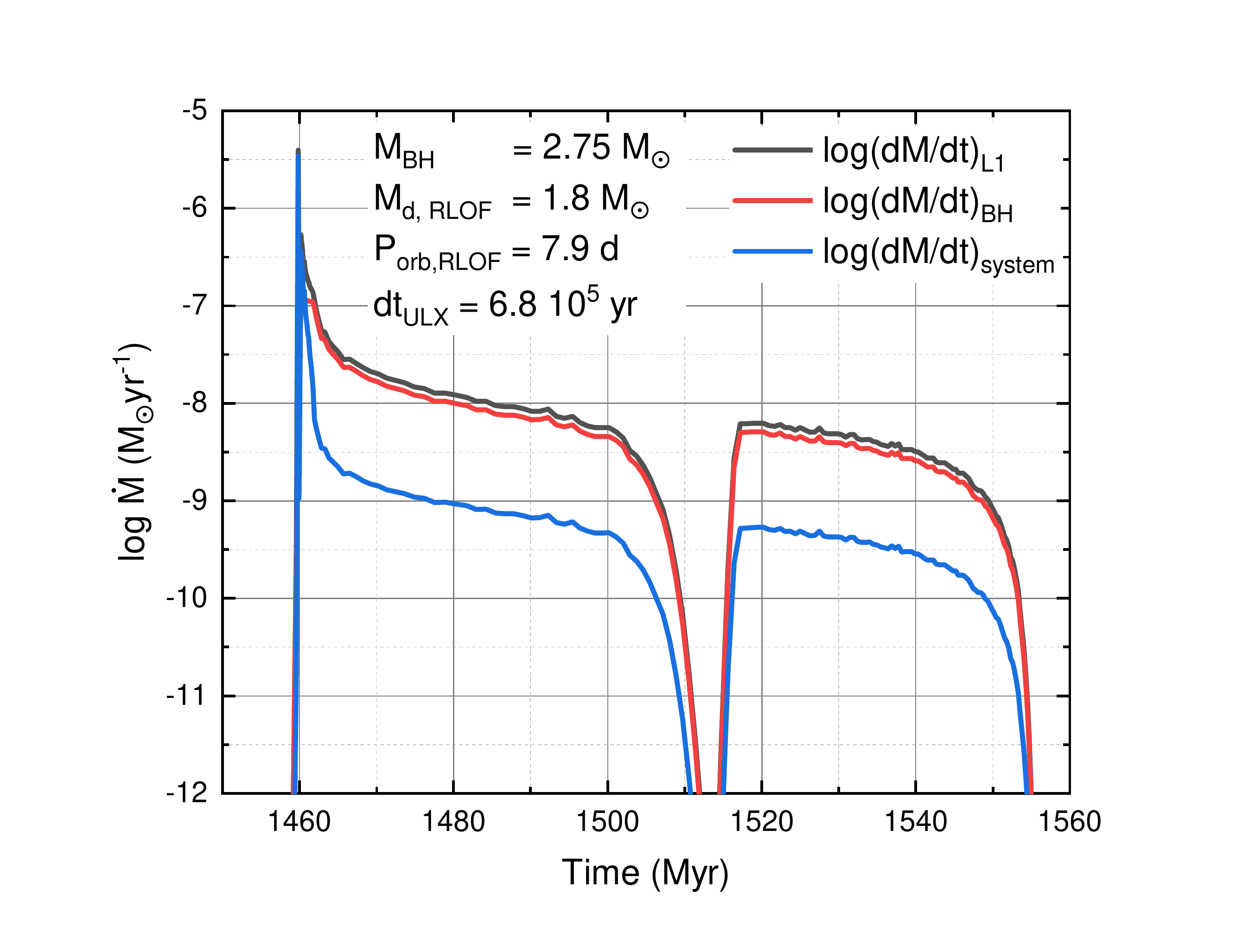}
\includegraphics[width=0.5\textwidth]{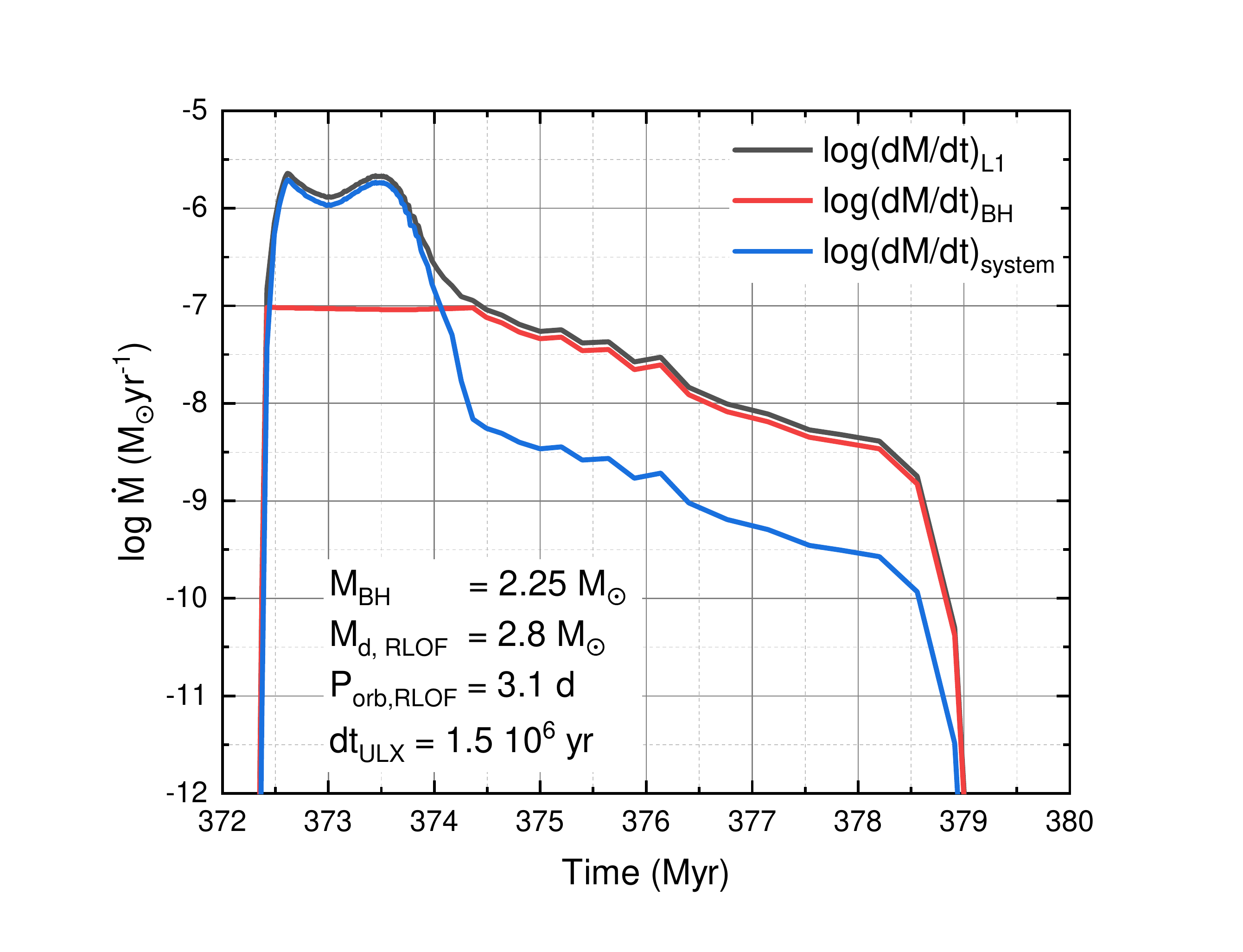}
\includegraphics[width=0.5\textwidth]{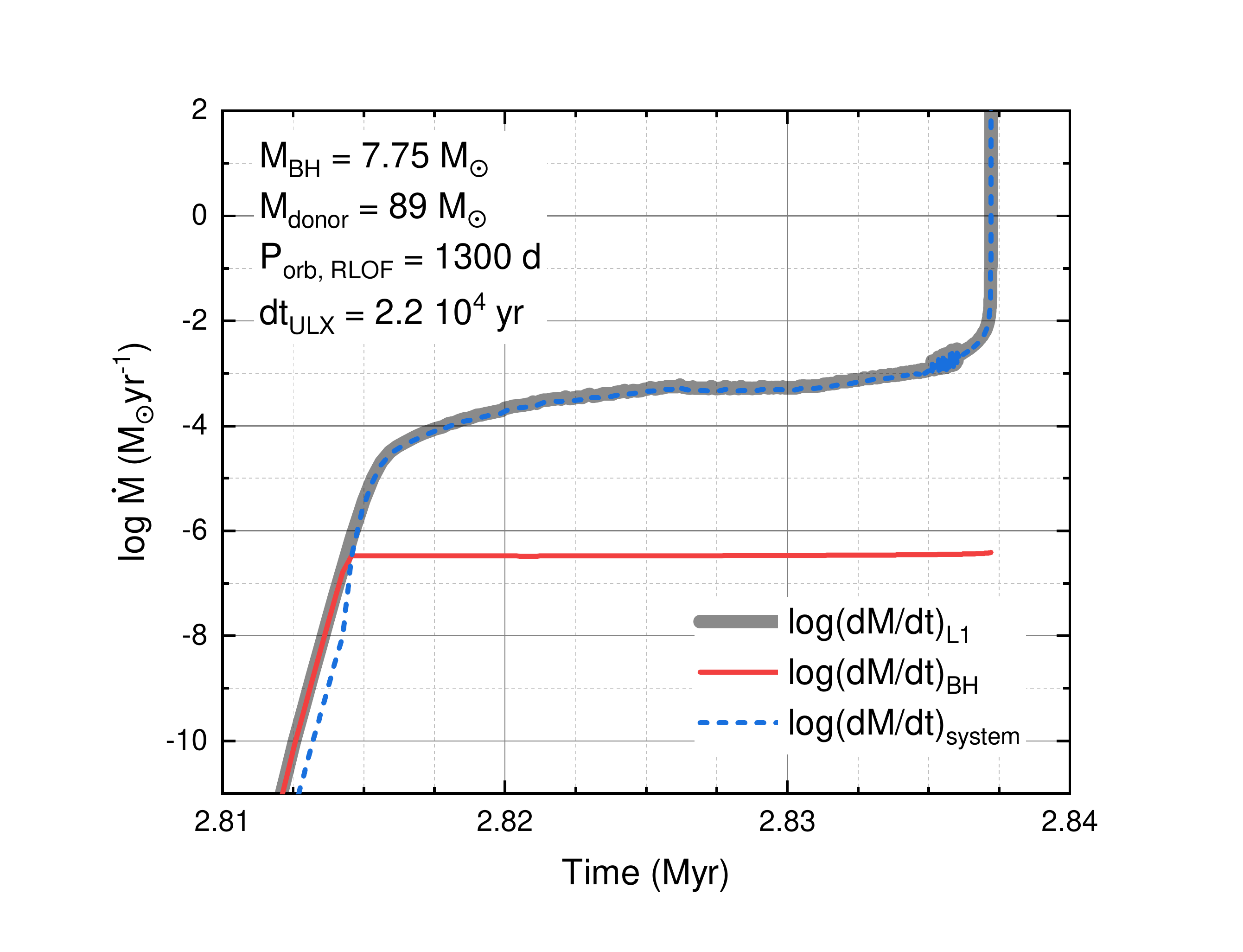}
\caption{\small
Rates of mass-loss by the donor and accretion
in the CBS with
$ M_\mathrm{BH}=2.75$\,\msun, $M_\mathrm{d}=1.8$\,\msun, $P_\mathrm{RLOF}$=7.9\,day (top left panel),
$ M_\mathrm{BH}$=2.25\,\msun, $M_\mathrm{d}$=2.8\,\msun, $P_\mathrm {RLOF}$=3.1\,day (lower left panel),
$ M_\mathrm{BH}=7.5$\,\msun, $M_\mathrm{d}=89$\,\msun, $P_\mathrm{RLOF}$=1300\,day (bottom right panel).
Black, red, and blue lines correspond to $\dot M_\mathrm{L_1}$), the rate of disk accretion onto
the BH $\dot M_\mathrm{BH}$ and the rate of mass loss from the system are taken equal to
0.1$\dot M_\mathrm {L_1} $. The binaries are ULX if
$ M_\mathrm{d}{>}M_\mathrm{Edd}$.
In the first case, the donor fills the Roche lobe at the stage of hydrogen-shell burning stage
on the RG branch.  The loss of matter is briefly interrupted when the radius of the donor surface
becomes less than the radius of maximum penetration of the convective shell during the previous
evolution (Kippenhahn et al. 1967).
In the second case, RLOF also occurs in the stage of the hydrogen-shell burning stage, but at an earlier phase
than in the first case, and the ULX stage lasts longer due to the lower initial donor mass.
In the case of a massive super-giant donor in a wide system (bottom  panel),
the ULX stage precedes the formation of a common envelope.
}
\label{fig:trc_2.5}
\end{figure}

\begin{figure}[]
\includegraphics[width=0.5\textwidth]{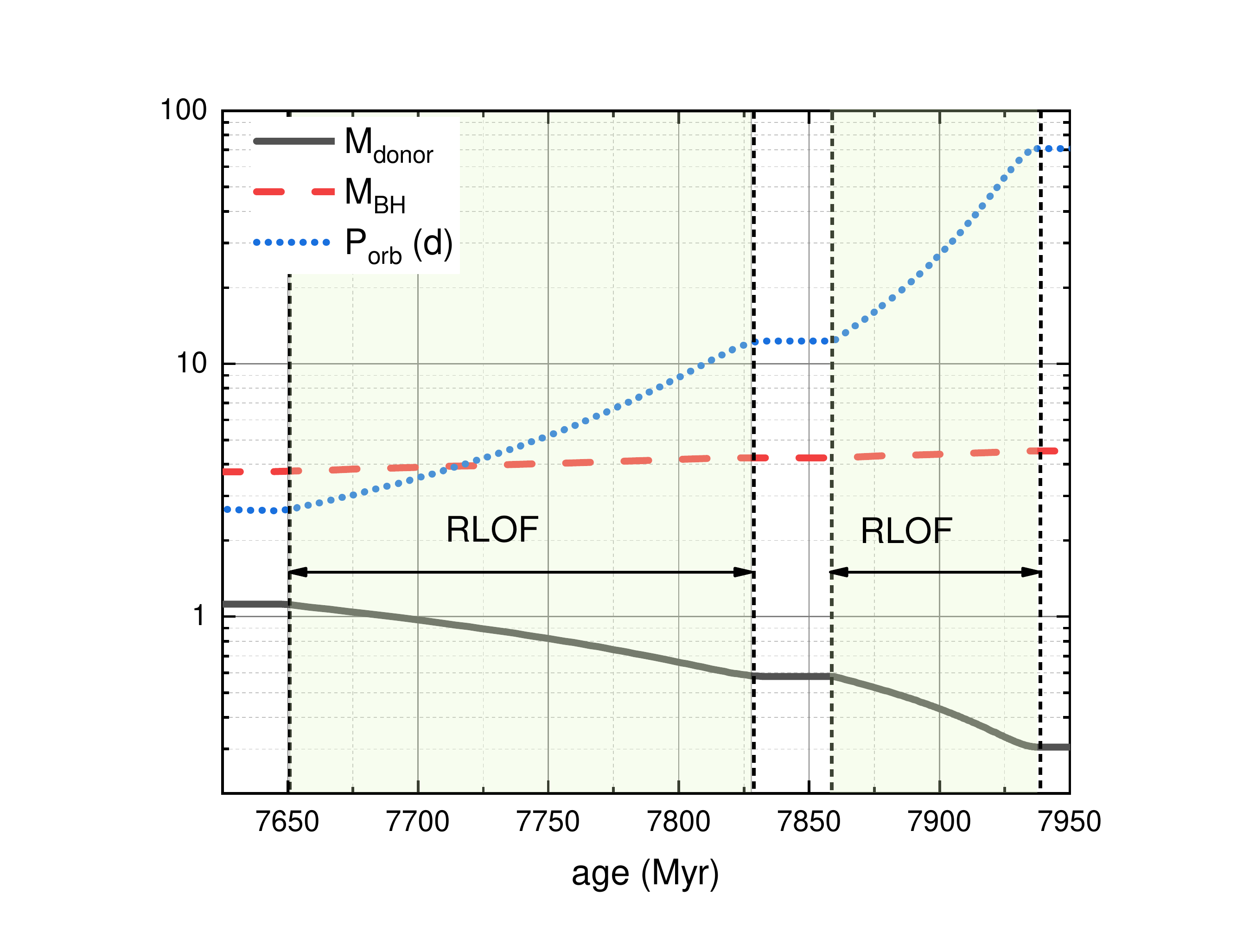}
\includegraphics[width=0.5\textwidth]{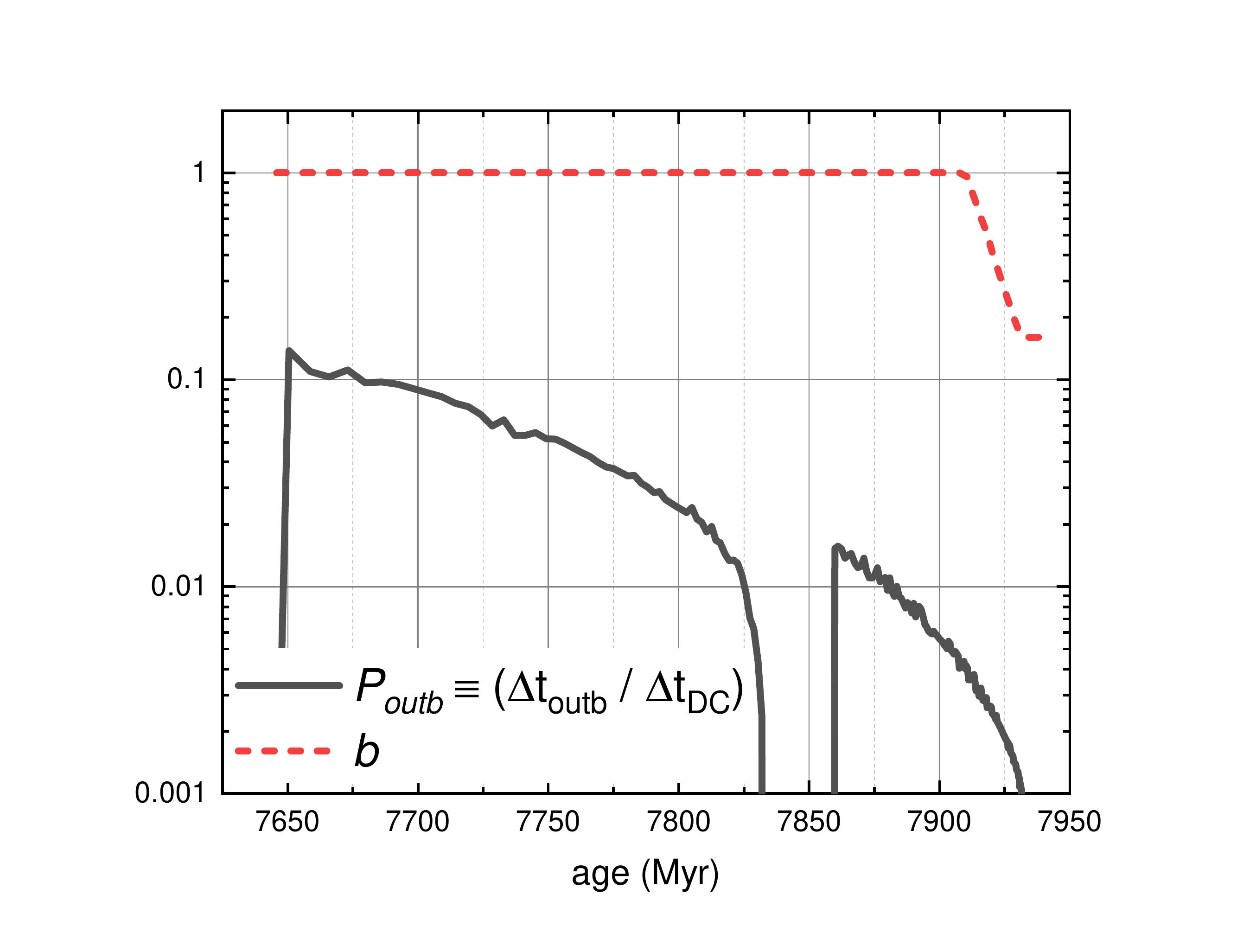}
\includegraphics[width=0.5\textwidth]{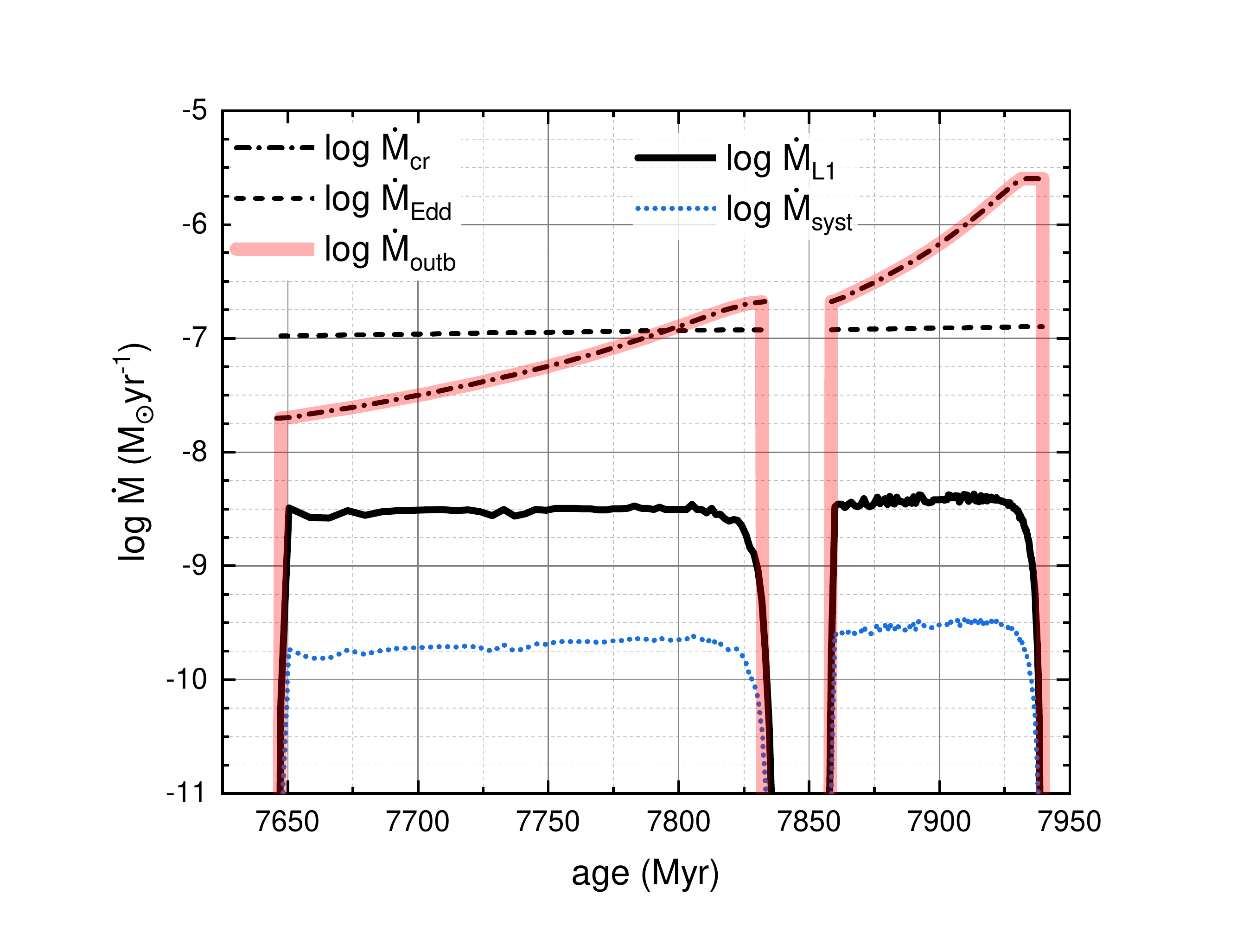}
\includegraphics[width=0.5\textwidth]{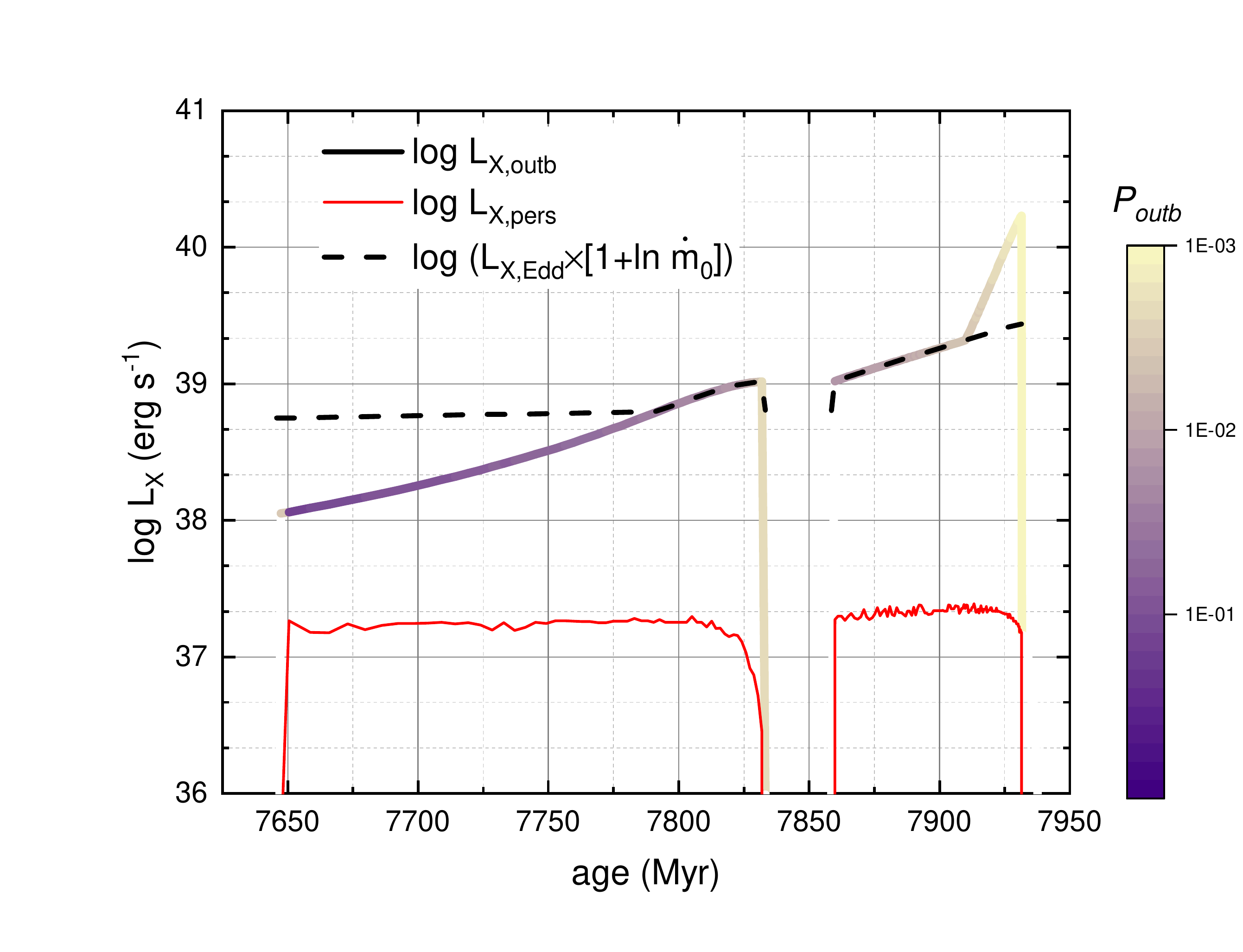}
\caption{
\small An example of variation of the parameters of a binary
system that has a stage of a transient
ULX. Top left panel: dependence of masses of components and orbital
period of the binary system on time.
Top right panel: variation of the beaming factor and
probability of detection of a transient source in the active state. The latter
is determined by the ratio of the time spent by the source in the outburst
$\Delta t_{outb}$ and the time of the duty cycle:
$P_{outb}=\Delta t_{outb}/\Delta t_{DC}$. Bottom left panel: dependence of the rate of mass flow via  L$_1$
and the rate of mass loss by the binary system $\dot M_\mathrm{syst}$ on time.
The dash-dotted line shows the critical value of the accretion rate
$ \dot M_{cr}$, below which the source is considered to be transient. Dotted line
shows Eddington accretion rate $\dot M_\mathrm{Edd}$.
Solid line --- accretion rate
onto c.o. during outburst ($\dot M_{outb} = \dot M_{cr}$).
Bottom right panel:
dependence of the X-ray luminosity during the outburst ($L_{X, outb}$) on time.
The line is colored by the probability of detecting of an active source.
Also shown are the X-ray luminosity $L_{X, pers}$ corresponding to the case
of a stable disk accretion at the same rate of mass loss by the donor and
Eddington luminosity $L_{X, Edd}$ with account for logarithmic multiplier
$(1+\ln \dot m_0)$, where $\dot m_0=\dot M_x/\dot M_{\rm Edd}$
(solid thin line and dashed line,
respectively). 
}
\label{fig:trc_old_age}
\end{figure}
Figure~\ref{fig:trc_2.5} shows examples of computed mass-loss
rates for the systems with different $M_{\rm BH}$, $M_d$, $P_{\rm orb}$
in which the evolution leads
to the formation of a ULX.
Figure \ref{fig:trc_old_age} shows an example of evolution of a system with
the initial BH mass $M_\mathrm{BH}=3.8$\,\msun\ and the visual donor star with
$M_\mathrm{d}=1.1$\,\msun,
$ P_\mathrm{orb}=2.7$\,day, in which a transient ULX is formed in case B
of mass exchange.
\end{document}